\documentclass[graybox,envcountchap,sectrefs,natbib]{svmono}
\usepackage{mathptmx} 
\usepackage{helvet}       
\usepackage{courier}    
\usepackage{type1cm}  
\usepackage{amsfonts}
\usepackage{amssymb}
\usepackage{xspace}
\usepackage{chapterbib}
\usepackage{makeidx}         
\usepackage{multicol}        
\usepackage[bottom]{footmisc}
\usepackage{graphicx}        

\begin{document}
\mainmatter
\setcounter{page}{0} 
\setcounter{chapter}{2} %

%
%
\let\jnl=\rmfamily
\def\refe@jnl#1{{\jnl#1}}%

\newcommand\aj{\refe@jnl{AJ}}%
\newcommand\actaa{\refe@jnl{Acta Astron.}}%
\newcommand\araa{\refe@jnl{ARA\&A}}%
\newcommand\apj{\refe@jnl{ApJ}}%
\newcommand\apjl{\refe@jnl{ApJ}}%
\newcommand\apjs{\refe@jnl{ApJS}}%
\newcommand\ao{\refe@jnl{Appl.~Opt.}}%
\newcommand\apss{\refe@jnl{Ap\&SS}}%
\newcommand\aap{\refe@jnl{A\&A}}%
\newcommand\aapr{\refe@jnl{A\&A~Rev.}}%
\newcommand\aaps{\refe@jnl{A\&AS}}%
\newcommand\azh{\refe@jnl{AZh}}%
\newcommand\gca{\refe@jnl{GeoCh.Act}}%
\newcommand\grl{\refe@jnl{Geo.Res.Lett.}}%
\newcommand\jgr{\refe@jnl{J.Geoph.Res.}}%
\newcommand\memras{\refe@jnl{MmRAS}}%
\newcommand\jrasc{\refe@jnl{J.RoySocCan}}%
\newcommand\mnras{\refe@jnl{MNRAS}}%
\newcommand\na{\refe@jnl{New A}}%
\newcommand\nar{\refe@jnl{New A Rev.}}%
\newcommand\pra{\refe@jnl{Phys.~Rev.~A}}%
\newcommand\prb{\refe@jnl{Phys.~Rev.~B}}%
\newcommand\prc{\refe@jnl{Phys.~Rev.~C}}%
\newcommand\prd{\refe@jnl{Phys.~Rev.~D}}%
\newcommand\pre{\refe@jnl{Phys.~Rev.~E}}%
\newcommand\prl{\refe@jnl{Phys.~Rev.~Lett.}}%
\newcommand\pasa{\refe@jnl{PASA}}%
\newcommand\pasp{\refe@jnl{PASP}}%
\newcommand\pasj{\refe@jnl{PASJ}}%
\newcommand\skytel{\refe@jnl{S\&T}}%
\newcommand\solphys{\refe@jnl{Sol.~Phys.}}%
\newcommand\sovast{\refe@jnl{Soviet~Ast.}}%
\newcommand\ssr{\refe@jnl{Space~Sci.~Rev.}}%
\newcommand\nat{\refe@jnl{Nature}}%
\newcommand\iaucirc{\refe@jnl{IAU~Circ.}}%
\newcommand\aplett{\refe@jnl{Astrophys.~Lett.}}%
\newcommand\apspr{\refe@jnl{Astrophys.~Space~Phys.~Res.}}%
\newcommand\nphysa{\refe@jnl{Nucl.~Phys.~A}}%
\newcommand\physrep{\refe@jnl{Phys.~Rep.}}%
\newcommand\procspie{\refe@jnl{Proc.~SPIE}}%

\newcommand{\Al}{$^{26}$Al\xspace}
\newcommand{\al}{$^{26}$Al\xspace}
\newcommand{\Be}{$^{7}$Be\xspace}
\newcommand{\be}{$^{7}$Be\xspace}
\newcommand{\bem}{$^{10}$Be\xspace}
\newcommand{\ca}{$^{44}$Ca\xspace}
\newcommand{\Ca}{$^{44}$Ca\xspace}
\newcommand{\cam}{$^{41}$Ca\xspace}
\newcommand{\Co}{$^{56}$Co\xspace}
\newcommand{\co}{$^{56}$Co\xspace}
\newcommand{\csm}{$^{135}$Cs\xspace}
\newcommand{\ct}{$^{13}$C\xspace}
\newcommand{\ci}{$^{57}$Co\xspace}
\newcommand{\Ci}{$^{57}$Co\xspace}
\newcommand{\ch}{$^{60}$Co\xspace}
\newcommand{\Ch}{$^{60}$Co\xspace}
\newcommand{\Cl}{$^{36}$Cl\xspace}
\newcommand{\li}{$^{7}$Li\xspace}
\newcommand{\Li}{$^{7}$Li\xspace}
\newcommand{\Fe}{$^{60}$Fe\xspace}
\newcommand{\fh}{$^{60}$Fe\xspace}
\newcommand{\fe}{$^{56}$Fe\xspace}
\newcommand{\Fr}{$^{57}$Fe\xspace}
\newcommand{\fr}{$^{57}$Fe\xspace}
\newcommand{\mg}{$^{26}$Mg\xspace}
\newcommand{\Mg}{$^{26}$Mg\xspace}
\newcommand{\mn}{$^{54}$Mn\xspace}
\newcommand{\Na}{$^{22}$Na\xspace}
\newcommand{\Ne}{$^{22}$Ne\xspace}
\newcommand{\Ni}{$^{56}$Ni\xspace}
\newcommand{\nh}{$^{60}$Ni\xspace}
\newcommand{\Nh}{$^{60}$Ni\xspace}
\newcommand\nuk[2]{$\rm ^{\rm #2} #1$}  
\newcommand{\pd}{$^{107}$Pd\xspace}
\newcommand{\pb}{$^{205}$Pb\xspace}
\newcommand{\tc}{$^{99}$Tc\xspace}
\newcommand{\Sc}{$^{44}$Sc\xspace}
\newcommand{\Ti}{$^{44}$Ti\xspace}
\newcommand{\ti}{$^{44}$Ti\xspace}
\def\aa{$\alpha$}
\newcommand{\about}{$\simeq$}
\newcommand{\cms}{cm\ensuremath{^{-2}} s\ensuremath{^{-1}}\xspace}
\newcommand{\degree}{$^{\circ}$}
\newcommand{\flux}{ph~cm\ensuremath{^{-2}} s\ensuremath{^{-1}}\xspace}
\newcommand{\fluxrad}{ph~cm$^{-2}$s$^{-1}$rad$^{-1}$\ }
\newcommand{\ga}{\ensuremath{\gamma}}
\newcommand{\gam}{\ensuremath{\gamma}}
\def\nn{$\nu$}
\def\ra{$\rightarrow$}
\newcommand{\Msol}{M\ensuremath{_\odot}\xspace}
\newcommand{\msol}{M\ensuremath{_\odot}\xspace}
\newcommand{\Msolppc}{M\ensuremath{_\odot} pc\ensuremath{^{-2}}{\xspace}}
\newcommand{\Msolpy}{M\ensuremath{_\odot} y\ensuremath{^{-1}}{\xspace}}
\newcommand{\msb}{M\ensuremath{_\odot}\xspace}
\newcommand{\Msun}{M\ensuremath{_\odot}\xspace}
\newcommand{\Rsun}{R\ensuremath{_\odot}\xspace}
\newcommand{\rsun}{R\ensuremath{_\odot}\xspace}
\newcommand{\Lsun}{L\ensuremath{_\odot}\xspace}
\newcommand{\lsun}{L\ensuremath{_\odot}\xspace}
\newcommand{\solar}{\ensuremath{_\odot}\xspace}
\newcommand{\zs}{Z\ensuremath{_\odot}\xspace}





%
%
%

\chapauthor{Maria Lugaro\footnote{Monash University, Victoria 3800, Australia} and Alessandro Chieffi\footnote{I.N.A.F., 00133 Roma, Italy}}
\chapter{Radioactivities in Low- and Intermediate-Mass Stars}
\label{Ch:stars}
\chaptermark{radioactivity in low/intermediate-mass stars}

Energy in stars is provided by nuclear reactions, which, in many
cases, produce radioactive nuclei. When stable nuclei are irradiated by a
flux of protons or neutrons, capture reactions push stable matter out of
stability into the regime of unstable species. The ongoing production of radioactive
nuclei in the deep interior of the Sun via proton-capture reactions is
recorded by neutrinos emitted during radioactive decay. These neutrinos
escape the inner region of the Sun and can be detected on Earth.
Radioactive nuclei that have relatively long half lives may also be detected in
stars via spectroscopic observations and in stardust recovered from
primitive meteorites via laboratory analysis. The vast majority of these
stardust grains \index{stardust}  originated from Asymptotic Giant Branch \index{stars!AGB} (AGB) stars. 
This is the final phase in the evolution of stars initially less massive than $\simeq$10~\Msun, during which nuclear energy is produced by alternate hydrogen and helium burning in shells above the core. The long-lived 
radioactive nucleus $^{26}$Al is produced in AGB stars by proton captures  at relatively high temperatures, above 60~MK. Efficient production of $^{26}$Al occurs in massive AGB stars ($>4:5$\Msun), where the base of the convective envelope reaches such temperatures. Several other long-lived radioactive nuclei, including $^{60}$Fe, $^{87}$Rb, and $^{99}$Tc, are produced in AGB stars when matter is exposed to a significant neutron flux leading to the synthesis of elements heavier than iron. Here, neutron captures occur on a timescale that is typically slower than $\beta$-decay timescales, resulting in a process known as $slow$ neutron captures 
(the $s$-process). However, when radioactive nuclei with half lives greater than a few days are produced, depending on the temperature and the neutron density, they may either decay or capture a neutron, thus branching up the path of neutron captures and defining the final $s$-process abundance distribution. The effect of these  \emph{branching points} is observable in the composition of AGB stars and stardust. This nucleosynthesis in AGB stars could produce some long-living radioactive nuclei in relative abundances that resemble those observed in the early solar system.

\section{The Missing Element}

The element with 43 protons in its nucleus, lying between molybdenum and
ruthenium, was known for a long time as the \emph{missing} element. Since
the 19$^{\rm th}$ century there had been many unsuccessful attempts at its discovery.
Finally, in 1937 Italian physicist Emilio Segr\'e and \index{Segr\`e, E.} 
chemist Carlo Perrier \index{Perrier, C.} found two isotopes of the missing element through
measurements of radioactivity from discarded cyclotron parts: they
observed several decay periods and proved they were occurring at Z=43.
Hence the missing element did not exist in nature because of its
instability against nuclear decay. The discoverers named the missing element 
technetium
(Tc), from $\tau\epsilon\chi\nu\eta\tau\acute{\omicron}\varsigma$, which
means artificial in Greek, since it was the first element produced
artificially.  Fifteen years later, it was shown that Tc is not only
made by men but also by stars. In 1952, astronomer Paul Merrill \index{Merrill, P.} observed
the absorption lines corresponding to the atomic structure of
Tc in the spectra of several giant stars.  Merrill was at first cautious about this
result. To start with, the element he identified did not even exist on
Earth. Second, up to then it was assumed, and not proved wrong, that
all stars had the same chemical composition. This was in agreement with
the accepted theory of the time that the elements were all produced
during the Big Bang and their abundances in the Universe were not
modified by any further process. Merrill's discovery in that
respect was truly revolutionary: given the relatively short half lives
of the Tc isotopes \index{isotopes!98Tc} (a few Myrs at most), the Tc lines were the first
indisputable demonstration that this radioactive element was made {\it in situ} in the stars where it was observed. This finding brought a
radical change in the way we understand the origin of the elements, and
the theory of stellar nucleosynthesis introduced in Chapter 2 began to
take shape and garnered authority. In this chapter we discuss the life of those
stars that, like our Sun, evolve twice through Red Giant stages. We describe
how they produce long lived radioactive nuclei, like Tc, in their
interiors, how the signature of such radioactivity is carried outside
the star, and how it can be observed.

\section{The Production of Radioactive Nuclei in Stellar Interiors}
\label{sec:basic}

In Subsection \ref{subsec:hburn} we first derive the four basic equations that
control the quasi-equilibrium configuration of a self-gravitating gas sphere,
namely, the hydrostatic equilibrium equation (that describes the balance 
between the pressure
gradient and gravity) and the energy transport equation (due to photons
and/or convection), plus the two associated continuity equations for  
mass and energy flux. Then, we
show that energy losses, which occur mainly from the stellar surface in 
stars of mass less than $\sim$10 \Msun, force the gas to contract 
and to heat, in accordance with the
virial theorem. The progressive increase of the central temperature
allows the activation of nuclear processes and we describe two
sequences that convert protons into $^4$He nuclei ($\alpha$ 
particles): the PP \index{process!pp} \index{process!CNO}
chain and the CNO cycle. Since proton capture 
inevitably pushes matter out of the stability, 
both these sequences
produce radioactive nuclei that decay by emitting neutrinos.

In Subsection~\ref{subsec:sun} we briefly describe the quest for solar
neutrinos and the various experiments that eventually allowed 
the demonstration that the lower than predicted neutrino flux from the Sun (the
so-called Solar Neutrino Problem) is the consequence of 
neutrino oscillations among their three different flavors.

\subsection{The Stellar Energy Source and Radioactive Isotopes}
\label{subsec:hburn}

A star is, in first approximation, a spherically symmetric, gaseous \index{stars!structure}
cloud contracting under its own gravity and progressively heating up
while losing energy from its surface in the form of photons. A 
strong temperature gradient, with the temperature decreasing from the 
centre to the surface, pushes the photon flux outward until the \emph{mean free 
path}\footnote{The average distance a particle travels between collisions.}
of the photons eventually becomes larger than their distance from the 
surface, allowing their escape. The Virial theorem links the energy 
gained by the gravitational field $\Delta\Omega$ to that absorbed by the 
gas $\Delta U$: $\Delta U=-{\Delta\Omega}/{3(\gamma -1)}$, where 
$\gamma$ is the ratio between two specific heats, that at 
constant pressure and that 
at constant volume, of the contracting gas. A stable quasi-equilibrium 
configuration exists for such a structure provided that $\gamma >$ 4/3. In 
this case, a fraction of the energy gained by the gravitational field must 
be liberated from the structure before the gas cloud can contract further. 
In the case of a perfect gas ($\gamma=5/3$) we obtain the classical result 
$\Delta U=-(1/2)\Delta\Omega$, stating that half of the gravitational 
energy liberated by each infinitesimal contraction is absorbed by the gas 
and half must be lost before an additional contraction can occur. The 
timescale over which the energy is lost from the system drives the timescale of 
contraction and keeps the structure in a quasi-equilibrium 
configuration. If, instead, $\gamma$ drops to 4/3, all the gravitational 
energy is absorbed by the gas and no time delay is required before a new 
contraction can occur. This is an unstable situation leading to  
collapse. In the evolutionary phases we discuss in this chapter 
$\gamma$ 
remains well above 4/3 and hence a stable quasi-equilibrium configuration 
is always assured.

The balancing forces required to maintain a stable stellar 
quasi-equilibrium configuration are due to pressure gradients and 
gravity. So, the first main equation of stellar structure describes the
equilibrium between these two forces, at any given distance from the
center of the star $r$:
\begin{equation}\label{hydroeq}  
$$$dP/dr=-GM\rho/r^{2}$$$
\end{equation}
where $P$ is the pressure, $G$ the gravitational constant, $M$ the cumulative mass inside $r$, and $\rho$ the density. Associated to this equation is a continuity equation for mass: 
\begin{equation}\label{masscon}
$$$dM/dr=4\pi r^{2}\rho .$$$
\end{equation}
By assuming, to zero order, that $\rho$ is constant within the star, the integration of eq. \ref{masscon} implies that:
\begin{equation}\label{rela0}
$$$\rho\propto~M/R^3$$$.
\end{equation}
Since the pressure at the surface of the star is much lower than
where the radius approaches zero, the center of the star, the equation of hydrostatic equilibrium, \ref{hydroeq}, basically says that 
\begin{equation}\label{rela1}
$$$P_c~\propto~M\rho/R_s$$$
\end{equation}
where $P_c$ is the central pressure and $R_s$ the stellar radius.
By inserting the relation \ref{rela0} into \ref{rela1} one obtains: 
\begin{equation}\label{rela2}
$$$P_c~\propto~M^2/{R_s}^4$$$.
\end{equation}
If the equation of state is  that of a perfect gas, i.e. $P\propto~\rho~T/\mu$, relation \ref{rela2} becomes:
\begin{equation}\label{rela3}
$$$T_c~\propto~\mu~M/R$$$
\end{equation}  
 where $T$ is the temperature and $\mu$ is the mean molecular
weight.
This equation provides an important basic relationship among central temperature ($T_c$), mass, and radius of a star, which only relies on the assumption of hydrostatic equilibrium.

The second major equation describing the structure of the quasi-equilibrium
configuration of a star determines the energy flux through the
structure. In stationary situations the energy is transported by
photons or electrons and application of the first Fick's law leads to
the well known equation: 
\begin{equation}\label{tflux}
$$$dT/dr=-3\kappa \rho L/(16ac\pi r^{2}T^{3})$$$
\end{equation}
where $\kappa$ is the opacity coefficient
representing the impenetrability of a gas to light, $L$ the luminosity
representing the amount of energy radiated per unit time, $a$ 
the radiation constant, and $c$ the speed of light. Furthermore, in
this case a continuity equation
\begin{equation}\label{econs}
$$$dL/dr=4\pi r^{2}\rho\epsilon$$$ 
\end{equation}
controls the conservation of energy, where $\epsilon$ represents the net
local energy budget, i.e., the sum of the nuclear energy production rate
$\epsilon_{\rm nuc}$, the neutrino energy loss rate $\epsilon_{\nu}$,
and the gravitational energy rate $\epsilon_{\rm g}$. Since the 
central temperature $T_c$ is much higher than the surface temperature, it is possible to obtain a basic relationship between central temperature, mass, luminosity  and radius of a star, i.e.:
 \begin{equation}\label{rela4}
$$${T_c}^4 \propto M L / R^4$$$. 
\end{equation}
By combining this relation with the previous relation \ref{rela3}, derived from hydrostatic equilibrium \ref{hydroeq}, one eventually obtains the fundamental relation between mass and luminosity:
\begin{equation}\label{mlrel}
$$$\rm L \propto \mu^4 M^3$$$. 
\end{equation}
Frequently, the energy produced locally cannot be transported quickly enough by radiation or conduction, and interior shells formally in an
equilibrium condition can become unstable in the sense that a displacement
from their equilibrium position is not fully counteracted by a restoring force.
Instead, matter is accelerated even further from its original
position and large scale motions of matter (\emph{convection}) is established.
Under these conditions energy is predominantly transported by
buoyancy-driven motions of bulk material due to their much larger mean 
free path with
respect to that of photons. A temperature gradient quite different
from that described by equation \ref{tflux} must then be used in these
regions. A direct consequence of these large scale motions is
that matter is mixed throughout an unstable region.

The system formed by the two basic equations related to hydrostatic 
equilibrium and energy transport plus the two associated continuity 
equations and the equation of state (supplemented by an opacity 
coefficient $\kappa=f'(\rho,T,{\rm chem.comp.})$ 
and a total energy generation coefficient $\epsilon=f(\rho,T,{\rm chem.comp.})$) constitutes the basic set of 
equations that describes the 
internal structure of a quasi-equilibrium (non rotating) stellar 
configuration at a given time. The temporal evolution of such a 
structure is determined by the rate at which energy is lost to the surroundings: 
the faster the energy is lost, the faster the structure evolves. 
Typical stellar luminosities  
range between $\sim 4\times10^{33}$ erg/s for a star of 1 \Msun\ and $\sim 
4\times10^{36}$ erg/s for a star of 6 \Msun\ and the associated lifetimes 
can be estimated by dividing the total amount of 
available energy by the loss rate L, i.e. the stellar luminosity.
If the only energy source was the gravitational field, the lifetime of a 
contracting gas cloud would be of the order of a few tens of millions of 
years (the Kelvin-Helmholtz timescale). Instead, as was known since 
the 1920s from radioactive dating of terrestrial rocks, that 
the age of the Earth is several Gyrs, much longer than the 
Kelvin-Helmholtz timescale. Thus, the Sun must be powered by different means. 
The lifetime of most stars is much larger than permitted 
by their gravitational energy reservoir alone. Instead of simple contraction,  
energy losses are replaced by the activation of nuclear 
fusion reactions among charged nuclear particles near the stellar core.

The efficiency of nuclear reactions, i.e., their \emph{rate}, depends on the 
abundances of the reactant nuclei and the cross section of each 
reaction averaged over a Maxwellian distribution of relative velocities 
between the target and the projectile nuclei.
For charged particle reactions the rate is mainly controlled by 
the Coulomb barrier generated by the number of protons in a nucleus.  
The nuclear reactions that activate at the lowest temperatures are 
those involving capture of protons, i.e., the nucleus of the lightest and most 
abundant element: hydrogen (H). 
Nature, however, does not allow the build up of stable nuclei made only of 
protons because in order to glue nucleons (i.e., proton and neutrons) 
together in a nucleus via the strong nuclear force, the repulsive 
electromagnetic force acting between protons needs to be diluted with a 
certain number of neutrons. The distribution of stable nuclei in the 
[N=number of neutrons, Z=number of protons] plane, the \emph{chart of 
nuclides}, clearly shows that the region where nuclei are stable
lies close to the N=Z line (the \emph{valley of $\beta$ stability}) up to the 
element Ca, and 
then bends slightly towards the neutron-rich side as the repulsion between 
higher number of protons needs to be diluted with more and 
more neutrons. Nuclei outside this valley are radioactive, i.e., 
unstable, and decay towards their closest stable daughter through 
$\beta$-decay weak interaction reactions (as described in Chapters~1, 2, 4, and 9). 
It follows that the build up of progressively heavier nuclei through 
the addition of protons naturally pushes the matter out of the stability 
valley, producing radioactive nuclei that decay back towards stability via 
$\beta^+$ decay.

A detailed analysis of the nuclear reactions involving the fusion of 
protons (H burning) foresees the existence of two processes. 
The PP chain activates at temperatures $\simeq$ 10 MK and 
operates through a sequence of proton captures and $\beta$ decays starting 
with a weak interaction p+p fusion. 
The processes involved in the PP chain \index{process!pp} are listed in Table ~\ref{tab:propro} together with 
the mass defect (Q$_{tot}$ values) in MeV and the energy carried away by neutrinos (Q$_{\nu}$). 
If the neutrino emission is described by an energy continuum, the maximum energy 
of this spectrum is reported. A direct build up of 
progressively heavier nuclei through successive proton captures stops very 
early, at $^3$He, because of the low cross section of the 
$^3$He+p reaction. Also, proton captures on 
the second most abundant isotope, $^4$He (or $\alpha$ particle, 
$N=Z=2$), cannot even begin, because nuclei with atomic mass number 
A = N+Z = 5 do not exist.\footnote{Their half lives
are of the order of 10$^{-22}$ s.}
In order to proceed with proton captures beyond 
$^3$He it is necessary 
to build up enough $^3$He nuclei to activate the 
capture of this nucleus by either another $^3$He nucleus, or 
$^4$He. The activation of $^3$He captures allows to overcome 
the non-existence of nuclei with A=5, though it still does not allow the 
build up of an appreciable amount of nuclei heavier than He. In fact, the product 
of the $^3$He+$^3$He reaction is an $\alpha$ particle plus two 
protons, while the product of the $^3$He+$^4$He reaction is \index{isotopes!7Be}
$^7$Be, whose fate, either proton or electron capture 
leads to the formation of $^8$Be, which very quickly decays in two 
$\alpha$ particles. In synthesis, the fusion of H mainly produces He, 
together with a number of radioactive nuclei that decay into 
their respective stable daughter nuclei emitting a neutrino. The  
energies and the number of neutrinos 
produced in these decays reflect the relative 
importance of the various PP-chain branches and the efficiency of 
the nuclear reactions in stars.

\begin{table}
\caption{The PP chain\label{tab:propro}}
\begin{tabular}{lcrcc}
\hline reaction &~& Q$_{tot}$ &~& Q$_{\nu}$ \\
          &~~& (MeV)     &~& (MeV)     \\
\hline
$p+p \rightarrow d+e^++\nu$            &~&  1.442 &~& 0.42(spectrum) \\ 
$d+p       \rightarrow ^3He$           &~&  5.494 &~&       \\ 
$^3He+^3He \rightarrow ^4He+2p$        &~& 12.860 &~&       \\ 
$^3He+^4He \rightarrow ^7Be$           &~&  1.587 &~&       \\ 
$^7Be+e^-  \rightarrow ^7Li+e^++\nu$   &~&  0.862 &~& 0.861(90\%)-0.383(10\%) (lines)\\ 
$^7Be+p    \rightarrow ^8B \rightarrow ^8Be+e^++\nu \rightarrow \alpha+\alpha$ &~& 18.209 &~& 14.060(spectrum) \\ 
$^7Li+p    \rightarrow ^8Be \rightarrow \alpha+\alpha$  &~& 17.347 &~&      \\ 
\hline  
\end{tabular} 
\end{table} 

The second process converting protons to $\alpha$
particles is the CNO cycle. \index{process!CNO} Given the high Coulomb barrier of the CNO
nuclei, this cycle becomes efficient at temperatures ($\rm T>20$ MK),
significantly higher than those relevant to the PP chain. The main section of this sequence is
characterized by the continuous conversion of C to N and viceversa. Let
us start, e.g., with the capture of a proton by a $^{12}$C (Table ~\ref{tab:cno}). The
outcome of this fusion is the radioactive nuclide $^{13}$N that
quickly decays $\beta^+$ into $^{13}$C. Efficient proton captures
by $^{13}$C lead to the synthesis of $^{14}$N. Proton captures by 
$^{14}$N produce $^{15}$O, a radioactive
nuclide that quickly decays in $^{15}$N. The fusion of a proton and a
$^{15}$N particle has, as the main outcome, a $^{12}$C nucleus
plus an $\alpha$ particle. The sequence sketched above is called the CN
cycle. If the temperature exceeds $\rm T \sim 25-30$ MK, also oxygen
enters the game and the full CNO cycle activates: $^{16}$O begins to
capture protons forming radioactive $^{17}$F that decays
to $^{17}$O. The capture of a proton by this particle leads to a
compound nucleus that preferentially splits into $^{14}$N and an
$\alpha$ particle, and that partly turns into $^{18}$F, which quickly
decays to $^{18}$O. Proton captures on $^{18}$O produce
preferentially $^{15}$N plus an $\alpha$ particle. The activation of
the channel $^{15}$N(p,$\gamma$)$^{16}$O closes the NO cycle,
processing material back into oxygen. The total abundance by number of
the CNO isotopes remains constant with time because the proton capture
on any of them (and the subsequent $\beta^+$ decays) just produce another
isotope in the same set.

\begin{table}
\caption{The CNO cycle\label{tab:cno}}
\begin{tabular}{lcrcc}
\hline reaction &~& $Q_{tot}$ \\
          &~~& (MeV)        \\
\hline
$^{12}C+p \rightarrow ^{13}N$          &~&  1.944 \\
$^{13}N   \rightarrow ^{13}C+e^++\nu$  &~&  2.220 \\
$^{13}C+p \rightarrow ^{14}N$          &~&  7.551 \\
$^{14}N+p \rightarrow ^{15}O$          &~&  7.297 \\
$^{15}O   \rightarrow ^{15}N+e^++\nu$  &~&  2.754 \\
$^{15}N+p \rightarrow ^{12}C+\alpha$   &~&  4.966 \\
$^{15}N+p \rightarrow ^{16}O$          &~& 12.127 \\
$^{16}O+p \rightarrow ^{17}F$          &~&  0.600 \\
$^{17}F   \rightarrow ^{17}O+e^++\nu$  &~&  2.761 \\
$^{17}O+p \rightarrow ^{14}N+\alpha$   &~&  1.192 \\
$^{17}O+p \rightarrow ^{18}F$          &~&  5.607 \\
$^{18}F   \rightarrow ^{18}O+e^++\nu$  &~&  1.656 \\
$^{18}O+p \rightarrow ^{15}N+\alpha$   &~&  3.981 \\
$^{18}O+p \rightarrow ^{19}F$          &~&  7.994 \\
\hline  
\end{tabular} 
\end{table} 

For $\rm T>25-30$ MK, the full CNO cycle
becomes efficient and quickly reaches a quasi-equilibrium in
which the abundance of each nucleus settles on a steady state value
determined by the balance between its production and destruction. For
example, the equilibrium abundance of $^{13}$C (assuming that
$^{13}$N decays instantaneously) is given by: \index{isotopes!13C}

$$ \frac{dY_{^{13}C}}{dt}= Y_{^{12}C}~Y_p~\rho~N_A~<\sigma~v>_{^{12}C+p}~-~Y_{^{13}C}~Y_p~\rho~N_A~<\sigma~v>_{^{13}C+p}=0  $$
where $Y_i$ refers to the abundance by number of a given species $i$, $t$ is 
time, $\rho$ is the density, $\rm N_A$ is Avogadro's number and 
$<\sigma~v>_{i+j}$ is the Maxwellian averaged product of the velocity $v$ times
the nuclear cross section $\sigma$ for a given 
reaction between nuclei $i$ and $j$.
The equilibrium condition immediately gives:

$$ \frac{Y_{^{12}C}}{Y_{^{13}C}}~=~\frac{<\sigma~v>_{^{13}C+p}}{<\sigma~v>_{^{12}C+p}}  $$
which means that the relative abundances between isotopes of the CNO cycle
depend only on the ratio between the respective cross sections for proton
capture. Typical isotopic and elemental ratios obtained in the
temperature range $\rm 30 \leq T \leq 100$ MK are reported in Table ~\ref{tab:ratios}. 

\begin{table}
\caption{Typical isotopic ratios from CNO cycling\label{tab:ratios}}
\begin{tabular}{lcccr}
\hline isotopic ratio &~& value &~& solar value \\
\hline
$Y_{^{12}C}/Y_{^{13}C}$  &~& $\simeq 4$                  &~&  89 \\
$Y_{^{14}N}/Y_{^{15}N}$  &~& $\simeq 4 \times 10^4-10^5$ &~& 272 \\
$Y_{^{17}O}/Y_{^{16}O}$  &~& $\simeq 10^{-2}-10^{-3}$                                  &~& 3.8 $\times 10^{-4}$ \\
$Y_C/Y_N$                &~& $\simeq 7 \times 10^{-3} - 2.5 \times 10^{-2}$            &~& 3.2 \\
$Y_N/Y_O$                &~& $\simeq 60-350$                                           &~& 0.13 \\
$Y_{^{18}O}/Y_{^{16}O}$  &~& $\simeq 2 \times 10^{-6}~{\rm for}~ T<50:60 MK$           &~& $2 \times 10^{-3}$ \\
$Y_{^{18}O}/Y_{^{16}O}$  &~& declines to $\simeq 5 \times 10^{-8}$ at T$\simeq$ 100 MK &~& $2 \times 10^{-3}$ \\

\hline  
\end{tabular} 
\end{table} 

The neutrinos emitted by the decay of radioactive nuclei synthesized
by the CNO cycle have characteristic energies different from those
emitted by the PP chain. Their detection would
provide precious information about the relative efficiency of the various 
reactions involved in the CNO cycle. 

In addition to the PP chain and the CNO cycle there is another sequence
of proton captures that can become efficient in stars, although it does 
not play a role in the energy budget. In the temperature range 40-50 MK the
proton captures listed in the upper part of Table ~\ref{tab:neal} quickly bring to 
their equilibrium values the abundances of $^{20}$Ne, $^{21}$Ne,
$^{22}$Ne, and $^{23}$Na, forming also in this case a \emph{NeNa}
cycle. For temperature in excess of 50 MK the \index{process!Ne-Na cycle} \index{isotopes!20Ne}
$^{23}$Na(p,$\gamma$)$^{24}$Mg channel competes with the \index{isotopes!24Mg}
$^{23}$Na(p,$\alpha$)$^{20}$Ne so that matter from the NeNa cycle
leaks towards more massive nuclei. At temperatures of order of 60 MK
also the proton captures listed in the lower part of Table ~\ref{tab:neal} fully activate so
that all the nuclei between $^{20}$Ne and $^{27}$Al reach their
equilibrium abundances. Is it worth noting that 
$^{26}$Al, a long-lived radioactive nucleus
with half life $7.17~10^5$ yr, is \index{isotopes!26Al}
included within this sequence. $^{26}$Al can be ejected 
into the interstellar medium by stellar outflows (winds) 
and its decay into $^{26}$Mg can be detected as diffuse 
$\gamma$-ray emission (Chapter 7, Section~4) 
when the metastable $^{26}$Mg relaxes towards its ground state. 
Moreover, it can be
included in dust grains that form around stars and decay  
within the already
formed minerals. This nucleus is thoroughly discussed in
Section~\ref{sec:al26}, Chapter 7 Section 4, and Chapters~4 and 9.  
Typical $Y_{^{26}Al}/Y_{^{27}Al}$
equilibrium ratios produced by H burning range between 
$3~10^{-2}$ at 60 MK and 0.8 at 100 MK. We refer the reader to the book by 
 \citet{cg68} for a
derivation of the basic stellar structure equations and detailed discussions of the physics involved in the study of 
stellar evolution. 

\begin{table}
\caption{The NeNaMgAl sequence\label{tab:neal}}
\begin{tabular}{lcrcc}
\hline reaction &~& Q$_{tot}$ \\
          &~~& (MeV)        \\
\hline
$^{20}Ne+p \rightarrow ^{21}Na$          &~&  5.979 \\
$^{21}Na   \rightarrow ^{21}Ne+e^++\nu$  &~&  3.548 \\
$^{21}Ne+p \rightarrow ^{22}Na$          &~&  6.739 \\
$^{22}Na   \rightarrow ^{22}Ne+e^++\nu$  &~&  2.842 \\
$^{22}Ne+p \rightarrow ^{23}Na$          &~&  8.794 \\
$^{23}Na+p \rightarrow ^{20}Ne+\alpha$   &~&  2.377 \\
\hline
$^{23}Na+p \rightarrow ^{24}Mg$          &~& 11.693 \\
$^{24}Mg+p \rightarrow ^{25}Al$          &~&  6.548 \\
$^{25}Al   \rightarrow ^{25}Mg+e^++\nu$  &~&  4.277 \\
$^{25}Mg+p \rightarrow ^{26}Al$          &~&  6.307 \\
$^{26}Al   \rightarrow ^{26}Mg+e^++\nu$  &~&  4.004 \\
$^{26}Al+p \rightarrow ^{27}Si$          &~& 12.275 \\
$^{26}Mg+p \rightarrow ^{27}Al$          &~&  8.271 \\
$^{27}Si   \rightarrow ^{27}Al+e^++\nu$  &~&  4.812 \\
$^{27}Al+p \rightarrow ^{28}Si$          &~& 11.585 \\
$^{27}Al+p \rightarrow ^{24}Mg+\alpha$   &~&  1.601 \\
\hline  
\end{tabular} 
\end{table} 

Although the sequences of nuclear reactions that power stellar luminosity are 
now considered to be well understood, a wise Galilean approach suggests 
to verify experimentally (whenever possible) their occurrence in stars. 
Our Sun provides a unique opportunity to 
accomplish such verification via the detection of neutrinos produced by 
radioactive decay in its deep interior.

\subsection{Solar Neutrinos -- a Unique Opportunity}
\label{subsec:sun}

The long-term stability of characteristic solar properties, in particular its luminosity and 
surface temperature, can be explained only if the solar \index{neutrino}
energy source is of nuclear origin and specifically involves proton 
captures, which 
are associated with abundant fuel and a long time scale. As discussed 
above, such an energy source inevitably results in the production of 
radioactive nuclei, which decay into their stable daughter nuclei through 
weak processes, hence emitting neutrinos. Modeling of the internal 
structure of the Sun predicts a central temperature at present of  
about 15 MK, and hence that the PP chain dominates (99.6\%) over the 
CNO cycle (0.4\%) converting H into $^4$He. \index{process!pp} The relative importance 
of the nuclear reactions in the PP chain in the Sun leads to the result that the 
majority (93\%) of the neutrinos produced should come from 
$\rm p(p,e^+\nu_e)d$ reactions (where d=deuterium, N=Z=1) and be of 
relatively low energy ($\rm E\leq0.42~MeV$, see Table ~\ref{tab:propro}), while only a 
minor fraction of the total neutrino spectrum is expected to be contributed by the decay of 
$^7$Be ($\simeq7\%$, $\rm E\simeq0.86~MeV$) and 
$^8$B($0.0075\%$, $\rm E<15~MeV$).

\bigskip \begin{figure}
  \centering
  \includegraphics[width=1.0\textwidth]{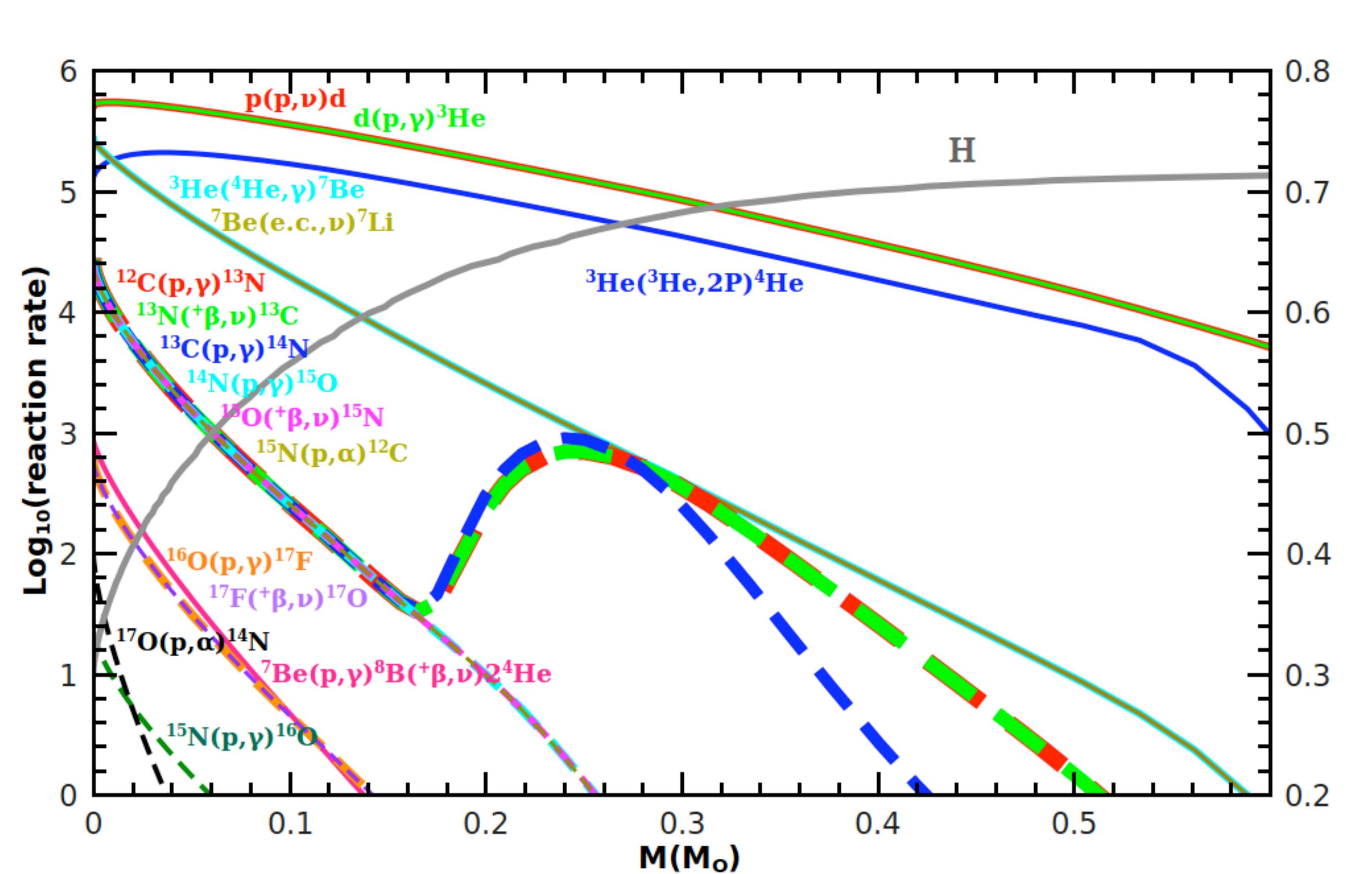}
  \caption{Rates of the reactions involved in the PP chain and the 
CNO cycle as a function of
  the mass coordinate in a solar-like stellar model having approximately 
the age of the Sun of 4.6 Gyr. The H abundance is also plotted and its range 
shown on the right-side y-axis.}
\label{fig:sunrates} 
\end{figure}

Figure~\ref{fig:sunrates} shows the rates of the nuclear reactions 
involved in the PP chain and CNO cycle as a function of the
mass coordinate for a 1 \Msun\ stellar model of solar \emph{metallicity}\footnote{The
term metallicity indicates the abundance of metals in a star, where
metals corresponds to all elements heavier than He. The metallicity of
the Sun is $\simeq$0.014 by mass fraction, where abundances are normalised to a total of
1, which means that 1.4\% of the solar matter is composed of elements heavier than 
He. The most abundant of these is oxygen, followed by carbon.}
and an age close
to the present age of the Sun, i.e., 4.6 Gyr. All the $\beta^+$ decays
are concentrated near the center where the bulk of the synthesis of
unstable nuclei takes place. The
relative importance of the various PP reactions, and hence of the
associated neutrino fluxes, depends on the rates of the nuclear
reactions involved, which, in turn, are a function of density,
temperature, and chemical composition. Since the internal structure of a
model of the Sun depends on the adopted and somewhat uncertain input
physics - e.g., the nuclear cross sections, the equation of state, 
the opacity, and
the chemical composition of the gas from which the Sun formed - the detection
of neutrinos from the Sun is fundamental not only to experimentally
verify the nuclear origin of the solar luminosity, but also to confirm
the overall reliability of the
modeling of the internal structure of the Sun and the adopted input
physics. \index{stars!structure}

It is therefore understandable that the quest for the solar neutrinos \index{neutrino}
started early, more than 40 years ago, with the Davis experiment \index{Davis, R.Jr}
(1967-1985) \citep[]{bcdr85,cleda98}. This experiment, based on the interaction  
between an electron neutrino and $^{37}$Cl, has a threshold energy of  
roughly 0.8 MeV and hence could essentially only detect the $^8$B 
neutrinos, which constitute a minor fraction of the neutrino flux  
from the Sun. The Davis experiment provided two basic results, one very 
encouraging and another very stimulating. First, it detected solar 
neutrinos, demonstrating beyond any doubt that proton captures are 
occurring in the interior of the Sun. Second, the detected neutrino flux 
was roughly one third of the predicted value. Such a result stimulated much 
further work and a large number of papers on this puzzle were written  
over the decades. The discrepancy was considered by many physicists as a 
strong indication that the basic modeling of the Sun was wrong, however, 
it must be recalled that it was confined to a very minor branch 
of the PP chain. The discrepancy became more serious with the advent of \index{GALLEX}
the GALLEX experiment  \citep[]{hahe98} - a collaboration among France, Germany, Italy, 
Israel, Poland, and USA, led by MPIK Heidelberg, 1991-1997 - and the \index{SAGE} 
SAGE experiment  \citep[]{abud99} - a Russian-American collaboration, 1990-2000. These 
modern sophisticated experiments were designed to detect the 
bulk of the neutrinos 
produced in the Sun, i.e., the low energy neutrinos produced by the p+p 
reaction. They confirmed both the detection of 
a neutrino signature from the Sun and the existence of a discrepancy 
between theoretical and observed fluxes. This result plunged the basics of 
solar modeling into a crisis because these experiments were sensible 
to the total number of electron neutrinos emitted by the Sun. This number 
is theoretically robust, depending only on the basic assumption 
that the solar luminosity comes from the conversion of protons into $\alpha$ 
particles, and not on the details of the 
modeling of the internal structure of the Sun or 
of the cross sections of the nuclear reactions involved. If 
the solar luminosity is powered by the conversion of protons into $\alpha$ 
particles, the total number of electron neutrinos emitted per second by 
the Sun must be $\rm 2.38 \times 10^{39}$ ($L_\odot$ in MeV/s) $/$ 25 
(energy provided per $\alpha$ nucleus in MeV) $\times$ 2 (number of $\nu_e$ 
produced per $\alpha$ nucleus) [$\rm \nu_e~s^{-1}$], which corresponds, at 
a distance of one astronomical unit, to a flux of $\rm 
6.78\times10^{10}~\nu_e~s^{-1}$. Instead, the neutrino flux detected 
by GALLEX and SAGE was only half of this prediction.

The solution to this puzzling discrepancy arrived when it was shown that 
neutrinos oscillate among three different species: $\nu_e$, $\nu_\mu$ and 
$\nu_\tau$. None of the Davis, GALLEX, or SAGE experimental set-ups could 
detect the $\nu_\mu$ and $\nu_\tau$ neutrinos so that all these 
experiments simply missed the fraction of neutrinos that reach the 
Earth as $\nu_\mu$ and $\nu_\tau$. The Sudbury Solar Neutrino 
Observatory (SNO) experiment  \citep[a Canadian, USA, and UK
collaboration that started
in 1999,][]{many02} was designed to detect all three neutrino flavors.
In 2001 this experiment finally showed that there is agreement between the 
observed and predicted neutrino fluxes and hence demonstrated that our
modeling of the interior of the Sun is basically correct. 
In 2008, the Borexino instrument finally also opened up the sub-MeV range for solar
neutrinos and detected $^7$Be neutrinos \citep{2008PhLB..658..101B}. For 
comprehensive reviews
on the solar neutrino problem and current status, see \citet{bape04,ba05,2010JPhCS.203a2081O}.


\section{Evolution to and through the First Giant Branch}
\label{sec:rgbhead}

In Subsection~\ref{subsec:rgb} we discuss the main evolutionary properties of
stars once they leave the long lasting phase of central H burning described
above, commonly referred to as the Main Sequence, and enter the phase known as the
First, or Red, Giant Branch (RGB). At the end of central H \index{stars!structure}
burning a star is composed of a H-exhausted core made primarily of He
and a H-rich envelope. Hydrogen burning shifts from the center to the
base of the H-rich mantle while the envelope expands causing the surface
of the star to reach radii 10-1000 times the solar radius and to cool
down to a few thousand K. This expansion triggers the formation of large
scale convective motions extending from the surface down to deep
regions in the star where partial H burning occurred during the Main Sequence. 
Some products of
this H burning are thus brought to the surface in a process known as
the 1$^{st}$ dredge-up. In this phase the He core grows in mass because the
H-burning shell continuously converts H-rich matter into He-rich matter
and deposits the ashes onto the He core. The temporal evolution of the He
core depends on its initial size, i.e., the size it had just after
the central H exhaustion, which is in turn mostly determined by the
initial stellar mass. If the mass of the He core is less than $\simeq
0.35 M_\odot$,
which occurs for initial stellar masses less than $\sim 2 M_\odot$, an
\emph{electron degenerate core} forms where matter reaches such extraordinarily
high density, up to $\simeq 10^6$ g/cm$^3$, that the dominant
contribution to its pressure arises from the Pauli exclusion principle,
which prevents the electrons from occupying identical quantum states.
This leads to an increase of the lifetime of this phase up to about 100 Myr,
and forces the subsequent He ignition to occur quite far from the
center. If the He core is instead more massive than 0.35 M$_\odot$, the
electrons remain far from the degeneracy regime.

In Subsection~\ref{subsec:li} we discuss the conditions under which
$^7$Li, the stable daughter of radioactive $^7$Be, may be produced, 
preserved, and brought to the stellar surface. 
This nucleus is typically destroyed in the PP chain (Table ~\ref{tab:propro}),
because its destruction rate is efficient at
temperatures lower than its production rate. A way to produce
 $^7$Li was proposed in 1971 by 
 \citet{cf71}: if freshly synthesized
$^7$Li is quickly brought to very low temperatures by mixing, 
then it can be preserved. If H burning occurs in a
convective environment it is in principle possible to find 
a high Li abundance on the surface of a star, as observed in some stars belonging to the 
First Giant Branch. However, these observations are in fact difficult to
explain because H burning occurs in a formally stable region 
well below the base of their
convective envelopes. Additional mechanisms
of mixing must be invoked to bring $^7$Li-rich material
into the convective envelope.

\subsection{The First Giant Branch}
\label{subsec:rgb}

During the Main Sequence phase of stellar evolution described in the 
previous section conversion of H into He via H burning in the 
centre of the star leads to a progressive increase of the mean molecular 
weight combined with a decrease of the amount of available fuel. The net 
result is a slight increase of the luminosity (because L scales with the 
fourth power of the molecular weight, equation \ref{mlrel} in Subsection~\ref{subsec:hburn}), 
and a mild 
expansion of the surface of the star because of the formation of a 
molecular weight gradient  \citep[]{sclc09}. 
Once H is exhausted in the central region of the star, the H-exhausted core, or 
He core, begins to contract on a gravitational timescale 
while the region of active nuclear burning smoothly shifts above the He 
core, where H is still abundant. Further evolution of the He core 
depends on its mass, which, in turn, depends on the initial total 
mass of the star. If the He-core is more massive than a threshold 
value of $\sim$ 0.35 \Msun, which happens for an initial total mass 
of $\sim$ 2 \Msun, its contraction induces strong heating 
of the He core itself, which quickly reaches a temperature of $\sim$ 100 
MK at which fusion reaction of $\alpha$ particles (He burning) is 
activated. If, instead, the core is less massive than 0.35 \Msun the high 
densities reached render the electron gas degenerate, hence supporting the 
structure against gravity without the need for additional contraction. 

This difference has a large impact on further evolution 
of the star because, in the latter case, the He core tends towards an almost 
isothermal configuration due to the large mean free path of degenerate 
electrons relative to that of photons. If the structure was 
isolated, as in the case of white dwarves, it would progressively cool 
down losing its stored energy through the surface. Instead, in the case 
discussed here, the degenerate structure heats up because it is surrounded 
by the H-burning shell, which continuously deposits freshly 
synthesized He onto the He core. The rate at which the maximum 
temperature increases with time in the degenerate He core is controlled by 
the growth rate of the He-core mass, which obviously coincides with the 
rate at which the H-burning shell converts H into He. Strong neutrino 
production  \citep[]{itad89} in the center of the electron degenerate core, due to 
the interaction of photons with the plasma and/or to the 
scattering of photons on electrons, carries away 
energy from the core pushing the location of the maximum temperature 
outward in mass. The off-center location of the maximum temperature is the result of 
the balance between energy loss due to the neutrino emission, which scales
directly with the density and pushes the maximum temperature outward, and 
the energy gain due to the compressional heating, which scales inversely 
with the density and pushes the temperature 
maximum back towards the center. The key stellar parameters that control the 
location of the maximum temperature are the CNO abundance and the initial 
mass of the star. The higher the CNO abundance, the faster the conversion 
of protons into $\alpha$ particles in the H-burning shell, the stronger 
the heating of the 
degenerate He core, and the closer the maximum temperature is to the center. 
The higher the initial mass of the star, the lower is the degree of electron 
degeneracy and the density in the He core, and hence the efficiency 
of neutrino production.

While the H-burning shell influences the evolution of the He core, the
growth of the He core influences the evolution of the H-burning shell as
well. In fact, the progressive heating of the core raises the temperature
at the surface of the core, where H burning occurs. This results in a
continuous positive feedback: the H burning shell deposits He onto the He
core, which therefore heats up. Such a heating leads to an increase of the
temperature and density in the H-burning shell, accelerating the H burning
rate and increasing the conversion rate of H into He, and therefore the
heating of the He core. As a consequence, the progressive increase of the
H burning rate determined by the growth of the He core mass forces the H
rich mantle of the star to expand and to cool. The cooling of the
stellar envelope triggers a large increase of the opacity because 
atoms become partially ionised. The temperature gradient steepens,
favoring the growth of convective instabilities that very quickly extend
over a major part of the H-rich mantle from the surface down to near the
outer border of the H-burning shell. A consequence of the growth of these
convective motions within most of the H rich mantle is a large increase of
the surface luminosity caused by the continuous increase of the H burning
rate coupled to the fact that the convective envelope does not absorb or
release energy but just transports it outward. \index{stars!structure}

Since convective motions play a fundamental role in the 
physical and chemical evolution of any star, we briefly sketch the basic
physical reason that leads to the growth of these large scale motions.
The equilibrium condition provided by counterbalancing pressure gradients
and gravity in stars does not necessarily imply stationary matter: a
bubble of stellar matter may be considered stable against
motion if a restoring force pushes it back towards its rest position
when, for any reason, it is slightly displaced from its equilibrium
location. Such a restoring force is simply given by Archimede's force,
i.e., it depends on the density contrast between that of the environment
and that of the bubble. If the density of an element of matter displaced
towards a lower/higher density region turns out to be even lower/higher 
than that of its new surroundings, the element will continue to 
\emph{raise}/\emph{sink} and move away from its rest
position, otherwise it will move back towards its equilibrium location.

Changes in the physical structure of the bubble during its motion play
an important role in determining its density and thus its behavior.
Mechanical equilibrium with the environment is certainly well verified
so that it can be safely assumed that the internal pressure within the
bubble istantaneously readjusts to that of the environment. More
difficult is to determine the amount of heat that the bubble can
exchange with the environment while moving. In the simplest case in
which the bubble does not exchange any heat with the surroundings
until it has covered a certain distance (adiabatic approximation), and
assuming that the region is chemically homogeneous, the critical
condition for the onset of large scale motions of the matter is that the
temperature gradient of the environment must exceed the adiabatic gradient 
(Schwarzschild \index{Schwarzschild criterion}
criterion). While the radiative temperature gradient remains less than
the adiabatic temperature gradient, an element of matter will remain
more/less dense than its surroundings if displaced towards less/more
dense regions (within stars these displacements are typically connected 
to movements outward/inward in mass), and hence it will experience a 
restoring force that will keep it anchored to its rest location. 
On the contrary, when the radiative
temperature gradient exceeds the adiabatic temperature gradient
stochastic motion of the matter is not hampered by a restoring force,
but it is amplified leading to the growth of large scale motions.
Hence, convective regions are associated with steep temperature
gradients, which typically occur either close to regions where
energy production is strongly concentrated, or in regions where the
mean free path of the photons, which scales with the inverse of the
opacity, becomes so small that radiation energy transport becomes
inefficient.

The determination of the temperature gradient in convective regions is
quite complex. Here it suffices to say that while in the interior of a
star the temperature gradient in a convective region remains very close to
the adiabatic gradient, in a convective envelope the temperature gradient
becomes much steeper (intermediate between the radiative and adiabatic
case) because the low density in the outer envelope makes energy transport
by convective eddies inefficient, so that both photons and eddies
contribute to the outwards transport of thermal energy. 

Since convective eddies have a very large mean free path with respect to
that of photons, convection is a very efficient
energy transport mechanism. In the specific case of the extended convective
motions that form above the H-burning shell in Red Giant stars, energy
transport is so efficient that virtually all the energy produced by the
burning shell is transmitted to the surface without essentially any
absorption by the convective layers. It follows that a star in the
H-burning shell evolutionary phase is forced to increase in size
to be able to get rid of the extra energy influx, while the drop of the
surface temperature is limited by the presence of a maximum temperature 
gradient: the adiabatic temperature gradient, which cannot be 
overcome by much in the largest fraction of the envelope mass.

The mere existence of stars in the RGB phase constitutes evidence of a)
the presence of an active H-burning shell, demonstrated by the breaking
of the mass-luminosity relation $\rm L \propto M^3$ that holds during
the Main Sequence phase, b) the presence of a maximum temperature
gradient, demonstrated by the only minor change of the surface temperature
along the RGB, c) the continuous increase of the energy production by the
H-shell burning, demonstrated by the continuous increase of the surface
luminosity, and d) the presence of an electron degenerate core (for
stars with initial mass less than $\simeq$~2~\Msun), demonstrated by the 
existence of a relatively long lasting, $\sim~10^8$~yr, and thus observable 
RGB phase, which would be prevented if the He core was 
gravitationallly contracting.

\begin{figure}
  \centering
  \includegraphics[width=1.0\textwidth]{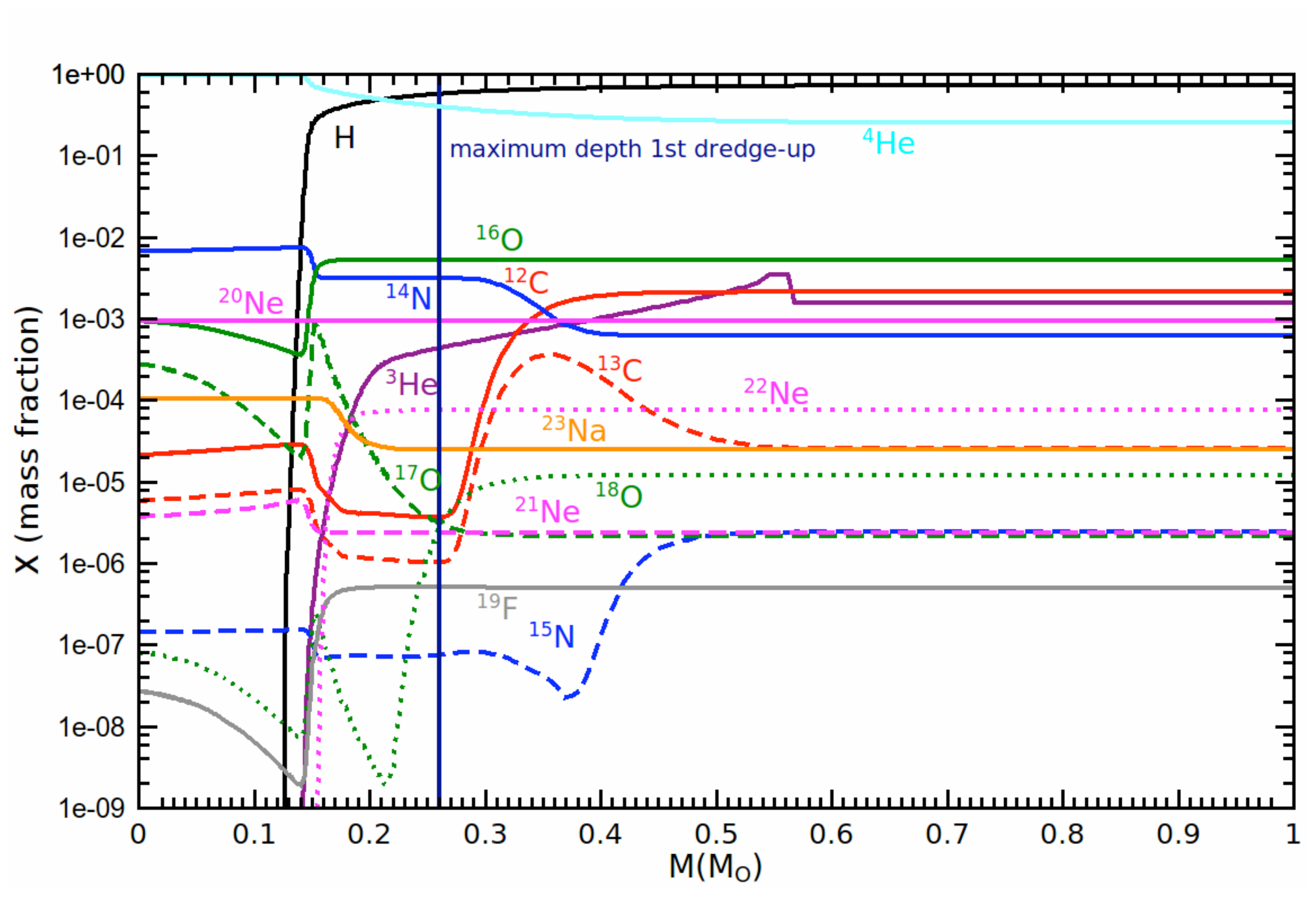}
  \caption{Snapshot of the abundances of several 
nuclear species in a solar-like stellar model  
  just before the onset of the 1$^{st}$ dredge-up. The maximum inward
penetration of the convective envelope during the 1$^{st}$ dredge-up
  is marked by the vertical solid blue line.}
\label{fig:1stdup} 
\end{figure}

Soon after the formation of the H-burning shell, the large scale motions
that grow in the H-rich envelope and rapidly extend from the surface down
to just above the top of the H-burning shell, bring to the stellar surface
matter partially processed by proton-capture reactions during the Main
Sequence phase. This mixing, referred to as the 1$^{st}$ dredge-up, modifies
the stellar surface composition. The amplitude of these modifications
depends on the initial stellar mass and metallicity, the general rule
being that the amplitude of the changes of the surface composition scales
with the initial stellar mass, directly up to 3 \Msun\ and then inversely
for higher masses, and inversely with the metallicity.
Figure~\ref{fig:1stdup} shows the abundance profiles of several nuclear
species as a function of the mass location for a solar-like stellar model
evolved to the RGB, just before the convective envelope deeply penetrates
into the star. The solid vertical line shows the maximum inward
peneration of the convective envelope. Since the convective motions reach
layers in which the local chemical composition has previously been
modified by nuclear burning, also the surface chemical composition is
modified by the mixing induced by these large scale motions. In particular
the surface He abundance is slightly increased by 5\%, $^3$He
increases by one order of magnitude, $^7$Li is destroyed, the
$^{12}$C$/$ $^{13}$C ratio drops from the solar value of 89 to
roughly 30, the $^{14}$N$/$ $^{15}$N ratio increases from the terrestrial 
value of 272 to $\simeq$ 500, while the oxygen isotopic ratios and those of heavier 
nuclei remain at their solar values. \index{isotopes!13C} \index{isotopes!14N}
In stars more massive than the Sun the oxygen isotopic ratios are also modified
with $^{16}$O/$^{17}$O decreasing by up to one order of magnitude, from
solar $\simeq$2700 to $\sim$250 and $^{18}$O/$^{16}$O increasing mildly
up to $\simeq$700 from the solar value of 500.
A detailed quantitative determination of these changes depends 
on the specific stellar model considered.

\bigskip 

The evolution of the star after the 1$^{st}$ dredge-up is characterized by the
H-burning shell progressively converting H from the convective envelope
into He, which is deposited onto the inert He core. The continuous mass
transfer from the envelope to the core progressively reduces the mass of
the envelope while its chemical composition does not change any more
because the temperature within the convective envelope is too low to
activate nuclear reactions.

The evolution along the RGB ends when the maximum temperature in the core 
is high enough, $\simeq$100 MK, to activate the burning of He via 
$3\alpha$ reactions, during which three $\alpha$ particles join into a 
$^{12}$C nucleus. If the pressure is 
dominated by degenerate electrons
the energy released by these reactions cannot be immediately
balanced by an expansion of the core. Hence, He ignition occurs through a 
series of \emph{flashes}, which progressively remove the degeneracy, shifting 
the burning towards the center. Once the electron degeneracy is fully 
removed, a quiescent central He-burning phase settles in.

All along the complex, and partly still mysterious, RGB evolutionary
phase that links central H to central He burning, radioactive nuclei are
produced by H-shell burning mainly via the CNO cycle. \index{process!CNO} Most of them,
however, have negligible lifetimes, so they could only be detected
through the neutrinos they emit. Unfortunately, no Red Giant star is
close enough to the Earth to allow the detection of neutrinos of nuclear
origin produced in its interior. However, there are two unstable nuclei,
$^7$Be and $^{13}$N, whose half life may be comparable or even
larger than some stellar timescales: for $^7$Be the half life is 
comparable to the envelope mixing \index{isotopes!13N} \index{isotopes!7Be}
turnover time, for $^{13}$N the half life is comparable to proton-capture 
timescale in extremely
metal-poor stars, because stars of lower metallicity are more compact 
and hotter, due to their lower opacity. 

In the next section we discuss specifically the abundance of
$^7$Li, the stable daughter of $^7$Be, in giant stars, which
could provide important clues about the presence of additional motions
extending below the base of the convective envelope.  This is important
because, as we described above, the modeling of large scale motions
within stars is still crude and their growth, timescale, and efficiency
not yet well understood.


\subsection{The Production of Li}
\label{subsec:li}

Lithium (Li) isotopes\footnote{Lithium has two stable isotopes 
$^6$Li and $^7$Li, of which $^7$Li is the more abundant 
representing 92\% of solar Li.} in stars are fragile as they are 
easily destroyed by proton-capture 
reactions once the temperature exceeds 3 MK. The destruction 
timescale drops from 100 Myr at 3 MK to only 0.3 Myr at 5 MK while their 
production through fusion reactions only occurs at much higher 
temperatures, between 10 MK and 25 MK. The lower limit is due 
to the fact that the synthesis of $^7$Li is initiated by the 
$^3$He($\alpha$,$\gamma$)$^7$Be reaction, which becomes 
efficient only at temperature of the order of 10 MK, while 
the upper value is due to activation of the 
$^7$Be(p,$\gamma$)$^8$B reaction, which overcomes the electron-capture reaction $^7$Be($\rm e^-,\nu$)$^7$Li 
above a temperature of the order of 25 MK. Hence, Li is 
efficiently produced in a temperature range where it is also  
efficiently destroyed and therefore there seems to be no room for Li 
production in a star. However, there are a number of Red Giant stars 
observed to be Li rich \citep[]{cgsbs00,bala05,ulpbal07}.

A possible way out of such a puzzling situation was recognized 
by Cameron and Fowler (1971) and is based on the idea that 
instabilities, such as convection, rotation-induced 
instabilities, thermohaline mixing, etc., may bring freshly made 
$^7$Be from its production site to more external regions, where the 
temperature is low enough to inhibit proton captures on $^7$Li, on 
a timescale shorter than that of electron capture of $^7$Be. Note 
that the electron-capture rate of $^7$Be shows a mild increase as the 
temperature decreases. \index{Cameron, A.G.W.} \index{Fowler, W.A.}

\begin{figure}
  \centering
  \includegraphics[width=1.0\textwidth]{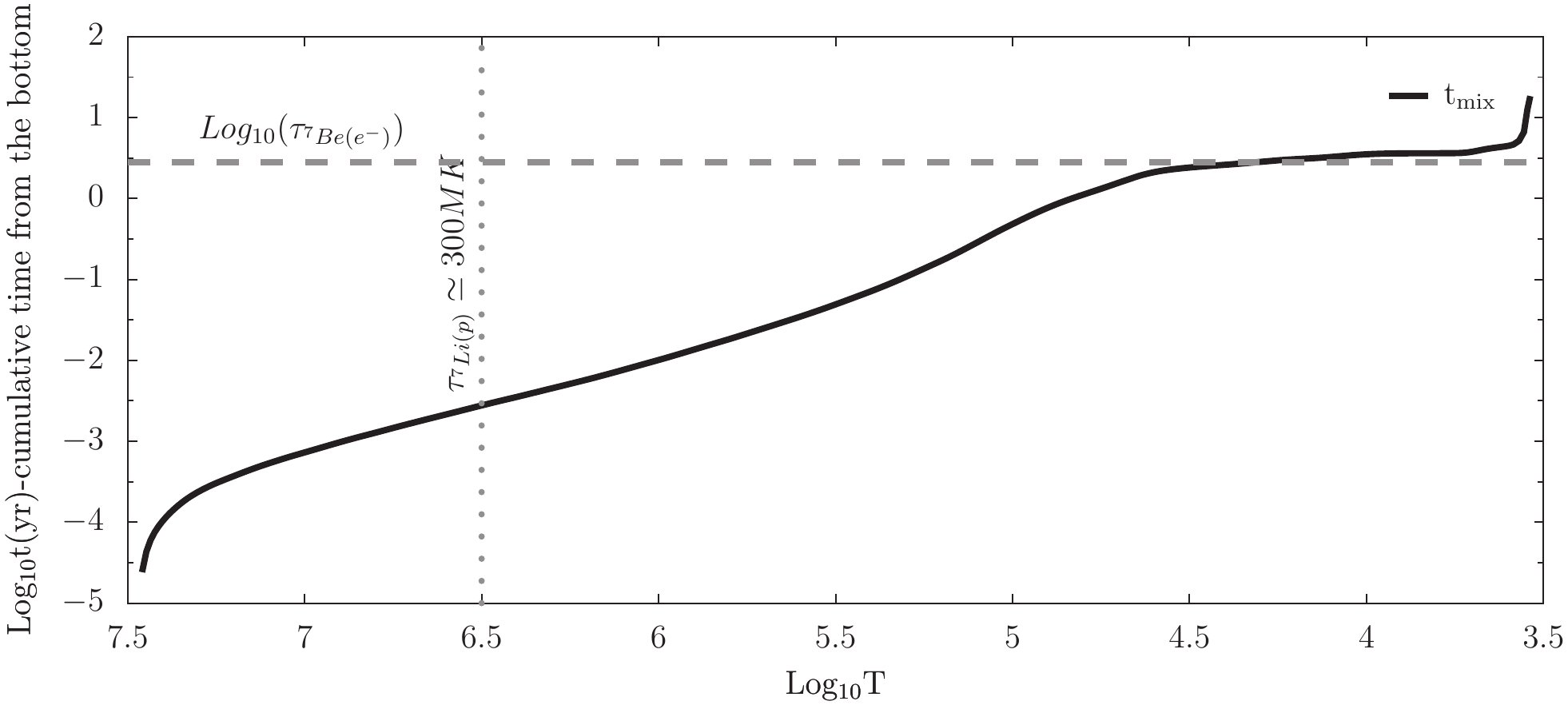}
  \caption{The cumulative turnover time of the convective eddies computed 
from the base of the
   convective envelope up to the surface as a function of temperature, T. 
The star is a 6\Msun\ star
   of solar metallicity some time after the beginning of the AGB phase. 
The horizontal dashed grey
   line marks the typical timescale of the 
$^7$Be($\rm e^-,\nu$)$^7$Li reaction while the 
   vertical dotted grey line shows the threshold temperature below which 
the timescale of the proton 
   capture on $^7$Li becomes larger than 300 Myr.}
  \label{fig:tempoconv} \end{figure}

A typical environment in which the Cameron-Fowler mechanism  \index{process!Cameron Fowler mechanism}
operates is during the Asymptotic Giant Phase (AGB) phase \index{stars!AGB}
(Section~\ref{subsec:AGB}), if the  
star is more massive than 4:5 \Msun. These stars develop large scale motions 
in the H-rich mantle that extend from the surface down to regions where 
the temperature is high enough ($>40$ MK) for some nuclear burning to 
occur (Hot Bottom Burning), in particular via the \index{process!hot bottom burning}
$^3$He($\alpha$,$\gamma$)$^7$Be reaction. 
Fig.~\ref{fig:tempoconv} shows the cumulative turnover time from 
the base of the convective envelope to the region of temperature T given 
in the abscissa for a 6 \Msun\ star of solar metallicity sometimes after the 
beginning of the AGB phase. The horizontal dashed grey line marks the 
typical timescale of the $^7$Be($\rm e^-,\nu$)$^7$Li reaction 
while the vertical dotted grey line shows the threshold temperature below 
which the timescale of the proton capture on $^7$Li becomes larger 
than 300 Myr. In this environment $^7$Be produced above 10 MK is 
succesfully transfered before decaying to a region where its daughter 
$^7$Li can survive.

\bigskip \begin{figure}
  \centering
  \includegraphics[width=.80\textwidth]{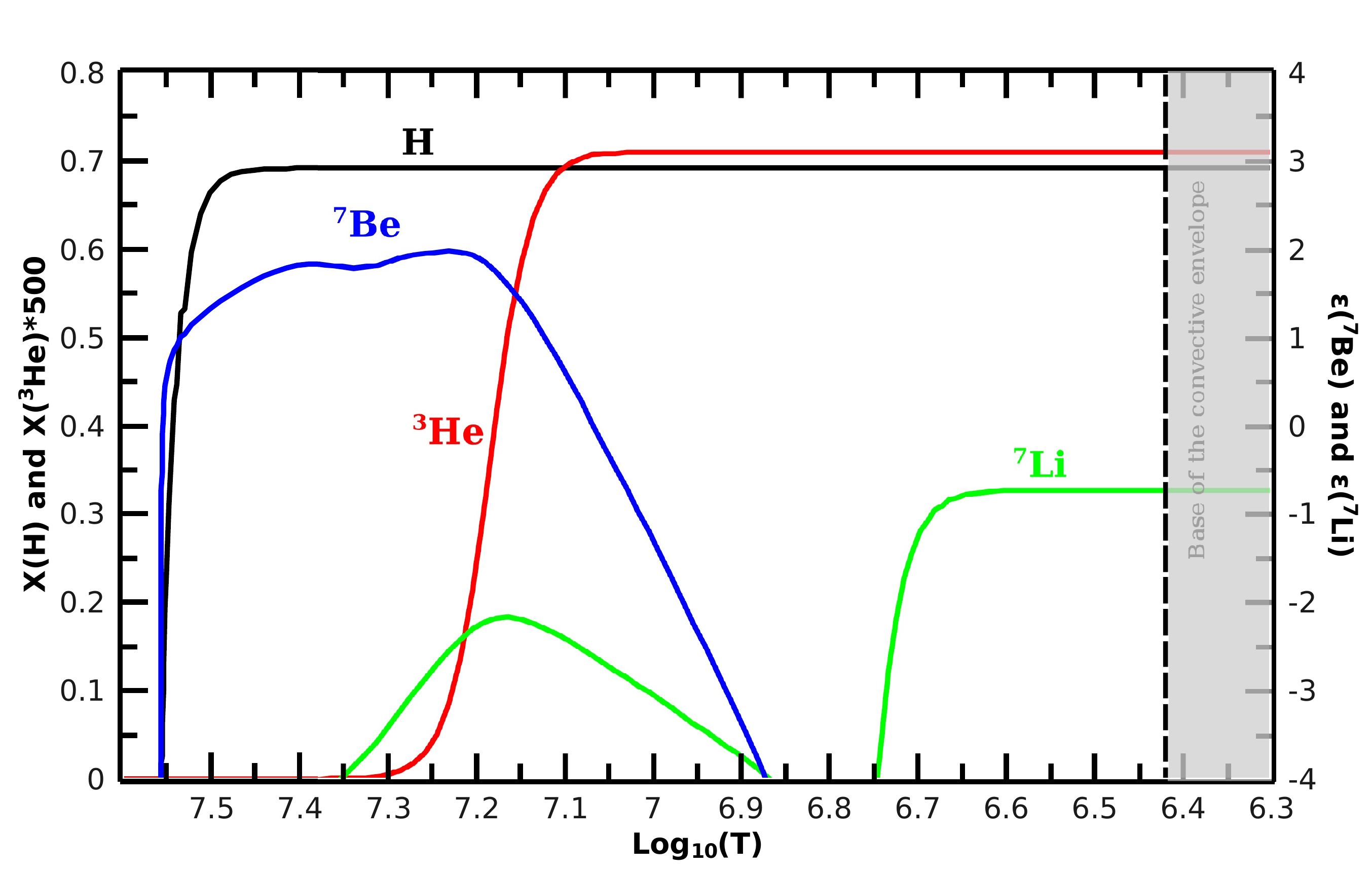}
  \caption{Abundance profiles for H, $^3$He, $^7$Be, 
and $^7$Li as function of the internal
  temperature in the region between the H-burning shell and the base 
of the convective envelope (grey area) 
  for a solar-like stellar model on the RGB. The abundances of 
both $^7$Be and $^7$Li are given in
  the widely adopted logarithmic scale in which 
$\rm \epsilon(X)=Log_{10}(N_X/N_H)+12$, where $\rm N_X$ and $\rm N_H$ 
  represent the abundances of element X and of hydrogen by number. 
In this scale the hydrogen abundance is equal to 12.
X(H) and X($^{3}$He)
represent the mass fraction of H and $^{3}$He.
}
  \label{fig:rgbstru} \end{figure}

An increase of the Li abundance at the surface of RGB stars is more 
difficult to achieve. Though the turnover time scale within the convective 
envelope is also in this case rather short ($\rm \simeq ~1~yr~at~\rm 
Log_{10}(L/L_\odot)\simeq2$), the temperature at the base of the convective 
envelope always remains well below 5 MK, too low to lead to an appreciable 
production of $^7$Be. Nonetheless, 
as mentioned above,
observations show the existence of 
a small number of Li-rich RGB stars. 
Fig.~\ref{fig:rgbstru} shows
the internal structure of the region around the 
H-burning shell from a solar-like stellar model on the RGB. 
Here, $^7$Be is 
synthesized well below the region where large scale motions of the matter
and hence mixing of the chemical composition occur. In this environment 
the Cameron-Fowler mechanism could operate only by assuming the presence 
of presently unidentified instabilities able to drive some mixing 
between the region rich in $^7$Be and the base of the convective 
envelope. The main constraint on this \emph{extra} mixing is that it 
must get close enough to the active H-burning shell to reach 
the layers where the $^7$Be production occur, but it must not enter \index{isotopes!7Be}
the region of the main nuclear burning. The reason is that 
the speed at which a star climbs along the RGB is regulated by the speed 
at which H is converted into He (see above) which, in turn, also depends on the 
amount of fuel that continuously enters the burning region. If this extra 
mixing reached the active burning region, it would inevitably bring 
fresh H into the burning region, therefore altering the rate at which H is 
consumed by the H-burning shell and hence the timescale of evolution 
along the RGB. This evolutionary timescale 
is observationally well established from counting the number 
of stars on the RGB in many Galactic globular clusters, 
and already very 
well reproduced by current models of these stars without extra mixing.

\bigskip \begin{figure}
  \centering
  \includegraphics[width=1.0\textwidth]{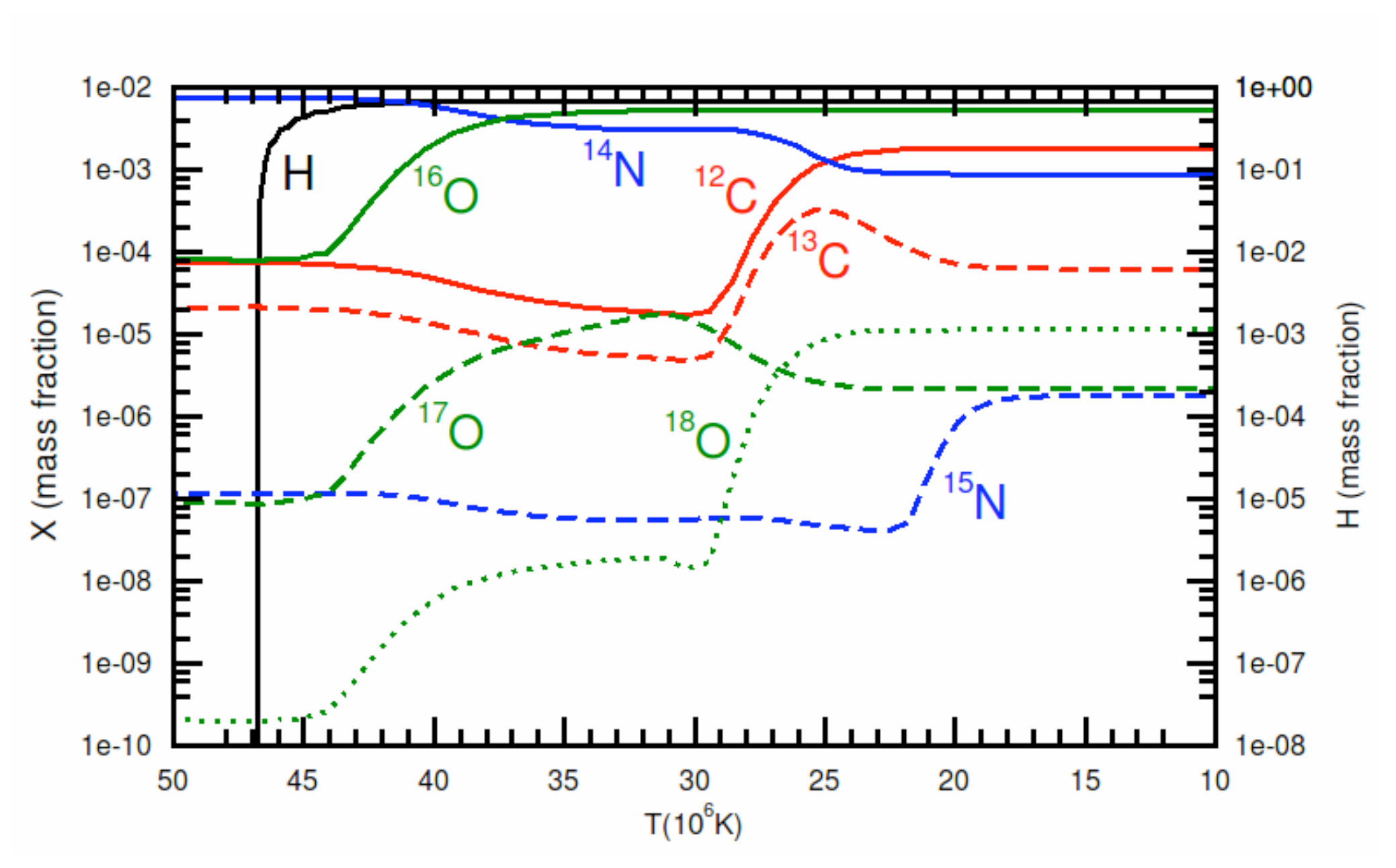}
  \caption{Abundance profiles of the CNO isotopes as function of the  
   temperature  on the RGB in a solar-like  stellar model.}
  \label{fig:cnoshell} 
\end{figure}

\bigskip \begin{figure}
  \centering
  \includegraphics[width=1.0\textwidth]{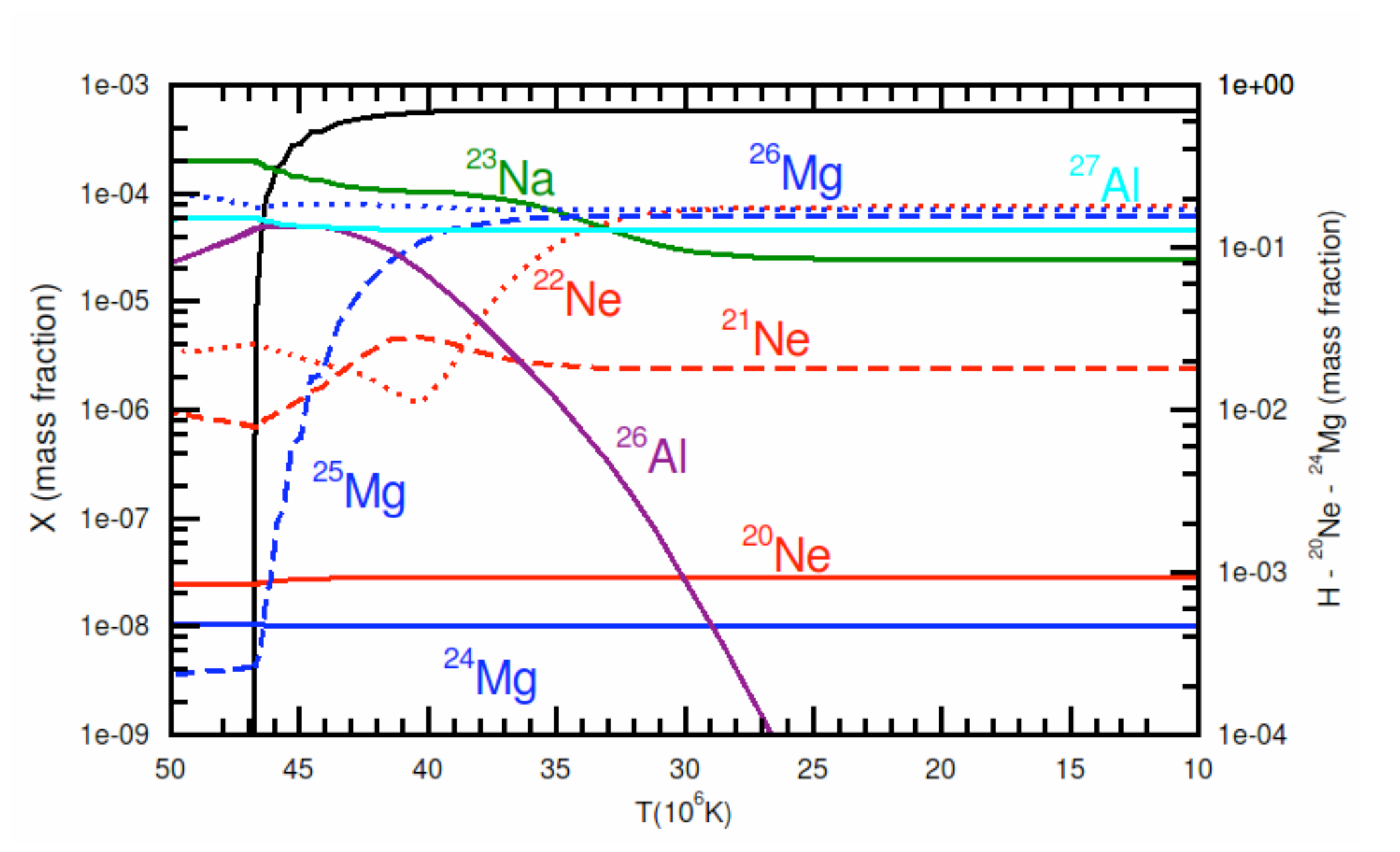}
  \caption{Abundance profiles of the Ne, Na, Mg and Al 
isotopes as function of the temperature  on the RGB in a solar-like
  stellar model.}
  \label{fig:nealshell} 
\end{figure}

There are other hints that point towards the presence of extra-mixing
phenomena in RGB stars (and perhaps in AGB stars too, as discussed in
Sections~\ref{sec:al26} and \ref{subsec:al26obs}). The observed surface
$^{12}$C/$^{13}$C ratio and N abundance are, respectively, too low
and too high with respect to the values predicted by the 1$^{st}$ dredge-up.
Extra mixing would naturally lower the first ratio and raise the N \index{isotopes!13C}
abundance, this being the signature of H burning (see Table \ref{tab:ratios}). 
A deeper mixing than
predicted by current models would also reduce the abundance of $^3$He
in the stellar envelope, which is increased by the 1$^{st}$ dredge-up, by
bringing this nucleus down to regions where it is destroyed. This
reduction is needed to avoid an increase of the $^3$He abundance in the
interstellar medium, which is not observed, due to the material expelled
by low-mass stars over the lifetime of the Galaxy. 

Figs~\ref{fig:cnoshell}
and~\ref{fig:nealshell} show the isotopic abundances of several 
nuclei up to Al within a solar-like star while climbing the RGB
(at $\rm Log(L/L_\odot)\simeq 3$). Note that we chose to use the
temperature as the abscissa instead of mass to better clarify the temperature at which
each nuclear species varies. The figures clearly shows that, for each
given depth reached by an extra mixing process, a few nuclei are expected
to be modified. For example, a drop of the oxygen abundance at the surface of
an RGB star due to an extra mixing process (the depth reached by the extra 
mixing should extend down to at least 40 MK in this case), would also
imply an increase of the surface abundances of both N and Na. Isotopes
like $^{18}$O and $^{22}$Ne are expected to be fully destroyed,
while the $^{12}$C/$^{13}$C ratio should drop and the
 \index{isotopes!22Ne} \index{isotopes!13C} \index{isotopes!14N}
$^{14}$N/$^{15}$N increase. Note that in any case 
it would be very difficult to
obtain a surface change of the Ne and the Mg abundances because their
most abundant isotopes, $^{20}$Ne and $^{24}$Mg, are not modified
unless the mixing reaches down to the location of main H burning. 

In summary, our modeling of mixing in stars is still
oversimplified and unrealistic as it is based on a simple buoyancy model.
Observational evidence of stellar abundances also involving radioactive
nuclei and their daughters points out that mixing of matter outside the
standard convective boundaries should occur in stars. These observations
can be used to improve our description of mixing phenomena in stars.


\section{Evolution in the Double Shell Burning Phase}\label{sec:agbhead}

We start Subsection~\ref{subsec:AGB} by describing the central He-burning
phase and the direct scaling of the mass of the convective core resulting
from He burning with the mass of the He core. The mass of the convective
core determines the size of the initial He-exhausted core, an important
parameter for subsequent evolutionary phases. As it happens previously
when H burning shifts from the centre to a shell, also when He is
exhausted in the core and He burning shifts from the centre to a shell, the
envelope is forced to expand and convective motions extend from the
external layers deeply inward into the star.  In stars more massive than 4:5
$M_\odot$ the convective envelope even penetrates within the He core
reducing its mass size and carrying to the stellar surface material
processed by nuclear reactions (2$^{nd}$ dredge-up). If the He-exhausted core
does not grow above $\sim$ 1.1 $M_\odot$, an electron degenerate core
forms again, this time made of C and O, on top of which are located two
burning shells: the He-burning and the H-burning shells. This marks the
beginning of the double burning shell phase: the Thermally Pulsing
Asymptotic Giant Branch (TP-AGB) phase. \index{stars!AGB}

The two key features of this phase are that (1) the two burning shells can
not be simultaneously active, but they alternate within a cycle in producing the
required energy and (2) He ignition within each cycle occurs through 
a thermal runaway
(or thermal pulse, TP) that ends when enough energy is injected in the
He-burning zone to convert its temperature and density profiles into a
configuration that allows stable burning. The frequency of these thermal
instabilities scales directly with the He-core mass.  Such an abrupt
injection of a quite large amount of energy ($\sim 10^{48}$ erg) induces
first the growth of a convective shell within the 
{\it intershell} zone, located between the two shells, and
second, soon after this
convective region is extinguished, the expansion of the base of the
H-rich envelope, which forces the convective envelope to penetrate well
within the intershell zone (3$^{rd}$ dredge-up).  The
combination of these two successive convective episodes allows nuclei
freshly synthetized by He burning to be carried up to the stellar
surface. Moreover, the temperature at the base of the convective
envelope scales directly with the He-core mass, and, in stars more
massive that 4--5 \Msun, reaches high enough values that H burning
activates (Hot Bottom Burning, HBB). \index{process!hot bottom burning}

In Subsection~\ref{subsec:superAGB} we discuss Super-AGB stars, i.e.,
stars with initial mass that locates
them in the interval between stars that
develop an electron degenerate core after He is exhausted in the center
and enter the AGB regime, and more massive stars that do not develop an
electron degenerate core. Super-AGB stars ignite carbon out of center in
semidegenerate conditions and go through a central C-burning phase.
However, the C-exhausted core is not massive enough to heat up to the Ne
burning, so an electron degenerate ONeMg core forms. These stars then go
through a thermally pulsing phase. The final fate of these stars depends on
the capability of their ONeMg core to reach the critical mass of $\sim
1.35~\Msun$ required to activate electron captures on $^{24}$Mg and
$^{20}$Ne. Stars with a core that does not reach this crital mass lose
all their H-rich envelope and end their life as ONeMg white dwarves, while
stars with a core that reaches this critical mass explode as 
\emph{electron-capture supernovae}. 
\index{supernova!electron capture}

We continue by briefly discussing mass loss during the AGB phase in
Subsection~\ref{subsec:mlossAGB}. The strong increase in surface
luminosity, coupled to luminosity variations and formation of dust grains
in the atmospheres of AGB stars, strongly enhances the mass-loss rate in this 
phase with the
consequence that all AGB stars lose their H-rich envelope, leaving behind
the naked electron degenerate core as a cooling CO white dwarf. \index{stars!white dwarf} Finally, in
Subsection~\ref{subsec:dustAGB}, we discuss the different species of dust
grains that form in the atmosphere of an AGB star. The key role here is
played by the C/O number ratio in the atmosphere because the strong bond
of the CO molecule results in trapping all of the atoms of the least 
abundant of the two
elements. In an oxygen-rich gas (O$>$C) the species of dust are oxide grains,
for example, Al$_3$O$_2$ (corundum) and many different types of silicates
(SiO, SiO$_2$, etc). In a carbon-rich gas (C$>$O), the species of dust are, \index{stardust!condensation} 
for example, SiC (silicon carbide) and C itself (graphite). Some of this
stellar AGB dust is now recovered from primitive meteorites, representing
a real speck of an ancient AGB star that we can analyse in the laboratory. 

\subsection{Asymptotic Giant Branch (AGB) Stars} \label{subsec:AGB} 

As anticipated at the end of Subsection~\ref{subsec:rgb}, 
once the central temperature in a RGB star 
exceeds 100 MK, He in the core starts being 
converted into $^{12}$C via $3\alpha$ reactions, and subsequently into 
$^{16}$O via 
$^{12}$C$(\alpha,\gamma)$ $^{16}$O reactions. The cross section of 
the $3\alpha$ reaction has a tremendous 
dependence on the temperature: it scales roughly as $\rm T^{23}$ in the 
range 100-300 MK, so that the energy produced by these reactions is very 
strongly concentrated towards the centre of the star where the temperature 
is at its maximum. The very large photon flux that forms in these conditions 
triggers the formation of large scale motions of the matter, which mix the 
material in the central part of the star (the convective core) in order to efficiently carry the energy outward. The mass of 
the convective core depends on the luminosity produced by the \index{stars!structure} 
$3\alpha$ reactions. This luminosity scales with the mass of the He core 
because the larger its mass, the larger is the amount of energy required to 
maintain the hydrostatic equilibrium (see Section~\ref{sec:basic}). Hence, the size of the 
convective core scales directly with the mass of the He core. The mass of the 
He core, in 
turn, scales directly with the initial mass of the star, thus, in conclusion, 
the mass of the convective core scales directly with the initial mass of 
the star. Analogously to the Main Sequence central H-burning phase, 
the energy production in the core is dictated by the mass of the 
star (see Section~\ref{subsec:hburn}). However, the role played by the total stellar 
mass in central H burning in the central He-burning phase is replaced by the He-core mass 
because the 
density contrast between the He core and the H-rich mantle is so large 
that the core does not feel the presence of the H rich 
mantle and evolves as if it was a naked He core.

In the meantime, the temperature at the He/H interface is high enough
that also an efficient H-burning shell is active leading to continuous
deposition of fresh He onto the He core. Moreover, large convective
motions develop (in most cases) in the outer H-rich envelope. The actual
extension and temporal variation of these convective regions depends on
the initial mass and chemical composition of the star.

At variance with H burning, no radioactive nuclei are produced by the
$3\alpha$ and the $^{12}$C$(\alpha,\gamma)$ $^{16}$O reactions
because they convert matter along the valley of $\beta$ stability. 
Radioactivity during He burning is produced instead via the sequence of 
reactions that convert $^{14}$N into $^{22}$Ne via a double $\alpha$ 
capture and the radioactive decay of $^{18}$F: \index{isotopes!18F} \index{isotopes!14N}
$^{14}$N$(\alpha,\gamma)$ $^{18}$F$(e^++\nu)$ $^{18}$O$(\alpha,\gamma)$ $^{22}$Ne.
In H-exhausted regions, 
$^{14}$N is by far the most abundant nuclear species after 
He because a main effect of the CNO cycle, which operated in 
the previous H-burning phase (see Section \ref{subsec:hburn}) is 
to convert most 
of the initial C and O, the two most abundant elements beyond H 
and He, into N. Hence, during He burning 
$^{22}$Ne becomes the most abundant \index{isotopes!22Ne}
isotope, after C and O, once $^{14}$N is fully consumed by $\alpha$ 
captures.

When He is exhausted in the centre, He burning moves smoothly outward in
mass leaving behind a CO core that begins to contract on a Kelvin-Helmholtz
timescale. Similar to the H-burning shell, also the He-burning shell
produces more energy than required to balance gravity 
because energy production is controlled by the size of
the underlying core, and not by the mass of the star. The CO core
increases progressively in mass because of the continuous deposition of
CO-rich material made in the He-burning shell and the He-burning shell
increases its energy production accordingly. As a consequence, the
overlying He+H-rich mantle is forced to expand substantially and to cool
down so much that the H-burning shell switches off. As during the RGB phase, 
this expansion progressively 
inhibits energy transport by radiation and large
scale motions of the matter progressively extend inward from the outer
envelope. In stars initially more
massive than 4--5 \Msun the convective envelope penetrates even inside
the He core (2$^{nd}$ dredge-up). The main consequences of this are a change of the
surface abundances and a reduction of the He-core mass. Similar
to what happens during the RGB, the formation of
an extended convective envelope forces the star to expand at roughly
constant surface temperature (because the onset of convective motions
fixes a maximum value for the temperature gradient, see Section
\ref{subsec:hburn}) and increasing luminosity. This phase is called \index{stars!AGB}
Asymptotic Giant Branch (AGB). The specific phase when the He-burning 
shell advances in mass eroding the border of the He core from within
is called Early Asymptotic Giant Branch (E-AGB). 

The competition between the advancing He-burning shell
and the sinking of the convective envelope during the 2$^{nd}$ dredge-up 
fixes the maximum mass that
the CO core ($\rm M_{\rm CO}$) reaches in this phase. If $\rm M_{\rm
CO}$ is larger than roughly 1.1 \Msun, the core heats up
to the C ignition temperature ($\sim 8 \times 10^8$~K), otherwise it
turns into an electron degenerate CO core able to self-sustain against
gravity without the need of additional contraction. The maximum
initial stellar mass for which an electron degenerate CO core forms is
of the order of 7--8 \Msun, for solar metallicity stars. While the
evolution of stars without an electron degenerate core is
dictated by the self gravity of the core, the evolution of stars with 
an electron degenerate core is controlled by the burning shells. In
the following we concentrate on the further evolution of the latter case,
i.e., the AGB, while Chapter 4 describes the further evolution of
the first case.

On the AGB three main regions may be identified: the electron degenerate 
CO core, a He-rich layer (also referred to as \emph{intershell} since it is 
located between the He- and the H-burning shells), and a H-rich mantle, 
most of which forms an extended convective envelope. As the He-burning 
shell approaches the border of the He core, it quenches because of the 
steep temperature drop associated with the drastic reduction of the mean 
molecular weight caused by the change from a He-dominated 
to a H-dominated chemical composition. Being less and less supported 
by the extinguishing He burning shell, the mantle is forced to shrink, 
heat up, and progressively re-activate the H-burning shell at its base. 
The H-burning shell starts to deposit fresh He onto the He shell  
forcing the 
intershell to heat up again. At this point a fascinating evolutionary 
phase begins in which nuclear burning and instabilities 
coexist, realizing a unique {\it habitat} in which a large 
number of nuclear species may be synthesized: the Thermally Pulsing AGB 
(TP-AGB) phase.

\bigskip \begin{figure}
  \centering
  \includegraphics[width=0.6\textwidth]{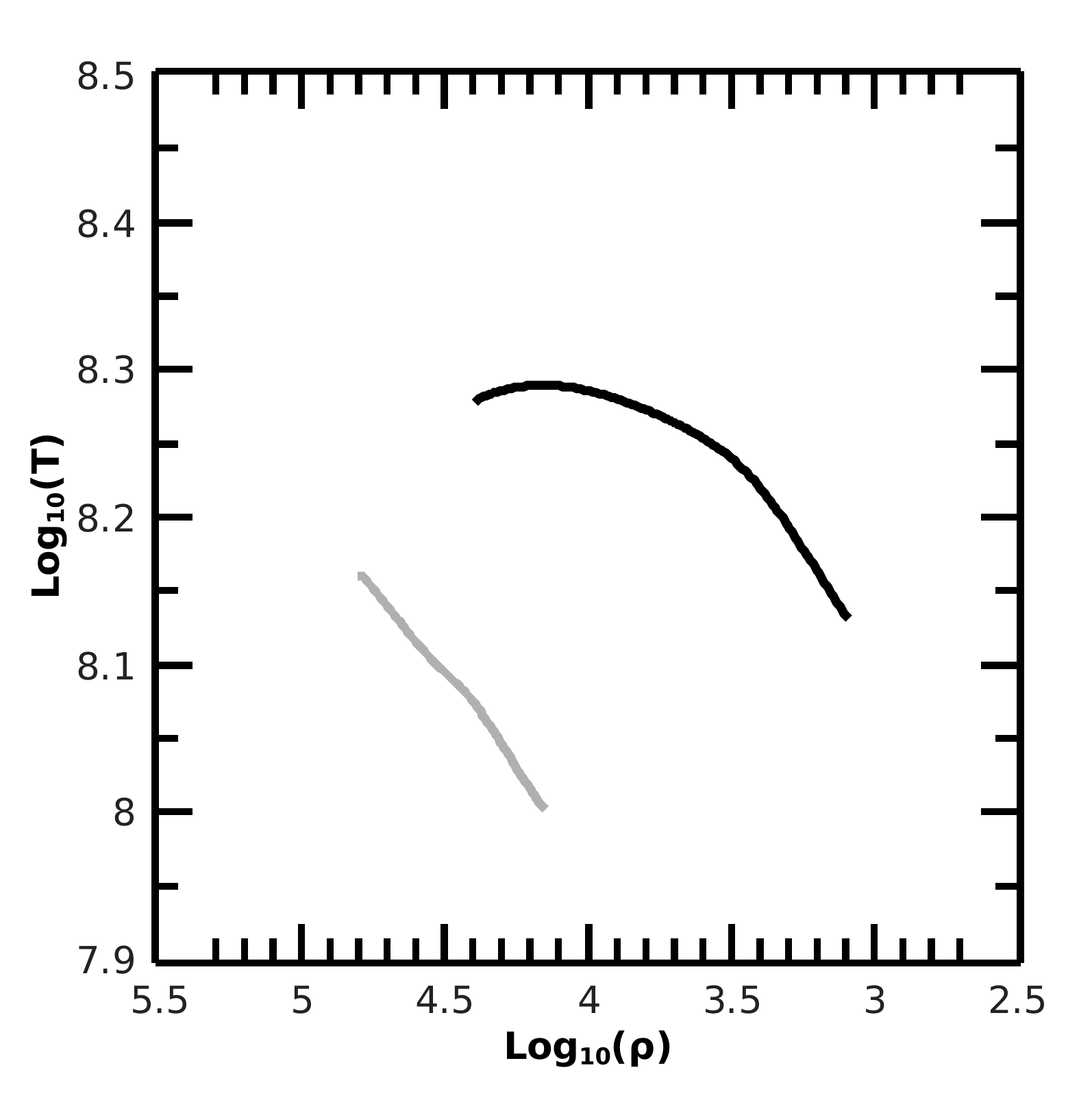}
  \caption{The gray thick solid line shows the 
typical Log(T)-Log($\rho$) profile in the intershell (in the range
  $10^{-3}<$X$_{\rm He}<0.9$) just prior the onset of a thermal pulse in a 
3 \Msun\ of solar metallicity while the black thick 
  solid line shows the typical profile in the same region 
at the end of the thermal runaway when the steady He burning occurs.}
  \label{fig:trocurve} 
\end{figure}

Quite schematically, the TP-AGB phase consists of a sequence of cycles each 
of which may be divided in two main phases: a quiescent H-burning phase 
during which the He-burning shell is inactive, and a  He-burning
phase during which the H-burning shell is inactive. Though the two shells do 
not operate simultaneously, they process roughly the same amount of mass 
per cycle so that the intershell mass changes (shrinks) slowly in time. The 
transition from the active He-burning phase to the active H-burning phase 
occurs quiescently in the sense that the energy provided by the H-burning shell 
progressively replaces that provided by the dimming He-burning shell. 
Instead, the transition from the active H-burning phase to the active He-burning phase 
occurs in a traumatic way, which is responsible for the peculiar sequence of 
events that characterizes the TP-AGB phase. \index{shell burning}

The reason for such a traumatic He ignition is that the pileup of fresh
He on top of an inert intershell leads to a T, $\rho$ profile in the
intershell that is controlled by the compressional heating caused by the
accretion of fresh He. This T, $\rho$ profile is quite different from the
typical one determined by the presence of an active burning shell. The
large amount of energy required to turn the T, $\rho$ profile from that 
determined by the accretion and that required by the steady He burning,
coupled to the very steep dependence of the cross section of the $3\alpha$
nuclear reaction on the temperature, determines the growth of a 
thermal runaway (or \emph{thermal pulse}, TP) in which a huge amount of
energy is released over a very short timescale. This runaway comes to an
end when enough energy has been deposited in the intershell to turn the T,
$\rho$ profile into a profile suited for quiescent He burning. 
As an example, Figure~\ref{fig:trocurve} shows as a gray  
line the typical Log(T),Log($\rho$) profile produced by the advancing 
H-burning shell just prior to $3\alpha$ ignition, while the black  
line shows the typical profile at the end of the thermal runaway during  
steady He burning. In this specific example 
roughly $\sim 10^{48}$ erg must be deposited in the 
intershell to perform the transition between the two configurations. 
This amount of energy is determined by the fact that the needed change of the 
$T,\rho$ structure in the intershell 
requires a reduction of the binding energy of the intershell.

The main effect of the rapid injection of energy into the intershell
during the TP is the production of a very strong energy flux, which
forces the growth of convective instabilities to efficiently carry the
energy outward.  This convective shell extends over most of the
intershell region and plays a fundamental role in reshuffling the
chemical composition within this region and hence influencing the
detailed nucleosynthesis that occurs at this stage (see next Sections).
Once the TP comes to an end, the convective shell disappears and the
quiescent He-burning shell phase begins. Another important side effect
of the rapid energy injection caused by the TP is the expansion of the
region above the He-burning shell, which forces a cooling of
this region. The consequence is that the H-burning shell switches off, 
and the temperature
gradient steepens. This favors the penetration of the convective
envelope down into the intershell so that 
nuclei freshly synthetized in the deep interior of the star are
efficiently brought up to the stellar surface (3$^{rd}$ dredge-up).
Similar to what
happens towards the end of the E-AGB phase, the He-burning shell
progressively runs out of power as it approaches the border of the He-rich 
layer, where the temperature drops below the value necessary for the He
burning. The overlying layers are forced to contract
and heat so that a H-burning shell activates again and a new cycle
starts.

\bigskip \begin{figure}
  \centering
  \includegraphics[width=.9\textwidth]{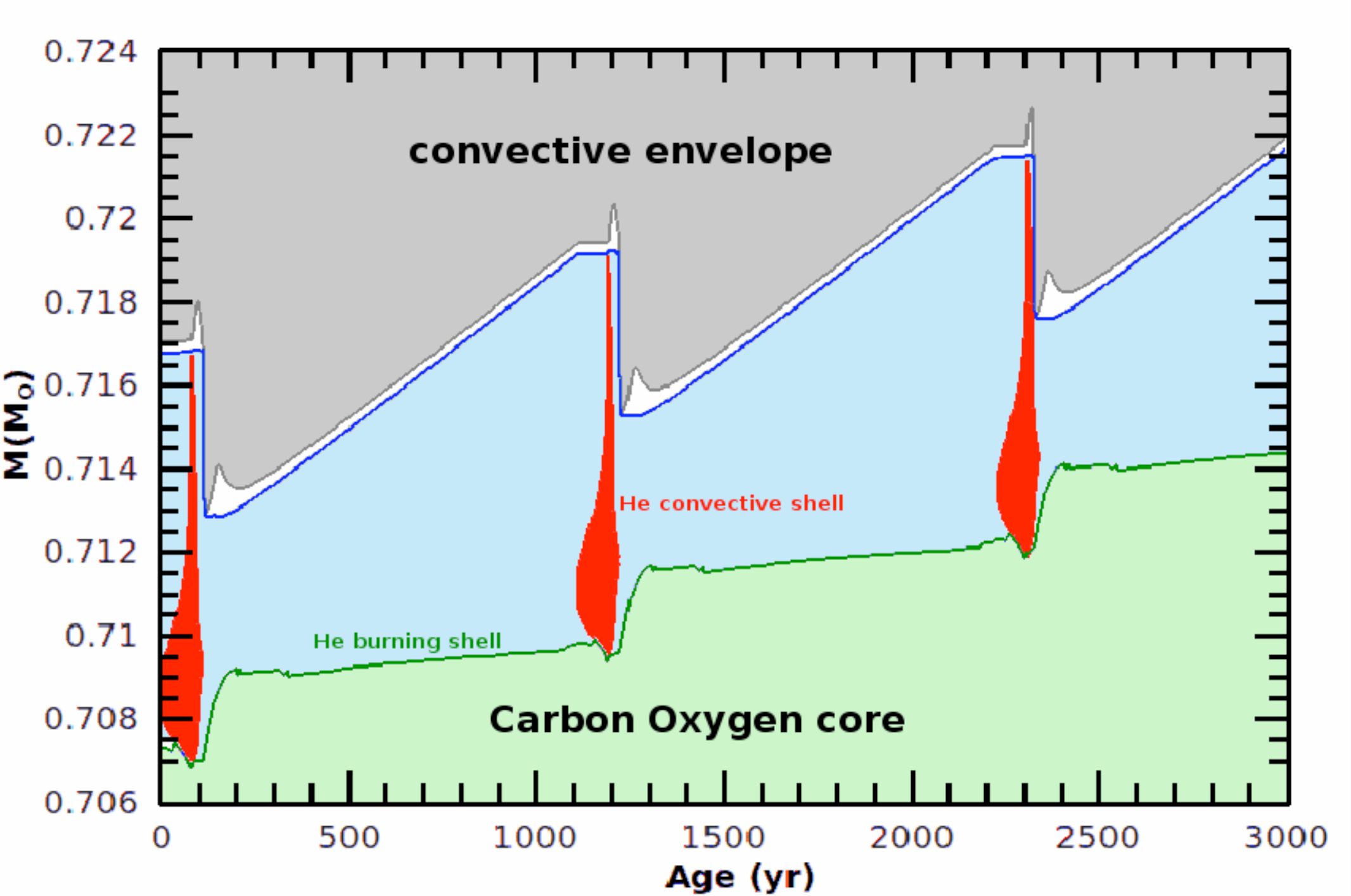}
\caption{Temporal evolution of the internal structure of a typical AGB
star through three consecutive TPs. The He convective shell is shown in
red while the convective envelope is grey. The H and He burning shells
are shown in dark blue and dark green, respectively, while the
intershell is light cyan. The timescale between the end of the He
convective shell episode and the beginning of the successive one has
been rescaled in order to improve readibility. In particular it has been
changed according to this formula: t=$\rm t_{\it end~conv~sh}$+(t-$\rm
t_{\it end~conv~sh}$) / ($\Delta t_{\it interpulse} \times 10^3$). In
this way the interpulse phase is rescaled to last $10^3$ yr.}
  \label{fig:convshell} 
\end{figure}

To visually illustrate the sequence of events making up a full TP cycle
and to make clear the pecularity of the TP-AGB phase, Fig.
\ref{fig:convshell} shows the temporal evolution of the internal structure
of a typical AGB star through three consecutive TPs. 

The quantitative characteristics of the TPs depend on the core and 
envelope masses, the general rule being that larger CO core masses 
correspond to higher frequencies of thermal pulses, higher  
temperatures, and shorter lifetimes of the He convective shell. Typical TP 
frequencies (determined after the first 20 TPs or so) range between 2 and 3 TPs 
per $10^5$ yr for a 3 \Msun\ star having $\rm M_{CO}\sim0.7\Msun$ and 
more than 50 TPs per $10^5$ yr for a 6 \Msun\ star having $\rm 
M_{CO}\sim0.96\Msun$, while the peak luminosity ranges between 
$\rm 1 and 10~10^8$\lsun.

Since the two burning shells process about the 
same amount of matter per cycle, the average growth rate of the CO core 
per cycle roughly equates that of the He core. This, coupled 
to the fact that He burning produces (per unit mass) roughly 10\% of the 
energy produced by H burning, and that the luminosity of these stars 
does not change appreciably between the two burning phases, allows us to 
estimate the relative burning lifetimes ($\rm t_{\rm He}/t_{\rm 
H}$). The amount of energy produced by He burning per cycle 
must balance the surface losses, i.e., $\rm \epsilon_{\rm He} \times \Delta 
M_{\rm He}=L_{\rm surface} \times t_{\rm He}$, where $\epsilon_{\rm He}$ 
represents the amount of energy liberated by the He burning per unit mass, 
$\rm \Delta M_{\rm He}$ the amount of mass processed by the He burning, 
$\rm L_{\rm surface}$ the luminosity of the star and $\rm t_{\rm He}$ the 
lifetime of the He burning phase. Similar for the H burning one may 
write that $\rm \epsilon_{\rm H}\times \Delta M_{\rm H}=L_{\rm surface} 
\times t_{\rm H}$. If the amount of mass processed is similar in the two 
cases (i.e., $\rm \Delta M_{\rm He}\simeq \Delta M_{\rm H}$), the relative 
lifetimes scale roughly as the two nuclear burning rates, i.e., $\rm t_{\rm 
He}/t_{\rm H} \sim 1/10$. The amount of He burnt during each 
He-burning episode is only partial, 
corresponding to about 25\%-30\% of the He present in the intershell.
Of this, roughly 25\% burns during the TP and the remainder during 
the quiescent He burning phase. The final nucleosynthetic result is that 
carbon is produced via the 3$\alpha$ reaction, but it is only  
marginally converted into 
oxygen. The typical composition of the intershell after this partial He burning 
is represented by matter 
made by roughly 75\% He and 23\% $^{12}$C, while the remaining few percent 
are made up of $^{22}$Ne (from conversion of $^{14}$N as detailed above) and 
of $^{16}$O. The $^{22}$Ne nuclei are of interest as they act as a neutron 
source in the TPs when the temperature reaches 300 MK.
The lifetime of the He convective shell varies 
between 100 yr and 10 yr for the 3 \Msun and 6 \Msun stellar models, 
respectively. Typically, the He burning shell is located 
between $7\times10^{-3}$ and $1.5\times10^{-2} \rsun$ from the center of 
the star, while the H-burning shell is located between $1\times10^{-2}$ 
and $2\times10^{-2} \rsun$. The intershell mass ranges roughly between 
$10^{-3}$ and $10^{-2} \Msun$. The surface radii of AGB  \index{stars!AGB} 
stars vary between hundreds to thousands times the solar radius.

The final fate of AGB stars is to lose all their H-rich mantle before the
electron degenerate core may grow to its most massive stable
configuration (i.e., the Chandrasekhar mass). Such a destiny is due to the
strong dependence of the mass-loss rate on the luminosity of the star
and on its surface chemical composition (see Section
\ref{subsec:mlossAGB}). The maximum mass size reached by the CO core,
which equates the mass of the newborn white dwarf, \index{stars!white dwarf} is
determined by the competition between the speed at which the
burning shells advance in mass and the efficiency of the mass loss that
erodes the H-rich mantle from the surface.

The occurrence of the 3$^{rd}$ dredge-up significantly affects the
evolutionary properties of an AGB star. First, it reduces the size of
the He core anticipating the quenching of the quiescent He burning phase
and hence its lifetime. Second, it slows down the overall growth rate of
the CO core and the He-rich shell.  Third, it carries to the stellar
surface a fraction of the material freshly synthesized by partial He
burning, i.e., C, $^{22}$Ne, and $slow$-neutron capture ($s$-process)
elements heavier than iron (see Section~\ref{sec:ncapAGB}), 
drastically modifying the
chemical composition of the star. In some cases, the star  
even changes
from the usual oxygen-rich (O$>$C) composition to carbon-rich (C$>$O),
with important consequences on the types of molecules and dust that can
form and the ensuing mass loss (see Section~\ref{subsec:dustAGB}).
Unfortunately, the question of the maximum depth reached by the
convective envelope during the 3$^{rd}$ dredge-up has always been highly   
debated and different results have been obtained over the years by
different authors for AGB stars over the whole mass interval from
1.5 \Msun~ up to the more massive thermally pulsing stars. The reason
is that, once the convective envelope enters the He core, a discontinuity in
the opacity (H is much more opaque than He) determines the formation of a
positive difference between the effective and adiabatic temperature 
gradients just at the
border of the convective envelope. This is an unstable
situation because the possible mixing of matter located just below the
base of the convective envelope with the H-rich convective mantle is an
irreversible process in the sense that these \emph{additional} mixed layers
become intrinsically convective because of the drastic increase of the
opacity due to the mixing. It is therefore clear that even small
different numerical techniques adopted by different authors may lead to
quite different results.

Furthermore, the occurrence of the 3$^{rd}$ dredge-up is important because it 
creates a sharp discontinuity between the convective envelope and the 
radiative intershell. Since a
sharp discontinuity is not a realistic configuration in these conditions, 
the occurrence of the 3$^{rd}$ dredge-up 
allows the possibility 
that some kind of \emph{diffusion} of protons occurs below the formal border of the
convective envelope when it reaches its maximum inward extension at the
end of the 3$^{rd}$ dredge-up smoothing out the discontinuity
(though this is not obtained by applying
the standard stability criteria for mixing). 
However, the shape, extent, 
and timescale over which the
diffusion of protons in the He/C intershell may occur is 
unknown, its modeling is still artificial and not based on
self-consistent computations. 

This diffusion allows the formation of regions where a
small amount of protons come in contact with matter that is 
predominantly composed 
of He and C, so that the ratio $\rm Y(H)/Y(C) \ll 1$, but does not 
contain any $^{14}$N, since this nucleus has been fully converted into
$^{22}$Ne in the previous TP. When these proton-enriched layers begin
to contract and to heat because of the quenching of the He burning, the
CN cycle activates, but it can not go beyond the synthesis of $^{13}$C
due to the low proton concentration. As the temperature increases to
roughly 90 MK, the $^{13}$C$(\alpha,{\textbf n})$ $^{16}$O
reaction becomes efficient and a significant neutron flux is produced. 
Hence, this diffusion plays a \index{process!neutron capture}
\index{isotopes!13C} \index{isotopes!22Ne}
pivotal role in the nucleosynthesis of species beyond the
Fe peak via neutron captures. 
A detailed description of the properties of
this neutron source and of its nucleosynthetic signature will be
presented in Section~\ref{subsec:sources}.
The
lack of $^{14}$N is crucial here, since this nucleus is a strong 
neutron poison and
its presence would inhibit neutron captures by Fe and the elements 
heavier than Fe.

As already discussed in Section~\ref{subsec:li}, 
typical temperatures at the base of the convective envelope do not exceed 
a few MK at most in the evolutionary phases prior to the AGB. Instead, 
another peculiarity of AGB stars is that during the H-burning phase 
the temperature at the base of the convective envelope may reach values in 
excess of several tens MK, and even exceed 100 MK, so that H-burning 
reactions activate within the 
convective envelope. \index{process!hot bottom burning}
The efficiency of this phenomenon, known as Hot Bottom Burning (HBB), 
scales directly with the temperature at the base of the envelope and hence 
with the CO-core mass, which in turn scales with the initial stellar mass. 
Hence, HBB is efficient in stars more massive than 4:5 \Msun, depending on the 
metallicity.
As the energy produced in the convective envelope sums to that produced by 
the H-burning shell, the core mass - luminosity relation changes (even 
strongly) in the presence of HBB. 
From a nucleosynthetic point of view the occurrence of an active H burning in a 
convective environment implies a redistribution of the processed material 
over all the convective zone, so the surface abundances turn towards the 
relative abundances typical of the H burning at high temperature. For 
example, an increase of the surface abundances of nuclei like 
$^{14}$N and $^{26}$Al (discussed in detail in 
Section~\ref{sec:al26}), a temporaneous increase of $^7$Li, a 
reduction of $^{12}$C and of the $^{12}$C/$^{13}$C ratio and the 
signatures of the CNO, NeNa and the MgAl sequences. \index{isotopes!14N} \index{isotopes!13C} \index{isotopes!26Al}

We refer the reader to the review paper by  \citet{herwig05} and to the 
book chapter on the evolution of AGB stars by  \citet{lw04} for a thorough 
presentation of the evolution of these cool giant stars.

\subsection{Super-AGB Stars} 
\label{subsec:superAGB}

In the previous section we identified stars that go through the double \index{stars!super-AGB}
shell burning of the TP-AGB phase as those that develop an electron 
degenerate CO core where carbon burning fails to occur. There is, 
however, another class of stars that experience the double shell burning 
phase: those with initial total mass between the maximum mass that forms 
an electron degenerate CO core where C does not ignite ($\rm M_{\rm 
up}$) and the minimum mass that does not form an electron degenerate CO 
core ($\rm M_{\rm mas}$). Stars more massive than $\rm M_{\rm mas}$ 
evolve up to the final core collapse as described in Chapter 4. 
Depending on the initial chemical composition and the adopted physics, 
$\rm M_{\rm up}$ ranges between 6--8\Msun, and $\rm M_{\rm mas}$ ranges 
between 9--12\Msun. It is important at this point to recall that these 
limiting masses are somewhat uncertain because they depend on 
the size of the convective core, the carbon to oxygen ratio left by the 
He burning, the efficiency of the 2$^{nd}$ dredge-up and the cross section 
of the $^{12}$C$+$$^{12}$C nuclear reaction. Unfortunately, all 
these quantities are still subject to severe uncertainties.

Stars falling between these two limits form a partially electron 
degenerate core, but are massive enough to ignite C in the core, lift 
the degeneracy, and go through the C burning in the core. They are not 
massive enough, however, to avoid the electron degeneracy of the ONeMg 
core left by C burning. The evolution of stars in this relatively small 
mass interval, called Super-AGB stars, has not been studied extensively 
up to now because of the difficulty in computing the C-burning phase
due to the removal of the degeneracy that occurs 
through a series of successive flashes, and the lack of massive computer 
power, which is needed to study this complex situation. This situation is rapidly changing and 
progress is currently under way on the computational modeling of 
Super-AGB stars  \citep[]{siess06,siess07}.

Since these stars form an electron degenerate core after core C burning, \index{stars!structure}
they also go through a double shell burning phase similar to the AGB phase experienced by their less 
massive counterparts. Because their degenerate cores 
are more massive, following the trend shown by AGB stars, the 
frequency of the thermal pulses is higher (up to 500 TPs per $10^5$ yr), 
the He peak luminosity is lower than in the normal AGB stars (up to  
$\sim 4 \times 10^{6}$\lsun), while the base of the convective envelope 
may reach temperatures as high as 110 MK, hence, H burning occurs 
within the convective envelope (Hot 
Bottom Burning). Similar to what happens in the more massive AGB stars, 
but quantitatively more pronounced, the luminosity produced in the 
convective envelope adds to that produced by the radiative H burning 
significantly altering the core mass - luminosity relation and the 
surface composition is modified by the signature of H burning.
The possible occurrence of the 3$^{rd}$ 
dredge-up would also shuffle the surface chemical composition with 
the typical products of the partial He burning, i.e. C, and $s$-process 
elements. 
The efficiency of the 3$^{rd}$ dredge-up is very uncertain also 
for these stars. In principle, one could 
expect a lower efficiency of the 3$^{rd}$ dredge-up because the amount of 
energy released by a TP is lower and the overall temperature is much 
higher than in a standard AGB star, so that it could be more 
difficult to expand the base of the mantle and to steepen the temperature 
gradient up to a value that would allow the convective envelope to penetrate 
the He core. Quantitative estimates of the yields of the nuclei 
specifically produced by 
the TP Super-AGB stars are in progress \citep[]{dl06}.

The final fate of a Super-AGB star depends on the competition between the 
advancing burning shells, which increase the size of the ONeMg core, and 
the mass loss, which limits its growth. Also an efficent 3$^{rd}$ dredge-up 
would contribute to limiting the growth of the core. 
Stars more massive than a critical 
value reach the threshold electron degenerate core mass for 
the onset of electron captures on $^{24}$Mg and $^{20}$Ne
after a certain number of TPs and eventually explode as \emph{electron-capture 
supernovae}. Stars less massive than the  
critical value, instead, end their life as ONeMg white dwarfs. \index{stars!white dwarf}
An estimate of the electron degenerate core mass 
above which electron captures become efficient in an ONeMg environment
can be determined by considering that the threshold energy for electron 
capture is 6 MeV for $^{24}$Mg and 8 MeV for $^{20}$Ne and that the 
mass of a fully electron degenerate core having a Fermi energy of the 
order of 6 MeV is $\simeq 1.35\Msun$. 
Thus, if the electron degenerate core grows to the threshold value of
$\simeq$ 1.35 \Msun\ or so, electron captures are activated on
$^{20}$Ne and $^{24}$Mg. 

This process removes electrons and hence
pressure from the center of the star, starting a runaway process that
leads to the core collapse and final explosion as electron-capture
supernova. The explosion of these electron-capture
supernovae is similar to that of core collapse supernovae (see Chapter~4),
with a few distinct features.  During the initial collapse of
the degenerate core, electron captures increase significantly the degree
of neutronization of the matter, i.e., raise the global neutron over 
proton ratio because of the capture of the electrons by the protons.
The nuclear species produced by explosive burning depend 
significantly on the neutron over proton ratio so that the higher the
degree of neutronization of the matter the higher the production of neutron-rich 
nuclei: in particular $^{58}$Ni becomes favored with respect to $^{56}$Ni. 
\index{isotopes!56Ni} \index{supernova}
Since the luminosity peak of a supernova correlates with the amount of
$^{56}$Ni produced during the explosion, a natural feature of these
electron-captures supernovae is a lower 
luminosity with respect to typical core collapse
supernovae.  Also, the
final kinetic energy of the ejecta is expected to be of the
order of $0.1\times10^{51}$ erg, roughly one order
of magnitude lower than in typical core collapse supernovae 
\citep[see, e.g.,][]{hmj08,wnjkm09}.

\subsection{Winds from AGB Stars} 
\label{subsec:mlossAGB}

An observed peculiarity of AGB stars is that they show strong
stellar winds, which carry material away from the surface of the star
into its surroundings. Nuclei newly synthetised during
the AGB phase and carried to the stellar surface by the 3$^{rd}$ dredge-up 
are shed into the interstellar medium so that AGB stars
contribute to the chemical make-up of their environments and of new
generations of stars. The mass loss rate due to winds in
AGB star increases as the star evolves along the AGB and can reach values as
high as 10$^{-4}$ \Msun/yr (to be compared, for example, to the solar mass loss rate of
10$^{-11}$ \Msun/yr) at the end of the AGB, which is known as the
\emph{superwind} \citep[]{iben83}. This is a strong and dense but slow wind,
with material leaving the star at relatively low speeds of 5-30 km/s.

The winds are caused by two main factors. \index{stars!stellar wind}
First, large quantity of dust form around AGB stars and 
radiation pressure acting on this dust contributes to driving the winds. 
The extended envelopes of red giant and AGB stars, where the temperature
drops down to $\sim$1,000 K, are an ideal location for the formation of a
large variety of molecules like CO, TiO, VO, as well as ZrO, when the gas has
been enriched in heavy elements such as Zr by the $s$-process and the
3$^{rd}$ dredge-up, and C$_2$, CN, and CH, when the gas has been
enriched in carbon by the 3$^{rd}$ dredge-up. In the case of refractory
elements, which have the property of condensing at high temperatures
directly from gas into the solid state, the gas condenses into tiny
particles, which then can grow into dust grains. 
Because of the large quantity of dust around them, AGB stars become
obscured toward the end of their life and can only be seen as mid-infrared
sources, since the dust absorbes the energy of the visual light coming
from the star and reemits it as infrared light.
Second, AGB stars are variable stars, 
meaning that their luminosity varies with time with changes occurring 
over relatively long periods $>$100 days. These luminosity 
variations are due to stellar pulsations, in the sense that the whole star 
expands and contracts. Pulsations produce changes in the stellar radius 
and temperature, which cause the variations in the stellar luminosity. 
When the pulsations attain a large amplitude they lead
to strong stellar winds and a large mass-loss rate.
Pulsation levitates matter above the photosphere and
increases the wind density by about two orders of magnitude \citep[]{wood1979,sedlmayr95,dorfi01}. 

The strong stellar winds driven by the
combined effects of radiation pressure acting on dust and pulsation
eventually erode the whole stellar envelope \citep[]{dupree86,wilson00}. 
Hence, the winds govern the
lifetime of AGB stars because when the envelope is almost completely
lost the star moves away from the AGB phase into the hotter post-AGB
phase. Toward the end of the post-AGB phase, the shell of material
ejected by the AGB star may become illuminated by the radiation coming
from the central star, and produce a planetary nebula. The former AGB
stars is now referred to as a \emph{planetary nebula nucleus} and 
finally turns into a cooling CO white dwarf. \index{stars!white dwarf}


\subsection{Dust from Giant Stars and the Origin of Stardust}
\label{subsec:dustAGB}

The specific dust species that form in the atmosphere of AGB  \index{stars!AGB} \index{stardust!condensation}
stars depends mainly on the $\rm C/O$ ratio. The difference in the type
of dust that can form in a carbon-rich or oxygen-rich gas is due to the
strong bond of the CO molecules: if O$>$C, all carbon atoms are locked
into CO and only oxygen-rich dust can form, viceversa, if C$>$O, all
oxygen atoms are locked into CO and only carbon-rich dust can 
form\footnote{This general rule is debated in the case of dust formation in
supernova ejecta, see Chapter 2, Section 2.5.3}. 
In an oxygen-rich gas    
(O$>$C) dust species 
are, for example, Al$_3$O$_2$ (corundum), CaAl$_{12}$O$_{19}$
(hibonite), MgAl$_2$O$_4$ (spinel), as well as many different types of
silicates (SiO, SiO$_2$, etc). In a carbon-rich gas (C$>$O), dust species are,
for example, SiC (silicon carbide), TiC (titanium carbide), and C itself
(graphite).

Formation of dust around AGB stars is well documented by spectroscopic
observations in the infrared  \citep[e.g.,][]{treffers74,speck00,speck09}
and predicted to occur by theoretical models
\citep[e.g.,][]{fleischer92,lodders95,gail99,fe02}. It is now
widely accepted that AGB stars are the most prolific
source of dust in the Galaxy. When summing up the contribution of the
different families of late red giant and AGB stars: i.e.,
spectroscopically, the M stars, the OH/IR stars\footnote{OH/IR stars
are cool red giants with strong hydroxyl (OH) masers and infrared (IR) emissions.}, and the
carbon stars, it results that $\sim$90\% of all dust of stellar origin
in the interstellar medium came from these sources \citep[]{whittet92}. 

\begin{table}
\caption{Meteoritic stardust grain types, populations, and origins$^a$\index{meteorites}\label{tab:gratypes}}
\begin{tabular}{lll}
\hline
Type & Population & Origin \\
\hline
oxide and silicate grains & I & AGB stars \\
 & II & AGB stars \\
 & III & undetermined \\
 & IV & supernovae \\
silicon carbide (SiC) & mainstream & AGB stars \\
 & Y & AGB stars \\
 & Z & AGB stars \\
 & X & supernovae \\
 & A+B & undetermined \\
 & nova grains & novae \\
silicon nitride & & supernovae \\
graphite & low-density & supernovae \\
 & high-density & AGB, post-AGB \\
diamond & & undetermined \\ 
\hline 
\end{tabular} \\
$^a$Table 5.2 (Chapter 5) gives complementary information to this table. 
\end{table} 

Thus, it is not surprising that the vast majority of stardust grains
extracted from meteorites (Chapter 2, Section 2.4. and Chapter 10, Section 2) show the signature of
an origin in AGB stars (Table \ref{tab:gratypes}. 
The main signatures of AGB nucleosynthesis imprinted in
meteoritic stardust grains are: (1) the O isotopic composition of the
majority of oxide and silicate grains showing excess in $^{17}$O and
deficits in $^{18}$O, and known as Population I and II of stardust oxide
grains  \citep[]{nittler97}, which match the O isotopic ratios observed
around AGB stars via spectroscopic observations of CO molecular lines
 \citep[e.g.,][]{harris87}, and (2) the distribution of the
 \index{isotopes!18O} \index{isotopes!13C} \index{isotopes!22Ne}
$^{12}$C/$^{13}$C ratios of $>$90\% of SiC grains showing a peak between
50 and 60 (solar value is 89) and known as the \emph{mainstream} SiC
population, which match the distribution derived from spectroscopic
observation of CO molecular lines in C-rich AGB stars  \citep[see Fig. 3
of][]{hoppe97a}. The Ne composition measured in stardust SiC grains -
corresponding to the Ne-E(H) component rich in $^{22}$Ne 
is also a clear signature of material from the intershell
of AGB stars, where $^{22}$Ne is abundant. Moreover, the elemental and
isotopic abundances of the heavy elements Kr, Sr, Zr, Ru, Xe (the Xe-S
component), Ba, Nd, Sm, W, and Pb present in trace amount and measured in
SiC grains clearly show the imprint of the $s$-process, which make
inevitable their connection to AGB stars. Smaller SiC Populations Y and Z
($\simeq$ 1\% each of the total recovered stardust SiC grains) are also
attributed to AGB stars, but of metallicity down to 1/3-1/5 of the solar value
\citep[]{hoppe97b, amari01a,zinner06}.

With regards to the remaining types and populations of stardust grains, 
core-collapse supernovae have been invoked as the origin site of 
Population X of SiC grains ($\sim$1\%) and the few recovered silicon 
nitride grains  \citep[]{nittler95}, showing excesses in $^{28}$Si and 
evidence of the early presence of $^{44}$Ti, as well as low-density 
graphite grains and 
Population IV of oxide and silicate grains  \citep[with excess in $^{18}$O 
and $^{18}$Si][see Chapter 4]{vollmer08,travaglio99}.  
Novae are invoked 
for a few SiC grains of unusal composition  \citep[excesses in $^{13}$C and 
$^{15}$N][see Chapter 5, Section 2]{amari01b}, while the origin 
of SiC grains 
of Populations A+B ($\simeq$5\% of all SiC grains, showing 
$^{13}$C/$^{12}$C $<$ 10) is still unclear  \citep[]{amari01c}. Oxide and 
silicate grains with deficits in both $^{17}$O and $^{18}$O, known as 
Population III, have been attributed to stars of metallicity lower 
than solar, however, the Si isotopic composition of the silicate grains 
belonging to this population is very close to solar, which does not 
support this interpretation. The origin of this population remains to be 
ascertained, together with the origin of high-density graphite grains and of 
the very abundant and extremely tiny ($10^{-9}$ m) meteoritic diamond grains, 
the majority of which probably formed 
in the solar system. For more details in meteoritic stardust see, e.g., 
 \citet{clayton04} and  \citet{lugaro05}.

Given compelling evidence that most stardust came from AGB stars, the 
composition of these grains can be used as a stringent constraint for 
theoretical models of AGB stars and, viceversa, the models can be used to 
identify the mass and metallicity range of the parent stars of the grains. 
Data from the laboratory analysis of stardust are usually provided with 
high precision, down to a few percent errors, and for isotopic ratios. 
In comparison, data from spectroscopic observations of stellar atmospheres 
usually are provided with 
lower precision, errors typically $>$50\%, and mostly for elemental abundances. 
Thus, the information from stardust grains represents a breakthrough in 
the study of AGB nucleosynthesis. Also, given that the abundances and 
isotopic compositions of elements heavier than Al and lighter than Fe, 
such as Si and Ti, are mostly unaltered by AGB nucleosynthesis, laboratory 
analysis of these elements in AGB stardust can be used to constrain in 
great detail the initial composition of the parent star of the grains, and 
in turn the chemical evolution of the Galaxy  \citep[e.g.,][]{zinner06}.

Meteoritic stardust provides us with abundant and precise information on 
radioactive nuclei in stars because the initial abundance of radioactive 
nuclei at the time of the formation of the grains is recorded by the 
signature of their radioactive decay inside the grains, which is easily 
derived from measurements of the excesses in the abundances of their 
daughter nuclei. An important example is that of $^{26}$Al, where the 
initial $^{26}$Al abundance in a stardust grain is revealed by excesses in 
$^{26}$Mg. This will be discussed in detail in Section~\ref{sec:al26}. 
In general, radioactive signatures in stardust have the potential to be used 
as clocks for the timescale of dust formation around stars and 
supernovae \citep{hoppe02}. \index{stardust} \index{meteorites}
Finally, stardust isotopic data provide a unique way to investigate the 
operation of the $s$-process in AGB stars, as will be discussed in 
Section~\ref{subsec:branchSiC}.

\section{Neutron-Capture Nucleosynthesis in AGB Stars}
\label{sec:ncapAGB}

In this section we show that:

\begin{itemize}
\item{}
{Free neutrons are produced in the TP-AGB
phase by the $^{22}$Ne($\alpha,n$)$^{25}$Mg reaction, 
which activates at $\sim
300 MK$ and operates during He burning in the intershell convective
region during thermal pulses, and the $^{13}$C($\alpha,n$)$^{16}$O 
reaction,
which activates at $\sim 90 MK$ and operates in a radiative
(and hence stable) region of the intershell during the H-burning phases. 
The free neutrons  
\index{process!s process} \index{process!neutron capture} \index{stars!AGB}
trigger the $s$-process, which produces half of
the cosmic abundances of the elements heavier than iron via neutron
captures mostly occurring on stable and long-lived radioactive nuclei.} 
\item{}
{Unstable isotopes with half lives higher than a few days can also
suffer neutron captures during the $s$-process, producing a wide variety
of \emph{branching points} on the $s$-process path, which define the details
of the abundance distribution produced by the $s$-process as a function of
neutron density and temperature.}
\item{}
{The overall $s$-process abundance distribution is 
defined by stable nuclei with a magic number of neutrons at the three 
$s$-process peaks at Sr, Ba, and Pb, and by the total amount of free 
neutrons available.}
\item{}
{Several long-lived unstable isotopes are produced by the $s$-process
(details in Section~\ref{sec:radioneutroncapheavy}). Among them 
\index{isotopes!99Tc} \index{isotopes!93Zr}
are $^{93}$Zr and $^{99}$Tc. Observations of monoisotopic stable Nb (the
daughter nucleus of $^{93}$Zr) and of Tc itself can be used as discriminant
between intrinsic (on the AGB) and extrinsic (with a former 
AGB binary companion) $s$-process-enhanced stars.}
\end{itemize}

\subsection{Neutron Sources in AGB Stars} \label{subsec:sources}

\begin{figure}
  \centering
  \includegraphics[width=\textwidth]{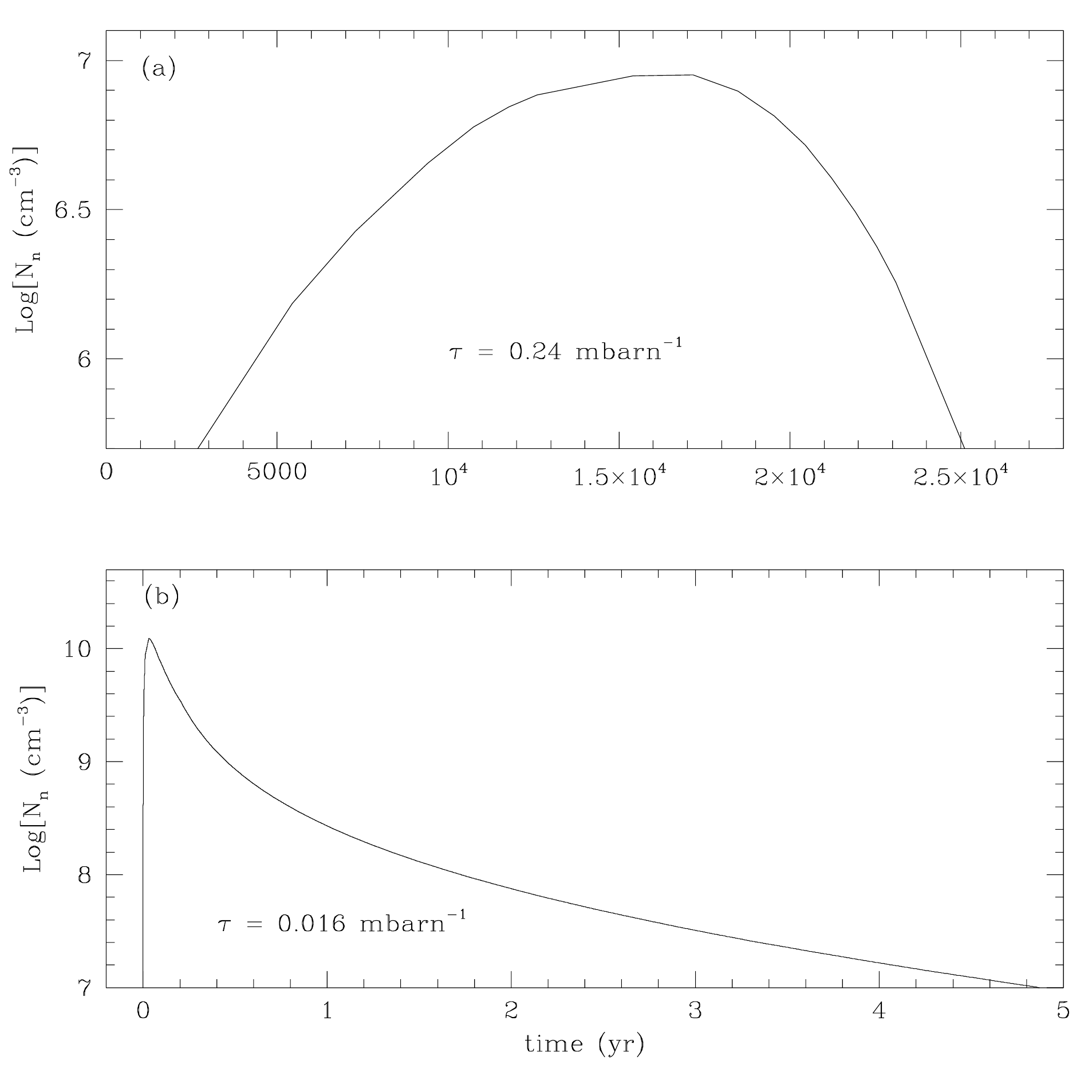}
  \caption{Neutron densities as functions of time corresponding to the 
activation of the two neutron sources in a 3 \Msun\ AGB star model of 
solar metallicity during the last interpulse-pulse cycle: (a) the $^{13}$C 
neutron source (the zero point in time represent the time from the start 
of the interpulse period, about 10,000 yr, when the temperature reaches 79 
MK); (b) the $^{22}$Ne neutron source (the zero in time corresponds to the 
time when the temperature in the TP reaches 250 MK).}
  \label{fig:neutrons} 
\end{figure}

In the double burning shell phase a nuclear reaction that may
produce a copious neutron flux is
$^{22}$Ne($\alpha$,n)$^{25}$Mg. $^{22}$Ne
is abundantly present in the intershell because it directly derives from
the initial abundance of O (the most abundant nucleus after H and He) as
a consequence of the operation of the CNO cycle first and of a double
$\alpha$ capture on $^{14}$N later. This means that this neutron
production channel is of \emph{secondary} origin, i.e., its efficiency scales
with the initial metallicity of the star. The relatively high Coulomb 
barrier of Ne (Z=10)
pushes the threshold temperature for $\alpha$ capture above 300 MK so
that this process can activate only within a hot He-burning region.
Since the temperature at the base of the He convective shell during 
thermal pulses scales
directly with the mass of the H-exhausted core, only stars initially
more massive than $\simeq$ 3 $\rm M_\odot$  \citep[]{iben75,iben78} can
efficiently activate this nuclear reaction. Panel b) in Fig.~\ref{fig:neutrons},
shows a typical profile of the neutron density versus time associated with
this neutron source. Its shape reflects the sharp rise of the temperature
caused by the growth of the thermal instability and the following quite
rapid decline due to the quenching of the instability. The high
activation temperature and its very short duration (a few years) lead to
a very high initial neutron density (reaching up to $N_n \simeq 10^{14}$
n/cm$^{3}$ in AGB stars of initial mass $\sim$ 6 \Msun) but to a small
total amount of $^{22}$Ne burnt per cycle, so that the total number
of neutrons released, i.e., the time-integrated neutron flux, or \emph{neutron 
exposure} $ \tau = \int^t_0 N_n v_{th} dt $, remains quite small, 
of the order of a few hundredth of 1/mbarn=1/$10^{-27}$ cm$^2$
(see Section~\ref{subsec:sprocess}). 
\index{process!s process} \index{process!neutron capture} \index{stars!AGB}
The signature of such an impulsive \index{process!neutron capture}
neutron flux on neutron-capture nucleosynthesis will be discussed in the
next section. We only remark here an important difference between
the neutron-capture nucleosynthesis occurring during AGB thermal pulses and 
that occurring in the He-convective shell of a massive star (other than the
fact that in the AGB case the exposure to neutrons occurs recurrently): 
the mass of the He-convective shell in AGB
stars is orders of magnitude smaller than that of a massive star so that
the smaller dilution induced by the mixing allows, in the former case,
many unstable nuclei to reach a higher equilibrium concentration. This
occurrence favors the synthesis of stable nuclei on the neutron-rich side  
of the valley of $\beta$ stability.

The problem with the $^{22}$Ne neutron source is that AGB stars observed 
to be enriched in $s$-process elements
have been identified as AGB stars of masses lower than $\sim$ 3 \Msun\
because a) their relatively low luminosities
 \citep[]{frogel90} match those of low-mass AGB models; b) their surface is
generally C enriched, an occurrence that rules out a significant HBB and
hence an initial mass greater than 3 \Msun; c) 
excesses of $^{25}$Mg, predicted to be produced by 
$^{22}$Ne($\alpha,n$)$^{25}$Mg, and of 
$^{26}$Mg, predicted to be produced by the twin channel
$^{22}$Ne($\alpha,\gamma$)$^{26}$Mg, with respect to $^{24}$Mg
are not observed  \citep[]{smith86,mcwilliam88}; d) the
high neutron density produced by the $^{22}$Ne channel, see Panel b)
in Fig.~\ref{fig:neutrons}, would favor the synthesis of neutron-rich nuclei 
like 
$^{96}$Zr and elements as Rb, at odds with spectroscopic observations
 \citep[]{lambert95,abia01} and the solar abundance distribution
 \citep[]{despain80}.
Thus, for the vast majority of $s$-enhanced AGB stars, another nuclear
fuel for the production of neutrons has to be invoked. 

Nuclei of
$^{13}$C are the best candidate for this role, given that the
$^{13}$C($\alpha,n$)$^{16}$O reaction activates at temperatures from
approximately 90 MK, which are easily reached in low-mass AGB stars. The
achievement of the threshold temperature is, however, a necessary
but not sufficient condition for a nuclear reaction to be
effective: an additional requirement is the presence of a
sufficient amount of reactants, in this case $^{13}$C. 
Models in
which no mixing is allowed in the layers in radiative equilibrium do
not naturally produce a significant concentration of $^{13}$C in the
intershell region
(see end of Section~\ref{subsec:AGB}). In fact, the $^{13}$C available in
the H-exhausted zone is that corresponding to the equilibrium value 
provided by the CNO cycle.
As a neutron source for the $s$-process, this $^{13}$C suffers 
two major problems: its 
abundance is too low to power a significant neutron flux and its ratio
with respect to $^{14}$N is too low ($^{13}$C/$^{14}$N$<<1$).
The $^{14}$N($n,p$)$^{14}$C 
reaction\footnote{This reaction
produces $^{14}$C, a radioactive nucleus with a half life of 5730 yr. This 
nucleus is not carried to the stellar surface by the 3$^{rd}$ dredge-up 
because it is destroyed by 
$^{14}$C($\alpha,\gamma)^{18}$O reactions during He-burning in the thermal 
pulse.}
\index{isotopes!14N} \index{process!neutron capture} \index{stars!AGB} \index{isotopes!13C} 
has a relatively high neutron-capture cross section of $\simeq$ 2 mbarn,
with respect to typical cross section of the order of 0.1-0.01 mbarn for
the light nuclei. Hence, it is a formidable poison that can even
completely inhibit the $s$-process.
Hence, the $^{13}$C neutron source 
represents a valid alternative to the $^{22}$Ne neutron source only if
additional $^{13}$C is produced in an environment 
depleted in $^{14}$N. A way out of this problem is to
assume that at the end of each 3$^{rd}$ dredge-up episode a small amount of
protons penetrates the intershell region.
The amount of protons engulfed
in the He/C rich intershell must be small ($\rm Y_p/Y_{\rm ^{12}C}<<1$)
because they must allow the conversion of $^{12}$C into $^{13}$C, but
not the conversion of $^{13}$C in $^{14}$N.
(Note that the intershell is essentially free of $^{14}$N
at the end of a thermal pulse because $^{14}$N nuclei have all been 
destroyed by $\alpha$
captures.) Once a small amount of protons has penetrated the intershell, the
progressive heating caused by the deposition of fresh He synthetized by
the H-burning shell induces the conversion of $^{12}$C in
$^{13}$C.
We can estimate the concentration of protons that
allows the build up of $^{13}$C, but not of $^{14}$N, by considering that
the production rate of $^{14}$N
equates that of $^{13}$C when the concentration of $^{13}$C rises
to a value of the order of 1/4 of that of $^{12}$C. 
Since the mass fraction of $^{12}$C in the
intershell is about 0.2, the two rates
equate each other for a $^{13}$C concentration of $\rm X_{\rm
^{13}C}\simeq0.20~(13/12)/4=5~10^{-2}$. If one requires the $^{13}$C
production rate to dominate that of $^{14}$N, the $^{13}$C
concentration must be reduced by at least a factor of 10, so that $\rm
X_{\rm ^{13}C}\simeq5~10^{-3}$. This abundance of $^{13}$C
corresponds to a proton concentration of the order of $\rm X_p=X_{\rm
^{13}C}/13=4~10^{-4}$.

A self-consistent scenario able to produce this small
amount of protons penetrating below the base of the convective envelope
has not been found yet: several mechanisms have been proposed 
 \citep[e.g.,][]{iben82,herwig97,langer99, denissenkov03} but none
of them can presently be considered as widely accepted. A discussion of
these alternative scenarios goes well beyond the purposes of the present
discussion. What matters, and what modelers often pragmatically assume, is
that a small amount of protons definitely penetrates in the intershell at the
end of 3$^{rd}$ dredge-up. The detailed features of the $^{13}$C
\emph{pocket} obtained with such a procedure are subject to large
uncertainties. 

Nonetheless the basic properties of the neutron flux that is obtained in 
this way are considered relatively well understood 
 \citep[]{gallino98,goriely00,lugaro03a}. The activation of the \index{isotopes!13C} 
$^{13}$C($\alpha,n$)$^{16}$O occurs well before the onset of the next 
thermal pulse and the $s$-process nucleosynthesis triggered by this 
neutron source occurs at low temperature in a radiative environment 
(see Section~\ref{subsec:sprocess}). Panel a) in Fig.~\ref{fig:neutrons} 
shows the temporal evolution of this neutron flux. The rather long 
timescale over which this neutron flux remains active is determined by 
the speed at which the H-burning shell accretes matter on the He core, 
which means a typical timescale of the order of $10^4$ yr. Given such a 
long timescale, $^{13}$C is totally consumed so that the total number 
of neutrons released is very large, with neutron exposures of the order 
of a tenth to a few mbarn$^{-1}$. The neutron density, instead, keeps to 
low values, up to $N_n \simeq 10^8$ n/cm$^{3}$. Let us finally remark 
that the neutron flux produced by the $^{13}$C neutron source is of 
\emph{primary} 
origin, i.e., independent on the initial stellar metallicity, since the 
$^{13}$C is made from 
$^{12}$C synthetized starting from the initial H and He.

\subsection{The $s$-Process in AGB Stars} 
\label{subsec:sprocess}

A fraction of the free neutrons produced in AGB stars by the $^{13}$C
and $^{22}$Ne neutron sources described above is captured by Fe seed
nuclei, leading to production of elements with large atomic mass
numbers up to Pb (A = 208) and Bi (A= 209) via the $s$-process.
\index{isotopes!13C} \index{isotopes!22Ne}  \index{process!s process} 
In general, a neutron flux that irradiates the surrounding matter
reproduces a situation analogous to that occurring during H burning,
where matter is irradiated by a flux of protons. While during a proton flux
matter is pushed out of the valley of $\beta$ stability toward the
proton-rich side, during a neutron flux matter is pushed out of the
valley of $\beta$ stability valley toward the neutron-rich side. Thus,
the presence of a neutron flux is inevitably associated to the synthesis
of radioactive nuclei that, sooner or later, decay back towards the
valley of $\beta$ stability.  

\begin{figure}
  \centering
  \includegraphics[width=0.6\textwidth]{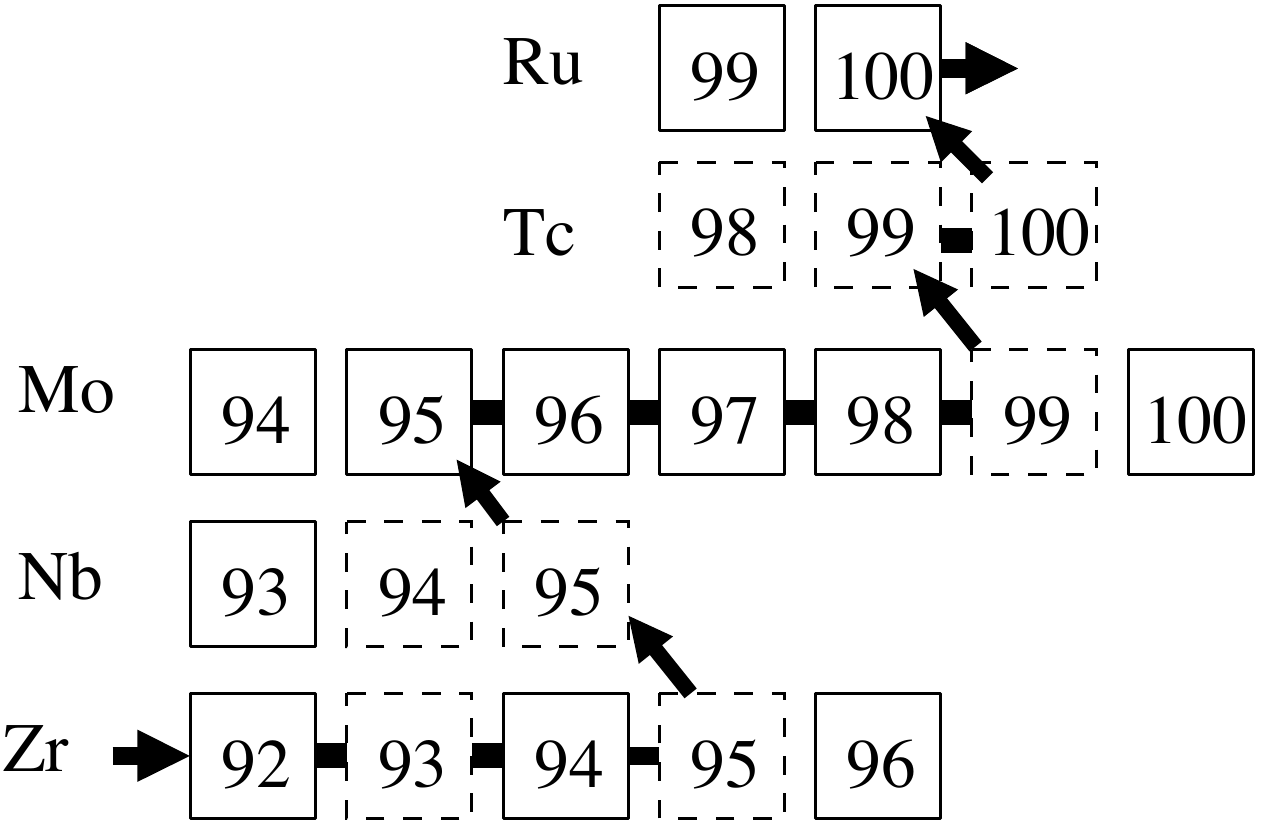}
  \caption{The main $s$-process path along the valley of $\beta$ stability 
from Zr to Ru is indicated by the thick solid line and arrows. Solid and 
dashed boxes represent stable and unstable nuclei, respectively. The 
radioactive nuclei $^{93}$Zr and $^{99}$Tc behave as stable during the $s$-process as their half lifes (of 
the order of 10$^5$-10$^6$ yr) are longer 
than the timescale of the $s$-process.} 
\label{fig:sprocesspath} 
\end{figure}

During the $s$-process, by definition, the timescale against $\beta$ 
decay 
of an unstable isotope is shorter that its timescale against neutron 
captures. Thus, neutron captures occur only along the valley of $\beta$ 
stability (Fig.~\ref{fig:sprocesspath}). For this condition to hold
neutron densities must be of the order of $N_n \sim 10^7$ 
n/cm$^3$. By comparison, during the $rapid$ neutron-capture process 
($r$-process), neutron densities reach values as high as $\rm 
10^{25}~n/cm^3$ so that neutron captures occur on a time scale 
less than a second (typically much shorter than that of radioactive 
decays) pushing matter towards very neutron-rich material. When the 
neutron flux is extinguished, the neutron-rich radioactive nuclei 
quickly decays back towards their stable isobars on the valley of 
$\beta$ stability. As presented in Chapter 4, the $r$-process is 
believed to occur in explosive conditions in supernovae.

In stellar conditions, though, neutron 
densities during the $s$-process can reach values orders of magnitude 
higher than $10^7$ n/cm$^3$.  Depending on the peak neutron density, as 
well as on the temperature and density, which can affect $\beta$-decay 
rates, conditions may occur for the neutron-capture reaction rate of an 
unstable isotope to compete with its decay rate. These unstable 
isotopes are known as \index{branching point}
\emph{branching points} on the $s$-process path. To calculate the fraction of 
the $s$-process flux branching off the main $s$-process path at a given 
branching point a \emph{branching factor} is defined as: 
$$ 
f_{branch}=\frac{p_{branch}}{p_{branch}+p_{main}}, 
$$ 
where $p_{branch}$ and $p_{main}$ are the probabilities per unit time
associated to the nuclear reactions suffered by the branching point
nucleus and leading onto the branch or onto the main path of the $s$-process, respectively.

There are several types of branching points: in the \emph{classical} case
$p_{main}$ corresponds to $\lambda$, i.e., the probability per unit time
of the unstable isotope to decay, and $p_{branch}$ corresponds to
$p_{n}$, i.e., the probability per unit time of the unstable isotope to
capture a neutron $< \sigma v > N_n$, where $N_n$ is the neutron density
and $< \sigma v >$ is the Maxwellian averaged product of the velocity
$v$ and the neutron-capture cross section $\sigma$.\footnote{Note that
$\sigma$ is usually given in mbarn, corresponding to $10^{-27}$ cm$^2$,
and that $< \sigma v >$ can be approximated to $\sigma \times
v_{thermal}$, where $v_{thermal}$ is the thermal velocity.
Neutron-capture 
cross sections for ($n,\gamma$) reactions throughout this chapter are 
given at a temperature of 350 million degrees, corresponding to an energy 
of 30 keV, at which these rates are traditionally given. Values reported 
are from the Kadonis database (Karlsruhe Astrophysical Database of 
Nucleosynthesis in Stars, http://www.kadonis.org/) and the JINA reaclib 
database (http://groups.nscl.msu.edu/jina/reaclib/db/index.php), unless 
stated otherwise.} A typical
example of this case is the isotope $^{95}$Zr in \index{isotopes!95Zr}
Fig.~\ref{fig:sprocesspath}, which has a half life of 64 days, and can
capture neutrons and produce the neutron-rich isotope $^{96}$Zr,
classically a product of the $r$-process, even during the $s$-process. 
When the branching point is a long-lived, or even stable isotope, but
its $\beta$-decay rate increases with temperature, the opposite
applies: $p_{branch}$=$\lambda$ and $p_{main}$=$p_{n}$.  In even more
complex situations, a radioactive isotope may suffer both $\beta^{+}$
and $\beta^{-}$ decays, as well as neutron captures. In this case, three
terms must be considered at denominators in the definition of the
branching factor above: $p_{n}$, and $\lambda$ for both $\beta^{+}$ and
$\beta^{-}$ decays. \index{decay!beta decay}

Branching points have been fundamental in our
understanding of the $s$-process conditions in AGB stars and will be
discussed in more detail in Section~\ref{sec:branchingpoints}. The
low neutron density associated with the $^{13}$C neutron source does not
typically allow the opening of branching points. On the other hand, the
high neutron density associated with the $^{22}$Ne neutron source activate
the operation of branching points on the $s$-process path, defining the 
details of the final abundance distribution.

It is possible to identify nuclei that can be produced only by the $s$-process (\emph{$s$-only} nuclei), 
which are shielded from $r$-process
production by a stable isobar, or only by the
$r$-process (\emph{$r$-only} nuclei), which are not reached by neutron
captures during the $s$-process as isotopes of the same element and same
atomic mass number A - 1 are unstable. Examples of $s$-only nuclei are
$^{96}$Mo and $^{100}$Ru shown in Fig. \ref{fig:sprocesspath}, which are
shielded by the $r$-only nuclei $^{96}$Zr and $^{100}$Mo, respectively. 
\index{isotopes!96ZMo} \index{isotopes!100Ru}
These, in turn, are not typically produced by the $s$-process as
$^{95}$Zr and $^{99}$Mo are unstable. Proton-rich nuclei which 
cannot be reached by
either the $s$- or the $r$-process must be produced via the \emph{$p$-process},
i.e., proton captures or photodisintegration of heavier nuclei, and are
labelled as \emph{$p$-only} nuclei (e.g., $^{94}$Mo in
Fig.\ref{fig:sprocesspath}). 

Models for the $s$-process have historically been tested against the solar 
system abundances of the $s$-only isotopes, as these were the first 
precise available constraints. Once a satisfactory fit is found to these 
abundances, the selected theoretical distribution can be used to determine 
the contribution from the $s$-process to each element and isotope. By 
subtracting this contribution to the total solar system abundance, an 
$r$-process contribution to each element can be obtained\footnote{The 
$p$-process contribution to elemental abundances is comparatively very 
small, $\simeq$1\%, except in the case of Mo and Ru, which have 
magic and close-to-magic $p$-only isotopes, where it is up to 
$\simeq$25\% and $\simeq$7\%, respectively.\index{process!p process}}    
 \citep[e.g.,][]{kaeppeler82,arlandini99}, which has been widely used to 
test $r$-process models, and to compare to spectroscopic 
observations of stars showing the signature of the $r$-process 
 \citep[]{sneden08}. For 
example, it is found that $\simeq$80\% of the solar abundance of Ba is due to the $s$-process, which is then 
classified as a typical $s$-process element, while 
$\simeq$5\% of the solar abundance of Eu is due to the $s$-process, 
which is then classified as a typical $r$-process element.
\index{abundances!Ba} \index{abundances!Eu} \index{process!s process}
 
Already B$^2$FH had attributed to the operation of the $s$-process the 
three peaks in the solar abundance distribution at magic numbers of 
neutrons N=50, the Sr, Y, and Zr peak, N=82, the Ba and La peak, and N=126, 
the Pb peak. This is because nuclei with a magic number of neutrons behave 
with respect to neutron-capture reactions in a similar way as atoms of 
noble gases do with respect to chemical reactions. Their energy levels, or 
shells, are fully populated by neutrons, in the case of magic nuclei, or 
by electrons, in the case of noble gases, and hence they are very stable 
and have a very low probability of capturing another neutron, in the case 
of magic nuclei, or of sharing electrons with another atom, in the case of 
noble gases. Nuclei with magic numbers of neutrons have small 
neutron-capture cross sections (of the order of a few to a few tens mbarn) 
with respect to other heavy nuclei, and they act as bottlenecks along the 
$s$-process path, leading to the observed abundance peaks. Nuclei located 
between the peaks, instead, have much higher neutron-capture cross 
sections (of the order of a few hundred to a few thousand mbarn). The 
neutron-capture chain in these local regions in-between magic nuclei 
quickly reaches equilibrium \index{nuclei!magic}
during the $s$-process. During a neutron-capture process the abundance 
$N_A$ of a stable isotope with atomic mass $A$ varies with time as:
$$ 
{d N_A \over dt} = production\,\,term\,-\, destruction\,\,term 
$$ 
$$ = 
N_{A-1} N_n \sigma_{A-1} \times v_{thermal} - N_A N_n \sigma_A \times 
v_{thermal}.  
$$
When replacing time with the neutron exposure $\tau$ one has: 
$$ 
{dN_A\over d\tau} = N_{A-1} \sigma_{A-1} - N_A \sigma_A, 
$$ 
which, in steady-state conditions ${dN_A\over d\tau} \rightarrow 0$
reached in between neutron magic nuclei, yields the simple rule to
derive relative $s$-process abundances away from neutron magic
numbers\footnote{For a detailed analytical description of the $s$-process refer to Chapter 7 of  
\citet{clayton68}.}:
$$ 
N_A \sigma_A \simeq constant. 
$$

It follows that the relative abundances of nuclei in-between the peaks are 
only constrained by their neutron-capture cross sections and do not 
provide information on the $s$-process neutron exposure. On the other 
hand, the relative abundances of the elements belonging to the three 
different peaks almost uniquely constrain the $s$-process neutron 
exposure. This is the reason behind the introduction and wide usage, both 
theoretically and observationally, of the $s$-process labels \emph{light $s$}
({\it ls}) and \emph{heavy $s$} ({\it hs}), corresponding to the average 
abundances of the $s$-process elements belonging to the first and second 
peak, respectively, as well as behind the importance of the determination 
of the abundance of Pb, representing the third $s$-process peak. 
In AGB stars, the high neutron 
exposure associated with the $^{13}$C neutron source drive the production of 
$s$-process elements even reaching up to the third $s$-process peak 
at Pb in low-metallicity AGB stars. On the other hand, the lower neutron
exposure associated with the $^{22}$Ne neutron source typically 
produces $s$-elements only up to the first $s$-process peak at Sr.

It is now ascertained that the $s$-process is responsible for the \index{abundances!s process}
production of about half the abundances of the elements between Sr and Bi 
in the Universe  \citep[see, e.g.,][]{kaeppeler82} and that it occurs in
AGB stars\footnote{Cosmic abundances of nuclei between Fe and Sr are
also contributed by the $s$-process, but in this case by neutron
captures occurring in massive stars during core He burning and shell C
burning  \citep[Chapter 4 and, e.g.,][]{raiteri92}.}. The
first direct evidence that the $s$-process occurs in AGB stars - and,
more generally, that nucleosynthesis is happening inside stars - was the
identification in the 1950s of the absorption lines of atoms of the
radioactive element Tc in the atmospheres of some cool giant stars. The
longest-living isotopes of Tc are $^{97}$Tc and $^{98}$Tc, with a half
\index{isotopes!98Tc} 
life of 4.0 and 4.2 million years, respectively.  Since these stars
would have taken billions of years to evolve to the giant phase, the
observed Tc could have not been present in the star initially. It
follows that the Tc must have been produced by the $s$-process inside
the stars. Actually, neutron captures do not produce $^{97,98}$Tc, but
the third longest-living isotope of this element: $^{99}$Tc 
(Fig.~\ref{fig:sprocesspath}), which has a
terrestrial half life of 0.21 million years. 
The presence of $^{99}$Tc in AGB stars
has been confirmed by measurements of the Ru isotopic composition in
stardust SiC grains, as will be discussed in
Section~\ref{sec:radioneutroncapheavy}.

The observation of Tc in giant stars has also been used to classify different 
types of $s$-process enhanced stars. If a given observed $s$-process enriched giant star shows the lines of 
Tc, then it must be on the AGB and 
have enriched itself of $s$-process elements. In this case it is 
classified as \emph{intrisic} $s$-process enhanced star and typically  
belongs to one of the reddest and coolest subclasses of the 
spectroscopic class M: MS, S, SC, and C(N), where the different labels 
indicate 
specific spectral properties - S stars show zirconium oxide lines 
on top of the titanium oxide 
lines present in some M stars and C(N) stars have more carbon than oxygen 
in their atmospheres - or the transition cases between those properties - 
MS is the 
transition case between M stars and S stars and SC is the transition 
case between S and C(N) stars. On the other hand, 
if an $s$-process enriched giant star does not show the lines of Tc, it is 
classified as \emph{extrinsic} $s$-process enhanced star. In this case its 
$s$-process enhancements have resulted from mass transfer from a binary 
companion, which was more massive and hence evolved first on the AGB 
phase. Stars belonging to the class of \emph{extrinsic} $s$-process enhanced 
stars range from Ba stars in the Galactic disk, to the older halo 
populations of carbon-rich CH and Carbon-Enhanced Metal-Poor (CEMP) stars 
 \citep[e.g.,][]{jorissen98,bond00,lucatello05}. Observations of Nb can also 
be used to discriminate intrisic from extrinsic $s$-process enhanced 
stars as Nb is 
destroyed during the $s$-process, but receives a radiogenic contribution 
over time due to the $\beta^{-}$ decay of $^{93}$Zr, with half life 1.5 
million yr, which is on the $s$-process path (see 
Fig.~\ref{fig:sprocesspath}).

\subsection{Branchings and the $s$-Process in AGB Stars} 
\label{sec:branchingpoints}

Branching points at radioactive nuclei have provided for the past 50 years 
\index{branching point} \index{process!s process} \index{stars!AGB}
important tools to learn about conditions during the $s$-process in AGB 
stars. This is because branching factors depend on the neutron density and 
can also depend on the temperature and density of the stellar material. 
This happens in those cases when the decay rate of the branching nucleus is 
temperature and/or density dependent. These branching points are referred 
to as \emph{thermometers} for the $s$-process. Traditionally, the solar 
abundances of isotopes affected by branching points were used to predict 
the neutron density and temperature at the $s$-process site using 
parametric models where parameters representing, e.g., the temperature and 
the neutron density were varied freely in order to match the observed 
abundances  \citep[e.g.,][]{kaeppeler82}. Later, detailed information on 
branching points became available from spectroscopic 
observations of stellar atmospheres and from laboratory analyses of 
meteoritic stellar grains. At the same time, models for the $s$-process in 
AGB stars have evolved from parametric into stellar models, where the 
temperature and neutron density parameters governing the $s$-process are 
taken from detailed computation of the evolution of stellar structure 
 \citep[]{gallino98,goriely00,cristallo09}. For these models branching points 
are particularly useful to constrain neutron-capture nucleosynthesis and 
conditions inside the thermal pulse because, typically, they open at 
high neutron densities during the high-temperature conditions that 
allow the activation of the $^{22}$Ne neutron source in the convective 
intershell region.

As the temperature, density, and neutron density vary with time in the 
convective intershell region, branching factors also change over time. 
For example, a 
classical branching point, where the branching path corresponds to 
neutron capture, progressively opens while the neutron density reaches its 
maximum, and then closes again while the neutron density decreases and the 
main $s$-process path is restored. Of special interest is that toward the 
end of the thermal pulse the neutron density always decreases 
monotonically with the temperature and thus with time (Fig.~\ref{fig:neutrons}) 
so that a \emph{freeze out} time can be determined for a given nucleus, which 
represents the time after which the probability that the nucleus captures 
a neutron is smaller than unity and thus the abundances are \emph{frozen} 
 \citep[]{cosner80}.  This can be calculated as the time when the neutron 
exposure $\tau$ left before the end of the neutron flux is 1/$\sigma$, 
where $\sigma$ is the neutron-capture cross section of the nucleus.
Hence, the abundances determined by
branching points defined by unstable isotopes with higher $\sigma$ freeze out
later during the neutron flux.


As a general rule of thumb, branching points that have the chance of 
being activated at the neutron densities reached in AGB stars are those 
corresponding to radioactive nuclei with half lives longer than at least a 
couple days. These correspond to similar half lives against capturing a 
neutron for neutron densities $\simeq 10^9-10^{11}$ n/cm$^3$, at AGB
$s$-process temperatures. Isotopes with half lifes longer than 
approximately 10,000 yrs can be considered stable in this context as the 
$s$-process flux in AGB stars typically lasts less than this time. We 
refer to these isotopes as long-lived isotopes and we discuss their 
production in AGB stars in detail in Section~\ref{sec:radioneutroncapheavy}. 
Very long-lived isotopes - half lives longer than $\sim$ 10 Myr - 
include for example $^{87}$Rb, and are considered stable in our context. 

A list of unstable isotopes at which branching points that become 
relevant in the $s$-process reaction chain in AGB stars is presented in 
Appendix~B of this book  as a complete reference to be compared against observational 
information and as a tool for the building of $s$-process networks.  Worth 
special mention are the branching points at $^{79}$Se, $^{85}$Kr, and 
\index{isotopes!79Se} \index{isotopes!85Kr} \index{isotopes!176Lu} \index{isotopes!87Rb} \index{isotopes!163Dy} \index{isotopes!179Hf}    
$^{176}$Lu for the involvement of isomeric states of these nuclei, at 
$^{151}$Sm, one among a limited number of branching points for which 
an experimental estimate of the neutron-capture cross section is 
available, at $^{86}$Rb, responsible for the production of the very 
long-living $^{87}$Rb, and at $^{163}$Dy and $^{179}$Hf, which 
are stable nuclei in terrestrial conditions that 
become unstable in AGB stellar interiors.

Taken as a whole, the list of branching points that may be operating 
during the $s$-process in AGB stars sets a powerful group of constraints 
on our theoretical $s$-process scenarios. They are particularly effective 
when each of them is matched to the most detailed available observations 
of its effects. For example, some elemental abundance ratios and isotopic 
ratios that are affected by branching points can be measured from a 
stellar spectrum via identificaton and analysis of different emission or 
absorption lines. In these cases, model predictions can be compared 
directly to stellar observations of $s$-process enhanced stars 
(Section~\ref{subsec:branchstars}). Isotopic ratios affected by branching 
points involving isotopes of refractory elements, but also of noble gases, 
have been or have the potential to be measured in meteoritic stardust SiC 
grains from AGB stars and provide unique constraints due to the large and 
expanding high-precision dataset available on the composition of 
stardust (Section~\ref{subsec:branchSiC}). 
The values of the solar abundance ratios of $s$-only isotopes affected by 
branching points (e.g., $^{134}$Ba/$^{136}$Ba, $^{128}$Xe/$^{130}$Xe, 
and $^{176}$Hf/$^{176}$Lu) 
must be matched by any $s$-process model. When these involve nuclei with 
peculiar structure, such as $^{176}$Lu, combined investigation of nuclear 
properties and $s$-process models drives progress in our understanding 
of both.

One advantage of the computation of branching points in AGB stars is that 
the activation of one branching point is almost completely independent 
from the activation of all the other branching points because the overall 
neutron flux is only very marginally affected by the details of the 
$s$-process path. Thus, it is possible to include in a $s$-process 
nuclear network only the branching points of interest for a specific 
problem, or a specific element, hence keeping it simple and saving 
computational time. \index{branching point}

One overall drawback of using branching points to understand the $s$-process
is that for the vast majority of the radioactive nuclei involved
there exist only theoretical or phenomenological determinations of their
neutron-capture cross sections
and of the temperature and density dependence of their decay
rates. This is due to the difficulty of producing experimental data for
radioactive targets (see Chapter 9) and means that there are 
always some uncertainties associated to model predictions of the
effect of branching points. These errors and their effect need to be
carefully evaluated in every single case.

\subsection{Signatures of s-Process Branching Points in Stars: Rb, Zr, Eu}
\label{subsec:branchstars}

\begin{figure}
  \centering
  \includegraphics[width=\textwidth]{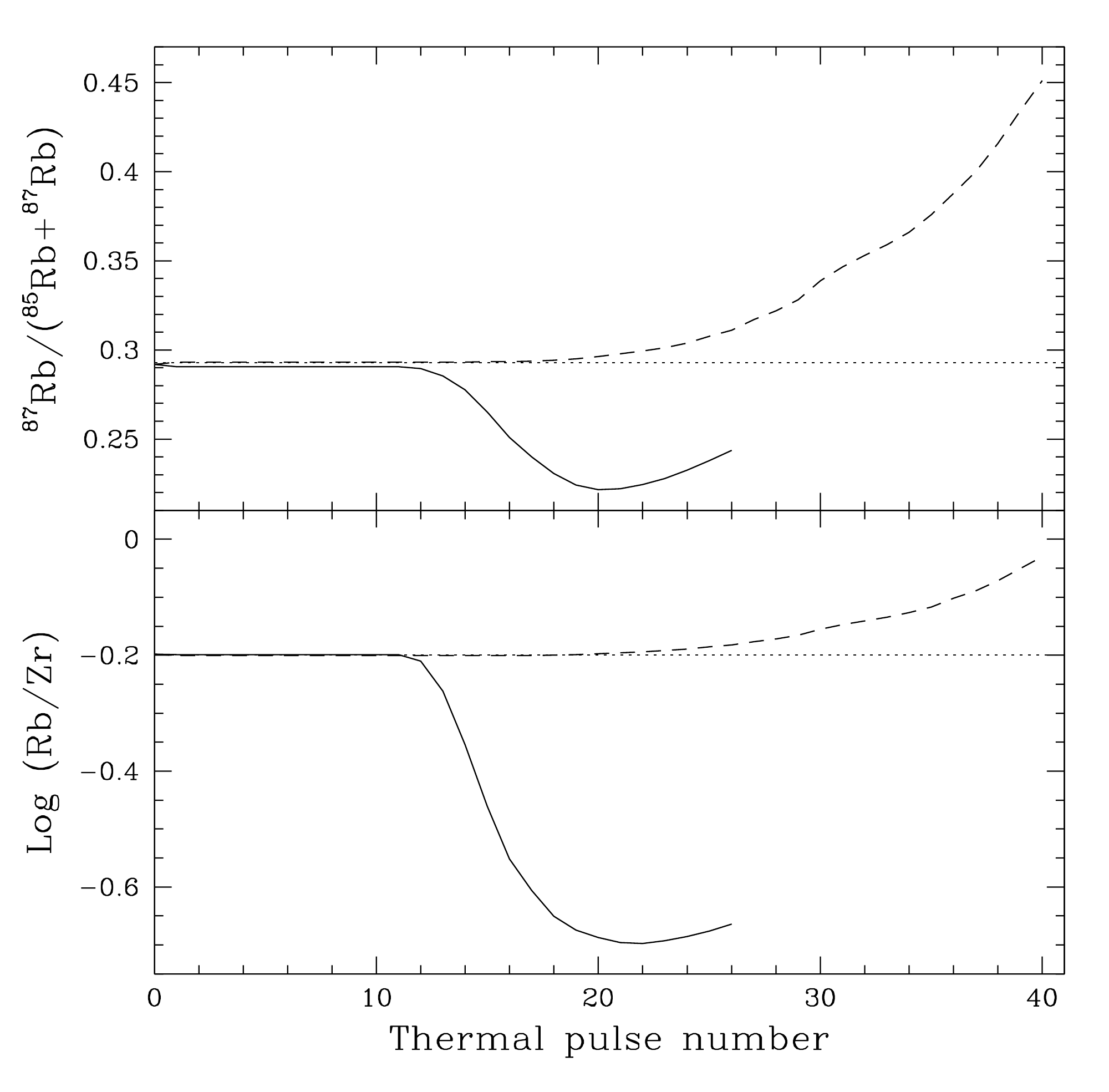}
  \caption{Ratio of $^{87}$Rb to the total abundance of Rb (top panel) and 
the Rb/Zr ratio (lower panel) computed in two solar metallicity AGB models 
 \citep[from][]{vanraai08a}. The dotted lines represent the initial solar 
ratios. The evolution lines represent a massive (6.5 \Msun) AGB model 
experiencing the activation of the $^{22}$Ne neutron source only (long-dashed lines), and a 
low-mass (3 \Msun) model experiencing the activation of the $^{13}$C 
neutron source only (solid lines, except for a marginal activation of the $^{22}$Ne 
neutron source in the latest thermal pulses leading to the small final 
increase in the $^{87}$Rb and Rb abundances) \index{isotopes!87Rb} .}
  \label{fig:rb87} 
\end{figure}

The abundance of $^{87}$Rb, which can be produced in AGB stars via 
\index{isotopes!87Rb} 
activation of the branching point at $^{86}$Rb, is a famous example of how 
detailed comparison of theoretical $s$-process abundances to the 
abundances observed in $s$-process enhanced stars provide a stringent test 
to our understanding of the $s$-process and AGB stars.  The abundance of 
$^{87}$Rb is particularly interesting because the element Rb can be 
spectroscopically identified and its abundance determined in AGB stars. 
Overall, Rb is an $r$-process element - only 22\% of its solar abundance 
can be ascribed to the $s$-process  \citep[]{arlandini99} - made up of two 
isotopes: $^{85}$Rb and the very long-lived $^{87}$Rb, which is treated 
as a stable isotope in this context.  Specifically, 92\% of solar 
$^{85}$Rb is made by the $r$-process because this nucleus has a relatively 
large neutron-capture cross section of 234 mbarn and thus it does not 
accumulate to high abundances during the $s$-process. On the other hand, 
$^{87}$Rb, as described in Appendix~B, has a magic number of neutrons 
N=50, and thus a relatively small neutron-capture cross section of 15.7 
mbarn. Hence, if it is reached by the $s$-process reaction chain via the 
activation of the branching points at $^{85}$Kr and $^{86}$Rb, it 
accumulates and is significantly produced.  It follows that when these 
branching points are activated during the $s$-process, the abundance 
$^{87}$Rb represents a fraction of the total abundance of $s$-process Rb 
larger than the initial solar fraction. This is illustrated in the top 
panel of Fig.~\ref{fig:rb87}. In the case of the massive AGB model, 
where the $^{22}$Ne neutron source is activated, the $s$-process occurs at 
high neutron density, and branching points are open, almost half of the 
final total abundance of Rb is made by $^{87}$Rb. In the case of the 
low-mass AGB model, instead, where the $^{13}$C neutron source is 
activated, the $s$-process occurs at low neutron density, and branching 
points are closed, only a quarter of the final total abundance of Rb is 
made by $^{87}$Rb.

The ratio of the abundance of Rb to that of a neighbouring $s$-process 
element, such as Sr, or Zr, whose overall abundance is instead not 
affected by the activation of branching points, can be determined in AGB 
stars and has been widely used as an indicator of the neutron density at 
which the $s$-process occurs. Observations of Rb/Zr ratios lower than 
solar in MS, S, and C stars have strongly supported the theoretical 
scenario where the main neutron source in these low-mass AGB stars is
the $^{13}$C($\alpha,n$)$^{16}$O reaction. This is because this neutron
source produces neutron 
densities too low to increase the Rb/Zr ratio above the solar value 
 \citep[see lower panel of Fig.~\ref{fig:rb87} and][]{lambert95,abia01}.

Massive AGB stars ($>$ 4:5 \Msun) have only recently been identified in
our Galaxy  \citep[]{garcia06,garcia07}. They belong to the group of OH/IR
stars \index{stars!OH/IR} \index{stars!AGB}
and they have been singled out as massive AGB stars on the basis
of their location closer to the galactic plane, which indicates that
they belong to a younger and thus more massive stellar population, and their longer
pulsation periods ($\simeq$400 days). Rb/Zr ratios in these stars are
observed to be well above the solar value, which has given ground to the
theoretical scenario where the main neutron source in these massive
AGB stars must be the $^{22}$Ne($\alpha,n$)$^{25}$Mg reaction, which
produces neutron densities high enough to increase the Rb/Zr ratio above
the solar value (see lower panel of Fig.~\ref{fig:rb87}).

Another indicator of the neutron density in AGB stars is the 
isotopic abundance of $^{96}$Zr, which is produced if the branching point 
at $^{95}$Zr is activated. Zr isotopic ratios were determined via 
observations of molecular lines of ZrO in a sample of S stars 
 \citep[]{lambert95}. No evidence was found for the presence of $^{96}$Zr in 
these stars. This result provides further evidence that the neutron 
density in low-mass AGB stars stars must be low. A low $^{96}$Zr abundance 
has been confirmed by high-precision data of the Zr isotopic ratios in 
stardust SiC grains, which are discussed in the following 
Section~\ref{subsec:branchSiC}.

Isotopic information from stellar spectra has also been derived for the 
typical $r$-process element Eu in old Main Sequence stars belonging to the 
halo of our Galaxy and enhanced in heavy neutron-capture elements 
 \citep[]{sneden08}. This has been possible because the atomic lines of Eu 
differ significantly if the Eu atoms are made of $^{151}$Eu instead 
\index{isotopes!153Eu} \index{process!r process} \index{isotopes!151Eu}   
of $^{153}$Eu, the two stable isotopes of Eu. In old stars showing overall 
enhancements of $r$-process elements, the total Eu abundance is roughly 
equally divided between $^{151}$Eu and $^{153}$Eu. This fraction is 
consistent with the solar fraction, and it is expected by Eu production 
due to the $r$-process  \citep[]{sneden02,aoki03a}. On the other hand, in two 
old stars showing overall enhancements in $s$- and $r$-process elements, 
roughly 60\% of Eu is in the form of $^{151}$Eu  \citep[]{aoki03b}. How these 
stars gained enhancements in the abundances of both $r$- and $s$-process 
elements is an unsolved puzzle of the study of the origin of the 
elements heavier than iron in the Galaxy  \citep[]{jonsell06}. This is 
because according to our current knowledge, the $r$- and the $s$-process 
are completely independent of each other, and occur in very different 
types of stars, core-collpase supernovae and AGB stars, respectively. 
During the $s$-process, the Eu isotopic fraction is determined mostly by 
the activation of the branching point at $^{151}$Sm, and the observed 60\% 
value is consistent with low neutron densities $\simeq 10^8$ n/cm$^3$ 
during the $s$-process. As more observations become available, the role of 
branching points during the $s$-process in AGB stars becomes more and more 
crucial to answering the questions on the origin of the heavy elements.

\subsection{SiC Grains from AGB Stars and Branching Points} 
\label{subsec:branchSiC}

Stardust \index{stardust} SiC grains from AGB stars represent a unique opportunity to
study $s$-process conditions in the parent stars of the grains through
the effect of the operation of branching points because SiC grains
contain trace amounts of atoms of elements heavier than iron, which
allow high-precision measurements of their isotopic ratios. Refractory
heavy elements, such as Sr, Ba, Nd, and Sm, condensed from the stellar
gas directly into the SiC grains while the grains were forming. Their isotopic
composition have been determined from samples of meteoritic residual
materials containing a large number of SiC grains using TIMS and SIMS
 \citep[see Chapter 10, Section 2
and][]{ott90,zinner91,prombo93,podosek04}. High-resolution \index{SIMS} SIMS has
also been applied to derive data in single stardust SiC grains for Ba
with the NanoSIMS  \citep[see Chapter 10, Section 2 and][]{marhas07}
and Eu with the SHRIMP  \citep[Sensitive High Resolution Ion
Microprobe,][]{terada06}. Isotopic ratios in a sample containing a large
number of SiC grains for many elements in the mass range from Ba to Hf
were also measured by ICPMS  \citep[Chapter 10, Section 2 and][]{yin06}.  

A general drawback of these experimental methods is that they do not allow 
to separate ions of same mass but different elements. Hence, 
interferences by isotopes of the same mass (isobars) are present, which is 
especially problematic for the elements heavier than iron where a large 
number of stable isobars can be found. Branching points, in particular, by 
definition affect the relative abundances of isobars, thus, with the 
methods above it is difficult to derive precise constraints on the effect 
of branching points on isotopic ratios. For example, the isobars $^{96}$Mo 
and $^{96}$Zr cannot be distinguished in these measurements, and thus 
it is not possible to derive information on the operation of the branching 
point at $^{95}$Zr.

Exceptions to this problem are the stable Eu isotopes, $^{151}$Eu and 
$^{153}$Eu, which do not have stable isobars and thus their ratio can be 
measured and used to constrain the neutron density and the temperature 
during the $s$-process in the parent stars of the grains via the branching 
points at $^{151}$Sm and $^{152}$Eu  \citep[]{terada06}, and the Ba isotopes, 
which are not affected by isobaric interferences because their isobars, 
the isotopes of the noble gas Xe isotopes, are present in very low amounts 
in the grains and are difficult to ionize and extract from the stardust 
(see specific discussion below in this section.) The Ba isotopic ratios, 
in particular the $^{134}$Ba/$^{136}$Ba and the $^{137}$Ba/$^{136}$Ba 
ratios, can be affected by branching points at the Cs isotopes  \citep[see 
below and][]{prombo93,marhas07}.
\index{isotopes!134Ba} \index{process!r process} \index{isotopes!136Ba} 

The application of RIMS (Chapter 10, Section 2) to the analysis of  
heavy elements in SiC grains has allowed to overcome the problem of
isobaric interferences, at the same time providing an experimental
method of very high sensitivity, which allows the measurements of trace
elements in single stardust grains  \citep[]{savina03a}. Since RIMS can
select which element is ionized and extracted form the grains, mass
interferences are automatically avoided. The Chicago-Argonne RIMS for
Mass Analysis CHARISMA has been applied to date to the measurement of Zr
 \citep[]{nicolussi:97}, Mo  \citep[]{nicolussi:98a}, Sr  \citep[]{nicolussi:98b},
Ba  \citep[]{savina:03b} and Ru  \citep[]{savina:04} in large single SiC grains
(average size 3 $\mu$m), providing high-precision constraints on the
operation of the $s$-process branching points that may affect the
isotopic composition of these elements. A detailed comparison between
data and models  \citep[]{lugaro03b} shows that AGB stellar models of low
mass and metallicity roughly solar, where the
$^{13}$C($\alpha,n$)$^{16}$O reaction is the main neutron source and the
$^{22}$Ne($\alpha,n$)$^{25}$Mg is only marginally activated, provide the
best match to all measured isotopic ratios affected by branching points.
This result is in agreement with the constraints mentioned above
provided by the isotopic data for Ba  \citep[]{marhas07} and Eu
 \citep[]{terada06} obtained via high-resolution SIMS.

For example, the $^{96}$Zr/$^{94}$Zr ratio is observed in all measured 
\index{isotopes!94Zr} \index{process!r process} \index{isotopes!96Zr} 
single SiC to be lower than solar by at least 50\%. Low-mass AGB models 
can reproduce this constraint due to the low neutron density associated 
with the main $^{13}$C neutron source, in which conditions $^{96}$Zr 
behaves like a typical $r$-only nucleus and is destroyed during the 
neutron flux. Massive AGB stars ($>$ 4:5 \Msun), on the other hand, experience high 
neutron densities and produce $^{96}$Zr/$^{94}$Zr ratios higher than 
solar. In more detail, the $^{96}$Zr/$^{94}$Zr ratio at the stellar 
surface of low-mass AGB stellar models reaches a minimum of $\simeq$90\% 
lower than solar after roughly ten 3$^{rd}$ dredge-up episodes, and then may 
increase again, due to the marginal activation of the $^{22}$Ne neutron 
source in the latest thermal pulses. This predicted range allows to cover 
most of the $^{96}$Zr/$^{94}$Zr of single SiC grains  \citep[see Fig. 5 
of][]{lugaro03b}.

Another interesting example is the $^{134}$Ba/$^{136}$Ba ratio, where both 
isotopes are $s$-only nuclei. During the low-neutron density flux provided 
by the $^{13}$C neutron source the branching point at $^{134}$Cs is closed 
and the $^{134}$Ba/$^{136}$Ba ratio at the stellar surface reaches up to 
$\simeq$ 20\% higher than the solar ratio after roughly ten 3$^{rd}$ 
dredge-up episodes. This value is too high to match the composition of 
single SiC grains. However, during the marginal activation of the 
$^{22}$Ne in the later thermal pulses, the branching point at $^{134}$Cs 
is activated, $^{134}$Ba is skipped during the $s$-process flux and the 
$^{134}$Ba/$^{136}$Ba ratio at the stellar surface is lowered down to the 
observed values roughly 10\% higher than the solar value  \citep[see 
Fig. 14 of][]{lugaro03b}.

The $^{137}$Ba/$^{136}$Ba ratio is another indicator of the neutron 
density because the activation of the chain of branching points along the 
Cs isotopes can produce $^{137}$Cs, which decays into $^{137}$Ba with a half 
life of 30 yr. Grain data do not show any contribution of $^{137}$Cs to 
$^{137}$Ba, indicating that the Cs branching points beyond $^{134}$Cs are 
not activated in the parent stars of the grains  \citep[see Fig. 14 
of][]{lugaro03b}. This, again, excludes massive AGB stars, with an important 
neutron contribution from the $^{22}$Ne neutron source, as the parent stars 
of the grains.

Another example of the signature of the $s$-process in meteorites is 
represented by very small variations, of the order of parts per ten 
thousand, observed in the osmium isotopic ratios of primitive chondritic 
meteorites. This fascinating anomaly looks like a \emph{mirror} $s$-process 
signature, meaning that they show exactly the opposite 
behaviour expected if the meteorite had a component carrying an 
$s$-process signature. They are thus interpreted as a sign of incomplete 
assimilation of stardust SiC grains within the meteorite 
 \citep[]{brandon05}. The branching points at $^{185}$W and $^{186}$Re make 
the $^{186}$Os/$^{188}$Os ratio a indicator of the neutron density for the 
$s$-process and the value for this ratio observed in chondrites suggest a 
low neutron density of N$_n = 3 \times 10^8 n$/cm$^3$  \citep[]{humayun07}, 
in agreement with other evidence discussed above.
\index{isotopes!188Os} \index{process!s process} \index{isotopes!186Os} 

Differently from refractory elements, the noble gases He, Ne, Ar, Kr, and 
Xe are not chemically reactive and can condense from gas into solid only 
at much lower temperatures than those around AGB stars. Still, they are 
found in SiC, even if in extremely low quantitites. Their atoms could have 
been ionized by the energy carried by the stellar winds during the AGB and 
post-AGB phase. These ions were then chemically reactive enough to be 
\emph{implanted} into already formed dust grains.

It has been possible to extract noble gases from meteoritic samples by
RIMS (see Chapter 11), laser gas extraction  \citep[]{nichols91} and
stepped-heating combustion of the sample to high temperatures, up to
2,000 degrees  \citep[]{lewis94}. In particular, for the heavy nobler gases
Kr and Xe, since their abundances are very low in stardust and the
stepped-heating experimental method does not provide high extraction
efficiency, it has been possible to extract their ions only from a large
amount of meteoritic residual material. The derived Kr and Xe isotopic
data is thus the average over a large number - millions - of grains.
Differential information as function of the grain sizes can still be
obtained by preparing the meteoritical residual in a way that selects
the size of the grains to be found in it.

The composition of Xe in SiC corresponds to the famous Xe-S component, one 
of the first signature of the presence of pure stellar material in 
primitives meteorites (see Chapter 2, Section 2.5), thus named because of 
its obvious $s$-process signature: excesses in the $s$-only isotopes 
$^{128,130}$Xe and deficits in the $r$-only and $p$-only isotopes 
\index{isotopes!124Xe}  \index{isotopes!128Xe} \index{isotopes!128Xe}  \index{isotopes!130Xe} \index{isotopes!134Xe}  \index{isotopes!133Xe} 
$^{124,126,136}$Xe (all with respect to the solar composition). The 
$^{134}$Xe/$^{130}$Xe ratio may be affected by the operation of the 
branching point at $^{133}$Xe during the $s$-process. This isotopic ratio 
in stardust SiC is very close to zero, indicating that the Xe trapped in 
SiC grains did not experience $s$-process with high neutron density 
 \citep[]{pignatari04}. This again allows the mass and 
metallicity of the parent stars of the grains to be constrained to 
low-mass AGB stars of roughly solar 
metallicity, in agreement with the conclusions drawn from the composition 
of the refractory elements.

The situation regarding the Kr isotopic ratios measured in SiC grains is 
much more complex. There are two branching points affecting the Kr 
isotopic composition: $^{79}$Se and $^{85}$Kr, changing the abundances of 
\index{isotopes!79Se}  \index{isotopes!85Kr} \index{branching point}
$^{80}$Kr and $^{86}$Kr, respectively, and both of them are tricky to 
model (see description in Appendix B). Moreover, the Kr atoms in stardust 
SiC appear to be consistent with implantation models of this gas into the 
grains only if these models consider two different components of implanted 
Kr  \citep[]{verchovsky:04}. One component was ionized and implanted in SiC 
at low energy, corresponding to a velocity of 5-30 km~s$^{-1}$, typical of AGB 
stellar winds, the other component was ionized and implanted at high energy, 
corresponding to a velocity of a few thousands km/s, typical of the winds 
driven from the central star during the planetary nebular phase. In the second 
situation, which is the case also for all the He, Ar, and Ne atoms found in 
SiC, the isotopic composition of the noble gases indicate that they must 
have come directly from the deep He-rich and $s$-process rich layers of 
the star, with very small dilution with the envelope material of initial 
solar composition. This is consistent with the fact that
at this point in time the envelope is very thin as most of the initial
envelope material has been peeled away by the stellar winds.

While the Kr AGB component is observed to be prominent in the small grains 
(of average size 0.4 $\mu$m) and shows low $^{86}$Kr/$^{82}$Kr and high 
$^{80}$Kr/$^{82}$Kr ratios, in agreement again with low neutron density 
$s$-process AGB models, the Kr planetary-nebula component is observed to 
be prominent in the largest grains (average size 3 $\mu$m) and shows high 
$^{86}$Kr/$^{82}$Kr and low $^{80}$Kr/$^{82}$Kr ratios, as expected 
instead in pure He-rich intershell material due to the higher neutron 
density $s$-process occurring in the final AGB thermal pulses 
 \citep[]{pignatari06}. Actually, it is difficult to reproduce the 
$^{86}$Kr/$^{82}$Kr up to twice the solar value observed in the largest 
grains even using the final pure $s$-process intershell composition of 
low-mass and 
solar metallicity AGB stellar models. This high $^{86}$Kr/$^{82}$Kr ratios 
may be the signature of high-neutron density $s$-process nucleosynthesis 
occurring in \emph{late} and \emph{very late} thermal pulses during the post-AGB 
phase  \citep[see, e.g.,][]{herwig99}, rather than during the AGB phase. 
Detailed $s$-process models are currently missing for this phase of 
stellar evolution.

In summary, the detailed information provided by stardust data on the 
isotopic ratios affected by branching points at radioactive nuclei on the 
$s$-process path has allowed us to pinpoint the characteristics of the 
neutron flux that the parent stars of stardust SiC grains must have 
experienced. The vast amount of information on the composition of light 
and heavy elements in SiC grains has allowed us to infer with a high 
degree of confidence that the vast majority of these grains came from 
C-rich AGB stars, i.e., C(N) stars, which have C$>$O in their envelope, the 
condition for SiC grain formation, of low mass and metallicity close to 
solar. In turn, the stardust data has been used to refine our theoretical 
ideas of the $s$-process in these stars confirming that $^{13}$C nuclei 
must be the main neutron source, while the $^{22}$Ne neutron source is 
only marginally active. \index{stardust}

\section{Nucleosynthesis of Long-lived Isotopes in AGB Stars} 

\subsection{$^{26}$Al}
\label{sec:al26}

The famous long-lived radioactive nucleus $^{26}$Al (with half life of 
\index{isotopes!26Al}  \index{process!hot bottom burning} \index{stars!AGB}
0.7 Myr), of interest from the point of view of $\gamma$-ray observations, 
meteoritic stellar grains, and the composition of the early solar system, 
can be produced in AGB stars via proton captures on $^{25}$Mg, i.e., the 
$^{25}$Mg($p,\gamma$)$^{26}$Al reaction, when the temperature is above 
$\simeq$60 MK  \citep[]{mowlavi00,vanraai08b}. As detailed in 
Section~\ref{subsec:AGB}, proton captures occur in AGB stars between thermal 
pulses in two different locations: (1) in the H-burning shell 
on top of the He-rich intershell, and (2) at the base of the 
convective envelope in massive AGB stars, above $\simeq$ 4 \Msun (in the 
process known as Hot Bottom Burning, HBB, Section~\ref{subsec:AGB}).

In setting (1), the intershell material is progressively enriched in 
$^{26}$Al as proton captures in the H-burning shell convert 80\% of 
$^{25}$Mg into $^{26}$Al. The efficiency of this 
conversion is determined by the fraction of 20\% of $^{25}$Mg+$p$ reactions 
producing the isomeric, rather than the ground, state of $^{26}$Al, which 
quickly decays into $^{26}$Mg with a half life of $\simeq$ 6 s. 
Most of the intershell $^{26}$Al abundance is destroyed by 
neutron captures before having the chance of being dredged-up to the 
stellar surface via the 3$^{rd}$ dredge-up. This is because the neutron-capture 
cross sections of $^{26}$Al, in particular the ($n,p$) and ($n,\alpha$) 
channels, are very efficient: $\sigma \simeq$ 250 and 180 mbarn, respectively.

More specifically, already during the interpulse period some $^{26}$Al is destroyed by 
neutron captures. This is because in the bottom layers of the ashes of H 
burning the temperature reaches 90 MK, high enough for the 
$^{13}$C($\alpha$,n)$^{16}$O reaction to occur using as fuel the $^{13}$C 
nuclei in ashes of H burning produced by CNO cycling. 
Then, neutron captures in the following 
thermal pulse destroy most of the $^{26}$Al that was left over in the 
H-burning ashes. First, the $^{13}$C nuclei that had survived in the top 
layers of the H-burning ashes are engulfed in the convective pulse, where 
the temperature quickly reaches 200 MK and the 
\index{isotopes!13C}  \index{isotopes!22Ne} 
$^{13}$C($\alpha$,n)$^{16}$O reaction is very efficiently activated. 
Second, the 
neutrons that may be released by the $^{22}$Ne($\alpha$,n)$^{25}$Mg 
reaction later on in the convective pulse, when the temperature is higher 
than roughly 250 MK, contribute to further destruction of $^{26}$Al. In 
this phase $^{26}$Al can be completely destroyed, depending on the 
temperature reached at the base of the convective pulse, which controls 
the efficiency of the $^{22}$Ne($\alpha$,n)$^{25}$Mg reaction. If the 
temperature reaches up to 300 MK, the $^{26}$Al abundance is decreased by 
two orders of magnitude in the He-rich intershell at the end of the 
thermal pulse.

\begin{figure}
  \centering
  \includegraphics[width=\textwidth]{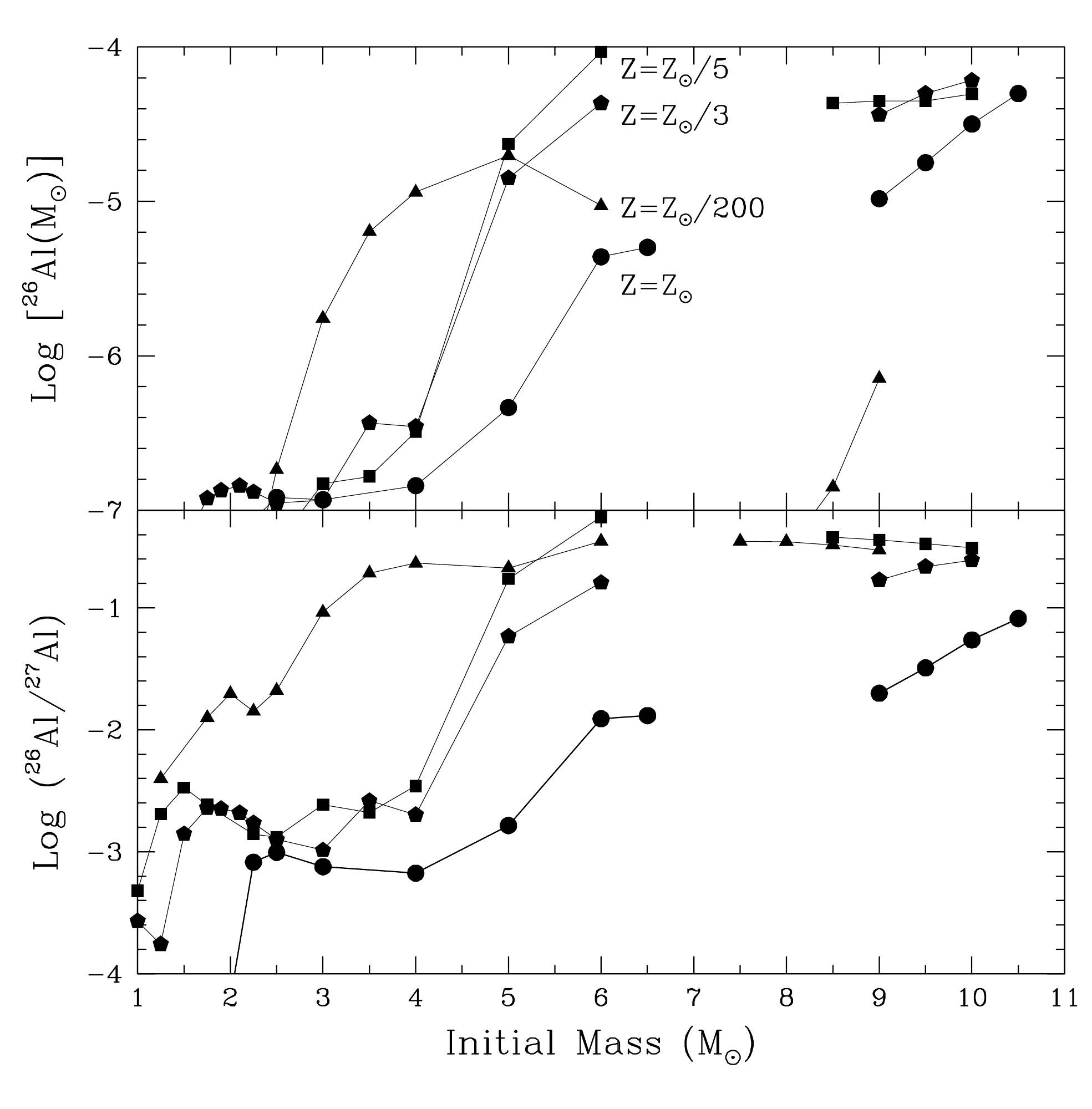}
  \caption{The yields of $^{26}$Al (top panel) for stellar models of 
different masses and metallicities (Z) from  \citet{karakas07} for 
AGB stars of masses up to 6 \Msun, and from  \citet{siess08} for 
the more massive Super-AGB stars. Yields are defined as the total mass of 
$^{26}$Al (in \Msun) lost in the wind during the whole evolution of the 
star (calculated as the average of the time-dependent envelope composition 
weighed on the mass lost at each time). The ratios of the yield of 
$^{26}$Al to the yield of $^{27}$Al are also shown in the bottom panel.}
  \label{fig:al26} \end{figure}

When the 3$^{rd}$ dredge-up occurs after the thermal pulse is extinguished, 
only a small mass of $^{26}$Al is carried from the intershell to the 
stellar surface, of the order of 10$^{-8}$ \Msun, mostly coming 
from a tiny region (roughly 10$^{-4}$ \Msun) at the top of the intershell, 
which was not ingested in the convective pulse and thus did not experience 
the availability of free neutrons. This small abundance of $^{26}$Al 
carried into the envelope translates into a small total contribution of 
the AGB winds to the abundance of $^{26}$Al in the interstellar medium 
(also defined as \emph{yield}) of 10$^{-7}$ \Msun, for AGB stars of masses 
between 1 \Msun\ and 4 \Msun, depending on the metallicity 
(upper panel of Fig.~\ref{fig:al26}), 
though allowing a noticeable increase in the $^{26}$Al/$^{27}$Al ratio at 
the stellar surface, up to a typical value of 2 $\times$ 10$^{-3}$ 
(lower panel of Fig.~\ref{fig:al26}).

The situation is very different for AGB stars of masses higher than 
approximately 4 \Msun. Proton captures occurring in setting (2), i.e., HBB 
at the base of the convective envelope, combined with the 3$^{rd}$ dredged-up 
of $^{25}$Mg produced from efficient activation of the 
$^{22}$Ne($\alpha$,n)$^{25}$Mg reaction in the thermal pulse, produce 
large amounts of $^{26}$Al. These are directly mixed to the stellar 
surface via the envelope convection resulting in yields up to 10$^{-4}$ 
\Msun, and $^{26}$Al/$^{27}$Al ratios up to 0.5 (Fig.~\ref{fig:al26}).  
During HBB the main channel for $^{26}$Al destruction is proton captures 
on $^{26}$Al itself, i.e., $^{26}$Al($p,\gamma$)$^{27}$Si reactions. Also 
in Super-AGB stars HBB produces large quantitites of $^{26}$Al 
 \citep[]{siess08}. 

Figure~\ref{fig:al26} shows the yields of $^{26}$Al and their ratio with
the yield of $^{27}$Al for a variety of AGB stars and Super-AGB of
different masses and metallicities. The plot shows how the
efficiency of $^{26}$Al production increases with stellar mass
and with decreasing metallicity of the stars. This is because the
efficiency of the HBB depends on the temperature at the base of the
convective envelope, which is higher for higher masses and lower
metallicities. For example, a 3.5 \Msun~ star of metallicity 200 times
lower than solar ejects the same amount of $^{26}$Al than a 6.5 \Msun\ 
star at solar metallicity. The reason is that the overall temperature
is controlled by the mass of the CO core, which scales directly with 
the initial mass and inversely with the initial metallicity (see 
Section~\ref{subsec:AGB}). In addition, the lower opacity in lower  
metallicity stars keeps the structure 
more compact and hence hotter. 

In the models of Super-AGB presented
in Fig.~\ref{fig:al26} the 3$^{rd}$ dredge-up is found to be negligible and
HBB produces $^{26}$Al via proton captures on the $^{25}$Mg initially
present in the envelope, without the contribution of $^{25}$Mg from the
intershell. Still, these stars produce a large amount of $^{26}$Al since
there is a large initial amount of $^{25}$Mg in the envelope due to the
large envelope mass.  The lower Super-AGB $^{26}$Al yield at metallicity
solar/200 is due to very high HBB temperatures, at which the rate of the
$^{26}$Al destruction reaction $^{26}$Al($p,\gamma$)$^{27}$Si is
significantly enhanced.

The yields predicted for the $^{26}$Al from AGB stars presented in
Fig.~\ref{fig:al26} are quite uncertain since there are several
stellar and nuclear uncertainties. First, there are uncertainties related to
the modelling of HBB. In fact, the temperature reached at the base of
the convective envelope, which governs the efficiency of the
$^{25}$Mg($p,\gamma$)$^{26}$Al reaction, depends on the modeling of the
temperature gradient within the convectively unstable region. Hence, 
different treatments of the convective layers may
lead to significantly different efficiencies of the HBB.
Second, the uncertainty in the efficiency of the 3$^{rd}$ dredge-up 
already discussed in Section~\ref{subsec:AGB} also affects the $^{26}$Al yields: 
in the low-mass models it affects the dredge-up of $^{26}$Al itself, in 
the massive models it affects the dredge-up of $^{25}$Mg, which is then 
converted into $^{26}$Al via HBB. Third, the mass-loss rate is another 
major uncertain parameter in the modelling of AGB stars. The mass-loss rate 
determines the stellar lifetime and thus the time available to produce 
$^{26}$Al and the final $^{26}$Al yield.

Another model uncertainty is related to the possible occurrence of extra 
mixing at the base of the convective envelope in the low-mass AGB models 
that do not experience HBB. Such extra mixing in AGB stars would be 
qualitatively similar to the extra mixing in red giant stars described in 
Section~\ref{subsec:li}. In the hypothesis of extra mixing, material travels 
from the base of the convective envelope inside the radiative region close 
to the H-burning shell, suffers proton captures, and is taken back up 
into the convective envelope. If the mixed material dips into the H-burning 
shell, down to  
temperatures higher than $\simeq$ 50 MK, then this mechanism could produce 
$^{26}$Al and contribute to some amount of this nucleus in the low-mass 
models  \citep[]{nollett03}. Unfortunately, from a theoretical point of view, 
there 
is no agreement on which mechanism drives the extra mixing and on 
the features of the mixing. Some constraints on it, however, can be 
derived from the composition of MS, S, SC, and C stars as well as 
meteoritic stellar grains, as will be discussed in detail in 
Section~\ref{subsec:al26obs}.

As for nuclear uncertainties, the 
rate of the $^{26}$Al($p,\gamma)^{27}$Si reaction is uncertain by three 
orders of magnitude in the temperature range of interest for AGB stars 
 \citep[]{iliadis01}, with the consequence that $^{26}$Al yields from AGB 
stars suffer from uncertainties of up to two orders of magnitude 
 \citep[]{izzard07,vanraai08b}. New experiments and approaches to estimate 
this rate are needed to get a more precise 
determination of the production of $^{26}$Al in AGB stars.

In spite of all these important uncertainties, current models do indicate 
that at least some AGB models produce a significant amount of $^{26}$Al. 
These models cover a small range of stellar masses, only those suffering 
HBB on the AGB phase. When the yields presented in Fig.~\ref{fig:al26} 
are averaged over a Salpeter initial stellar mass function, the result is 
that AGB stars globally do not provide an important contribution to the 
present abundance of $^{26}$Al in the Galaxy. This contribution sums up to 
only 0.24\% of the contribution from massive star winds and core-collapse 
supernovae  \citep[]{lc06,lugaro08}. Adding up the contribution of Super-AGB 
stars only marginally increases the contribution of AGB stars to Galactic 
$^{26}$Al to 0.85\% of the contribution coming from the more massive stars 
 \citep[see also][]{siess08}.

\subsection{Evidence of $^{26}$Al in AGB Stars} 
\label{subsec:al26obs}

It may be possible to determine the abundance of $^{26}$Al in AGB stars 
using molecular lines of Al-bearing molecules. This was carried out by 
 \citep{guelin95} for the nearest carbon star, CW Leo, using rotational lines 
of AlF and AlCl molecules with different Al isotopic composition. One 
observed line was tentatively attributed to $^{26}$AlF, and from its 
observed strength an upper limit of 0.04 for the $^{26}$Al/$^{27}$Al ratio 
was inferred. No $^{26}$AlCl lines were detected, which led to an upper 
limit of 0.1. These values cannot be reached by solar metallicity AGB 
models (Fig.~\ref{fig:al26}), however, this detection has not been 
confirmed so it is doubtful if it can represent a valid model 
constraint. \index{astronomy!molecular-line spectroscopy}

\begin{figure}
  \centering
  \includegraphics[width=\textwidth]{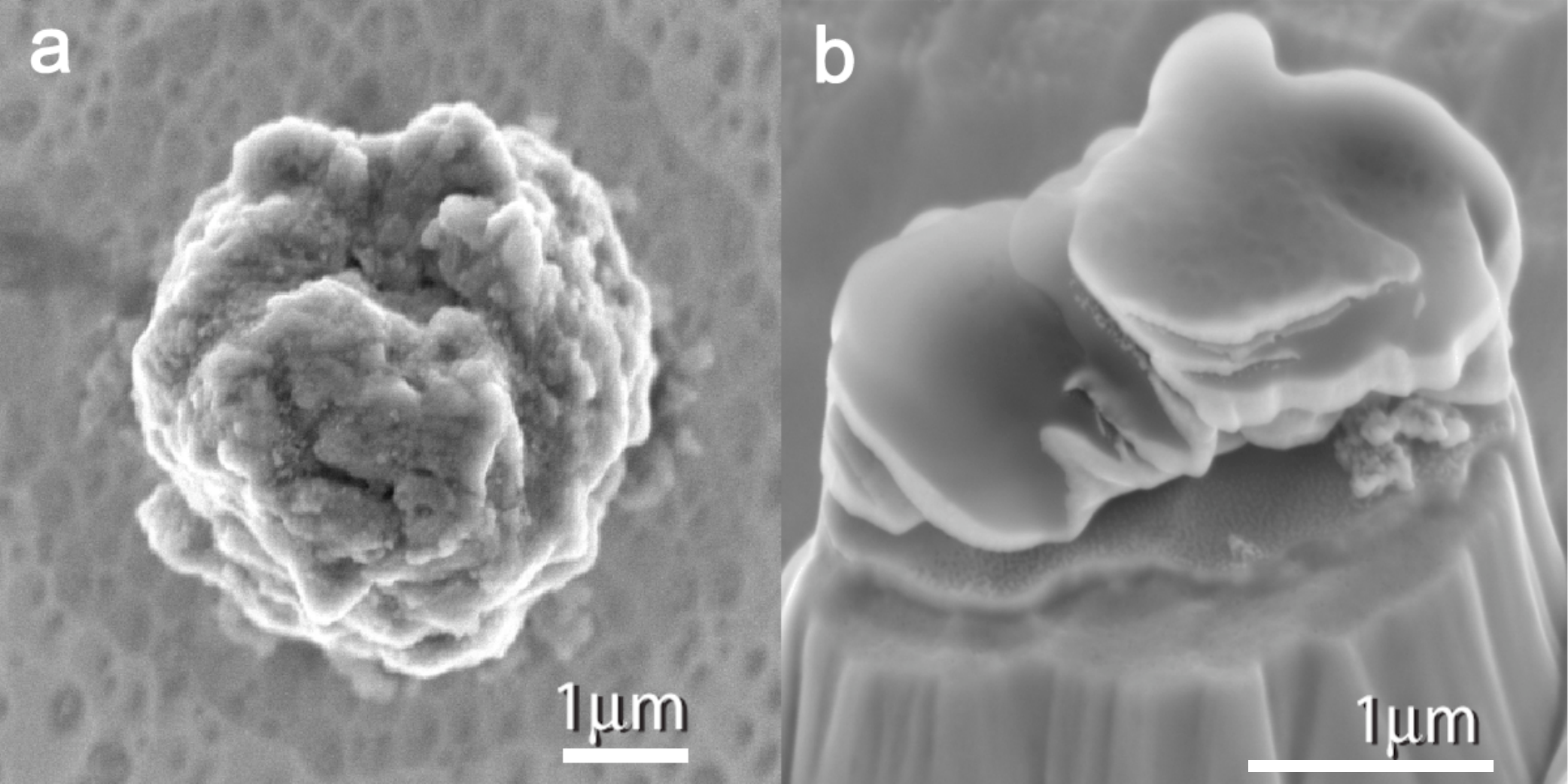}
  \caption{Scanning electron microscope images of dust grains from AGB stars.
(a) 4-$\mu$m-sized silicon carbide (SiC) grain; ubiquitous excesses in
$^{26}$Mg in such grains indicate prior presence of $^{26}$Al. (b)
2-$\mu$m-sized hibonite (CaAl$_{12}$O$_{19}$) grain KH15 \citep{2008ApJ...682.1450N}. 
The grain is sitting on a gold pedestal created by ion-probe sputtering
during isotopic analysis. Excesses of $^{26}$Mg and $^{41}$K indicate
that the grain originally condensed with live $^{26}$Al and $^{41}$Ca.
\index{stardust}}
  \label{fig:SiCgrain} 
\end{figure}

The main observational evidence of $^{26}$Al in AGB stars comes, instead, 
from stardust (see Fig.~\ref{fig:SiCgrain}). Aluminium is one of the main component of most oxide 
stardust grains recovered to date and the initial amount of $^{26}$Al 
present in each grain can be derived from excesses in its daughter nucleus 
$^{26}$Mg. Magnesium is not a main component in corundum (Al$_3$O$_2$) and 
hibonite (CaAl$_{12}$O$_{19}$) grains, hence, in this cases, $^{26}$Mg 
excesses are all attributed to $^{26}$Al decay. In the case of spinel 
(MgAl$_2$O$_4$) grains, instead, Mg is a main component of the mineral and 
thus the contribution of $^{26}$Al to $^{26}$Mg needs to be more carefully 
evaluated by weighing the contribution of the two components. 
Specifically, there are two atoms of Al per each atom of Mg in spinel, 
which corresponds to a roughly 25 times higher ratio than in the average 
solar system material.

The $^{26}$Al/$^{27}$Al ratios are observed to be different in the 
different populations of oxide and silicate grains  \citep[see, e.g., Fig. 8 
of][]{nittler97}. Population I grains cover a wide range of $^{26}$Al initial 
abundance, from no detection to $^{26}$Al/$^{27}$Al $\simeq$ 0.02. The 
presence of $^{26}$Al is used to discriminate Population I oxide grains 
coming from red giant or from AGB stars, since $^{26}$Al is expected to be 
present only in the winds of AGB stars. The $^{26}$Al/$^{27}$Al ratios of 
Population II grains lie at the upper end of the range covered by 
Population I grains, and reach up to $\simeq$0.1  \citep[see also Fig. 6 of][] 
{zinner07}. This is qualitatively consistent with the strong 
$^{18}$O deficits observed in the Population II grains, since both 
signatures are produced by H burning. The mysterious Population III show 
low or no $^{26}$Al, which may indicate that these grains did not come 
from AGB stars. Finally, Population IV grains from supernovae show 
$^{26}$Al/$^{27}$Al ratios between 0.001 and 0.01 (see Chapter 4). 

The $^{26}$Al/$^{27}$Al ratios together with the $^{18}$O/$^{16}$O ratios 
in Population I and II oxide and silicate grains provide an interesting 
puzzle to AGB modellers. Low-mass AGB models do not produce 
$^{26}$Al/$^{27}$Al ratios high enough and $^{18}$O/$^{16}$O ratios low 
enough to match the observations. Massive AGB models can produce 
$^{26}$Al/$^{27}$Al ratios high enough via HBB, however, in this case the 
$^{18}$O/$^{16}$O ratio is too low ($\sim$10$^{-6}$) to match the 
observations (see Section~\ref{subsec:hburn}). Grains with $^{18}$O/$^{16}$O $<$ 10$^{-4}$ 
may have been polluted by 
solar material during the laboratory analysis, which would have shifted 
the $^{18}$O/$^{16}$O ratio to higher values with respect to the true 
ratio of the grain. This argument was invoked to attribute a massive AGB 
stars origin to a peculiar Population II spinel grain, named OC2 
 \citep[]{lugaro07}. However, also the $^{17}$O/$^{16}$O ratio presents a 
problem for this and similar grains because at the temperature of HBB this 
ratio is always much higher than observed 
 \citep[]{boothroyd95,lugaro07,iliadis08}.

The extra-mixing phenomena mentioned in Secs.~\ref{subsec:li} and 
\ref{sec:al26} have been hypothesized to operate in low-mass AGB stars 
below the base of the formal convective envelope to explain the 
composition of Population I and II grains with $^{26}$Al/$^{27}$Al ratios 
greater than $\simeq 10^{-3}$. This idea has been investigated in detail 
by  \citet{nollett03} using a parametric model where the temperature 
(T$_p$), determined by the depth at which material is carried, and the 
mass circulation rate (M$_{circ}$) in the radiative region between the 
base of the convective envelope and the H-burning shell are taken as two 
free and independent parameters. This model was originally proposed to 
explain observations of AGB stars and grains showing deficits in $^{18}$O 
 \citep[]{wasserburg95}. 

In the case of SiC grains, Al is present in the grains as a trace 
element in relatively large abundance, while Mg is almost absent.  Again, 
this means that $^{26}$Mg excesses represent the abundance of $^{26}$Al at 
the time when the grains formed. Mainstream SiC grains from AGB stars show 
$^{26}$Al/$^{27}$Al ratios between $10^{-4}$ and $\simeq$ $2 \times 
10^{-3}$. Models of C-rich AGB stars, i.e., the low-mass models in the 
lower panel of Fig.~\ref{fig:al26}, which do not suffer HBB and hence 
can reach C/O$>$1 in their envelopes, match the observed upper value but do 
not cover the observed range down to the lower values 
 \citep[]{zinner07,vanraai08b}. It is difficult to interpret the lowest values 
as pollution of solar material as in this case one would expect a trend of 
decreasing $^{26}$Al/$^{27}$Al ratios with the total Al content in the 
grains, which is not observed. As discussed in 
Section~\ref{sec:al26}, the main problem modelers have to consider in this 
context is the large uncertainty in the $^{26}$Al$+p$ reaction rate 
 \citep[]{vanraai08b}. The rate uncertainty results in an uncertainty of one 
order of 
magnitude in the $^{26}$Al/$^{27}$Al ratios predicted for C-rich AGB 
stars. If the current upper limit of the rate is employed in the models, 
then only the lower values observed in SiC can be matched and we may need 
to invoke extra-mixing phenomena also to explain the Al composition of SiC 
grains. On the other hand, if the lower or recommended values of the rates 
are used, then only the higher values observed in SiC can be matched.

In conclusion, observational constraints of $^{26}$Al in AGB stars 
provide the potential to investigate some of the most uncertain input 
physics in the modelling of AGB nucleosynthesis: mixing phenomena and 
reaction rates.

\subsection{$^{60}$Fe} 
\label{sec:fe60}

The other famous long-lived radioactive nucleus $^{60}$Fe  \citep[with
a recently revised half life of 2.6 My,][]{rugel09}, of interest from the point of view of
$\gamma$-ray observations, meteoritic stellar grains, and the
composition of the early solar system, can be produced in AGB
stars \citep[]{wasserburg06,lugaro08} via the neutron-capture chain
$^{58}$Fe($n,\gamma$)$^{59}$Fe($n,\gamma$)$^{60}$Fe, where $^{59}$Fe is 
a branching point, and destroyed via
the $^{60}$Fe($n,\gamma$)$^{61}$Fe reaction, whose rate has been
measured experimentally by  \citep{uberseder09}. This is the same chain
of reactions responsible for the production of this nucleus in massive
stars (see Chapter 4, Section 3.1.2). Given that $^{59}$Fe is an unstable 
nucleus with \index{isotopes!60Fe} \index{stars!AGB} \index{isotopes!59Fe}  \index{isotopes!58Fe} 
a relatively short half life of 44 days and with a neutron-capture cross
section $\sigma \simeq$ 23 mbarn  \citep[]{rauscher00}, neutron densities
of at least 10$^{10}$ n/cm$^{3}$ are needed for this branching point to
open at a level of 20\%, allowing production of the long-living
$^{60}$Fe.  If the neutron density is higher than 10$^{12}$ n/cm$^{3}$,
then 100\% of the neutron-capture flux goes through $^{60}$Fe. 

From the description of the neutron sources in AGB stars 
(Section~\ref{subsec:sprocess}), it is clear that $^{60}$Fe can only be produced 
in the convective thermal pulses, where the neutron burst released by the 
$^{22}$Ne neutron source can reach the high neutron density required to 
open the branching point at $^{59}$Fe. Hence, the production of $^{60}$Fe 
in AGB stars is almost completely determined by the activation of the 
$^{22}$Ne neutron source.  The $^{13}$C neutron source may instead destroy 
some $^{60}$Fe in the intershell  \citep[]{wasserburg06}.

\begin{figure}
  \centering
  \includegraphics[width=\textwidth]{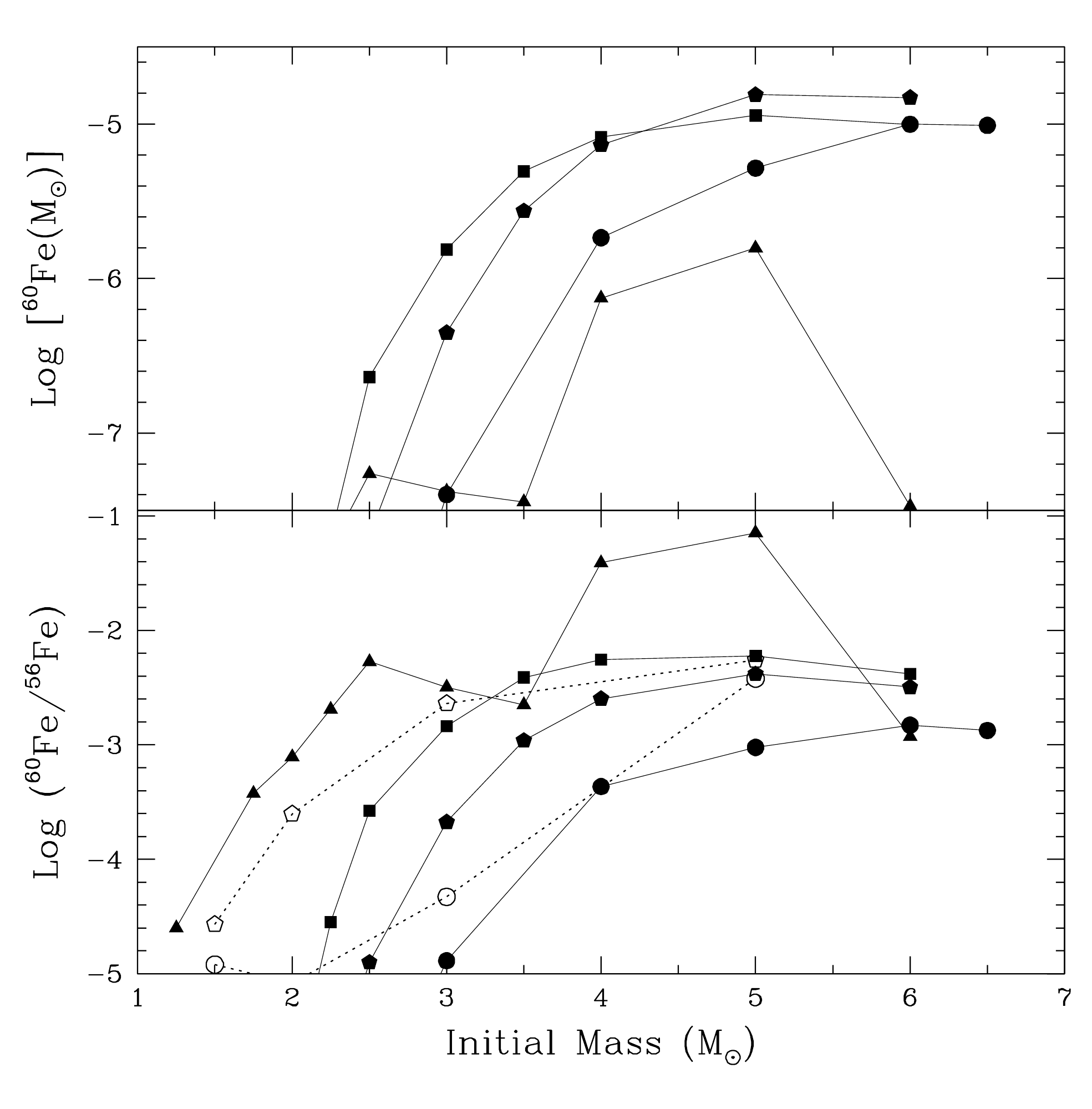}
  \caption{The yields of $^{60}$Fe (top panel) from  \citet{karakas07} 
(see caption of Fig.~\ref{fig:al26} for definition of a yield) and the ratio 
of the yield of $^{60}$Fe to the yield of $^{56}$Fe (bottom panel) for 
stellar models of different masses and metallicities  \citep[full symbols from]
[]{karakas07}. The symbols representing the different metallicities are the 
same as in Fig.~\ref{fig:al26}.
For comparison, the ratios of the abundances 
of $^{60}$Fe and $^{56}$Fe at the end of the AGB evolution computed by 
 \citet{wasserburg06} are also shown as open symbols.}
  \label{fig:fe60} 
\end{figure}

The AGB yields of $^{60}$Fe, and their ratios with the yields of 
$^{56}$Fe, are shown in Fig.~\ref{fig:fe60}. As the temperature at the 
base of the convective thermal pulses increases with increasing the 
stellar mass and decreasing the metallicity, the amount of $^{60}$Fe 
delivered to the interstellar medium increases, reaching up to 10$^{-5}$ 
\Msun, a value comparable to that delivered by a supernova of $\simeq$ 
20 \Msun\  \citep[]{lc06}.
Ratios of the $^{60}$Fe and $^{56}$Fe abundances at the 
end of the AGB 
phase from the AGB neutron-capture models of  \citep{wasserburg06} also 
plotted in Fig.~\ref{fig:fe60}.

As for $^{26}$Al, also in the case of $^{60}$Fe stellar and nuclear 
uncertainties affect the results presented in Fig.~\ref{fig:fe60} (and 
different choice in the model inputs are responsible for variations in 
the results obtained by different authors). First, the overall mass 
carried to the envelope via the 3$^{rd}$ dredged-up is 
essential to the determination of the envelope $^{60}$Fe abundance in 
AGB stars. This is because $^{60}$Fe is made only via neutron captures 
in the He-rich intershell and needs to be mixed into the envelope in 
order to show up at the stellar surface and to be carried to the 
interstellar medium by the winds.  Hence, the $^{60}$Fe yield is 
directly related to the efficiency of the 3$^{rd}$ dredge-up. For example, 
models experiencing little or no 3$^{rd}$ dredge-up produce a null $^{60}$Fe 
yield. This important point applies to all long-living radioactive 
nuclei produced in AGB stars, except for the case of $^{26}$Al, which is 
made via HBB directly within the envelope. Second, the mass-loss rate affects 
the result as it determines the stellar lifetime and thus the number of 
thermal pulses and 3$^{rd}$ dredge-up episodes. 

Nuclear physics inputs that contribute important uncertainties to the 
production of $^{60}$Fe are the rate of the neutron source reaction 
$^{22}$Ne($\alpha$,n)$^{25}$Mg, which determines how many neutrons are 
produced in the thermal pulses, and the neutron-capture cross section of 
$^{59}$Fe, which is estimated only theoretically  \citep[]{rauscher00} as the 
short half life of this nucleus hampers experimental determinations.

\subsection{$^{36}$Cl and $^{41}$Ca} \label{sec:cl36ca41}

Two more long-lived radioactive nuclei lighter than iron are of special 
interest because they are observed to be present in the early solar system 
and can be made by neutron captures in the intershell of AGB stars: 
$^{36}$Cl (with half life of 0.3 Myr) and $^{41}$Ca (with half life of 0.1 
Myr). Differently from $^{60}$Fe, production of these nuclei does not 
require the activation of branching points, since $^{36}$Cl and 
$^{41}$Ca are made by neutron captures on $^{35}$Cl and $^{40}$Ca, 
respectively, which are stable nuclei with relatively high solar 
abundances. Neutron captures also destroy $^{36}$Cl and $^{41}$Ca via 
different channels, the predominants being $^{41}$Ca($n,\alpha$)$^{38}$Ar, 
with $\sigma \simeq$ 360 mbarn and $^{36}$Cl($n,p$)$^{36}$S, 
with $\sigma \simeq$ 118 mbarn.
\index{isotopes!41Ca}  \index{isotopes!36Cl} \index{process!neutron capture}

Neutrons coming from the $^{22}$Ne neutron source are responsible for the 
production of $^{36}$Cl and $^{41}$Ca. As there are no branching points 
involved, this is not due to the high neutron density of this neutron 
flux, as it is for the production of $^{60}$Fe, but to the fact that 
neutrons released by the $^{22}$Ne in the thermal pulse affect the 
composition of the whole He-rich intershell material, where large 
initial quantities of the seed nuclei $^{35}$Cl and $^{40}$Ca are 
available. On the contrary, neutrons released by the $^{13}$C neutron 
source affect a small fraction of the intershell material, being the 
$^{13}$C pocket roughly 1/10$^{th}$ to 1/20$^{th}$ of the 
intershell (by mass) in the current models.

In general, to produce neutron-rich isotopes of elements lighter than iron
by the $s$-process a small number of neutrons captured by seed nucleus are
needed: only one in the cases of $^{36}$Cl and $^{41}$Ca. Hence, final
abundances are determined to a higher level by the availability of seed
nuclei, rather than that of free neutrons. For the light nuclei a
production flux from the lighter to the heaviest elements does not occur
(strictly speaking it is not correct to apply the $s$-process terminology
in this case), instead, the nucleosynthetic process is very
localized: neutron captures on the sulphur isotopes, for example, do not
affect the abundances of the chlorine isotopes and so on.  This is because
neutron-capture cross section of nuclei lighter than iron are much smaller
(by as much as 3 orders of magnitude) than those of typical nuclei heavier
than iron. Hence, to produce nuclei heavier than iron by
the $s$-process, instead, including the relatively large number of long-living
radioactive nuclei lying on the $s$-process path discussed in
Section~\ref{sec:radioneutroncapheavy}, a production flux from the lighter to
the heavier elements occurs, where many neutrons are captured by the iron
seeds and it is possible to reach up to the heaviest elements. Hence, the
number of free neutrons plays a dominant role in this case.

\begin{figure}
  \centering
  \includegraphics[width=\textwidth]{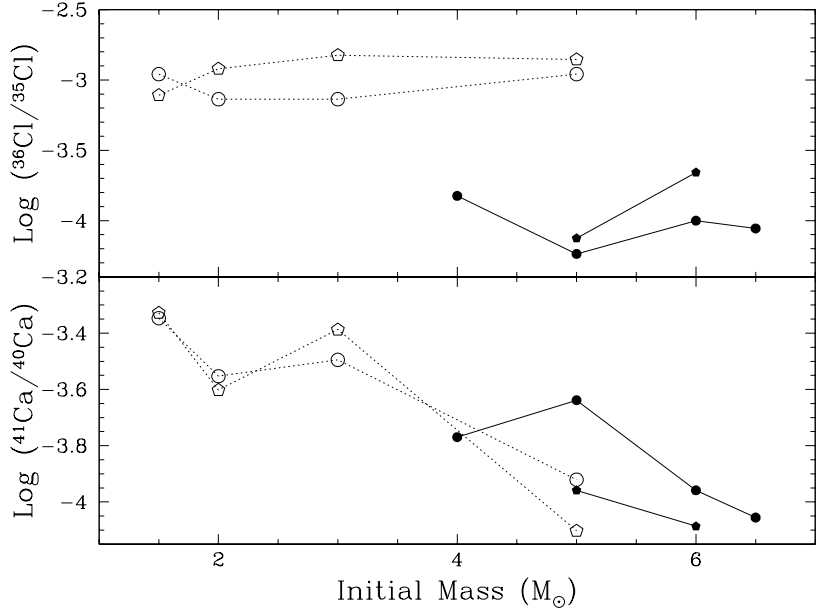}
  \caption{$^{36}$Cl/$^{35}$Cl and $^{41}$Ca/$^{40}$Ca abundance ratios at 
the end of the AGB evolution computed by  \citet{wasserburg06} (open 
symbols) and by van Raai, Lugaro, \& Karakas (unpublished results, 
full symbols). The symbols representing the different metallicities are the 
same as in Fig.~\ref{fig:al26}.}
  \label{fig:cl36ca41} 
\end{figure}

The $^{36}$Cl/$^{35}$Cl and $^{41}$Ca/$^{40}$Ca abundance ratios at the 
end of the AGB evolution computed by  \citet{wasserburg06} and by  
van Raai, Lugaro, \& Karakas  \citep[which are based on the same codes and 
stellar models of][]{karakas07} are plotted in 
Fig.~\ref{fig:cl36ca41}. 
As in the case of $^{60}$Fe, the main model uncertainties affecting these 
results is the efficiency of the 3$^{rd}$ dredge-up, the mass-loss rate, and the 
rate of the $^{22}$Ne($\alpha$,n)$^{25}$Mg reaction. 

Moreover, while experimental estimates for the neutron-capture cross 
section of $^{36}$Cl and $^{41}$Ca are available 
 \citep[e.g.,][]{desmet06}, a difficult problem is to provide a reliable 
set of values for the electron-capture rate of $^{41}$Ca, in 
particular as it is expected to vary significantly for different 
temperatures and densities relevant to stellar conditions (Chapter 9).   
As most electron captures in the $^{41}$Ca atom occurs on 
electrons belonging to the electron shell closest to the nucleus (the K 
shell), when the temperature increases to 100 MK and all electrons have 
escaped the atom leaving the nucleus bare, the half life of $^{41}$Ca 
increases by almost three orders of magnitude. However, if, still at a 
temperature of 100 MK, the density increases to 10$^4$ g/cm$^3$, 
electrons are forced nearby $^{41}$Ca nuclei and the half life decreases 
back to its terrestrial value. The only set of theoretical data for this 
reaction are those provided by  \citet{fuller82}. Moreover, the 
temperature and density dependence of the electron-capture rate of 
$^{41}$Ca has never been properly implemented in AGB stellar models, in 
particular it has not yet been solved coupled to convective motions, 
both in the thermal pulses and in the stellar envelope, where material 
is constantly carried from hotter denser regions to cooler less dense 
regions and viceversa. Given these considerations, we are far from an 
accurate determination of the abundance of $^{41}$Ca made by AGB stars,
although there is observational evidence of its presence in these stars
from $^{41}$K enrichments in hibonite stardust grains attributable 
to {\it in situ} 
decay of now-extinct $^{41}$Ca \citep{2008ApJ...682.1450N}.

In summary, and in relevance to the early solar system composition of 
\index{solar system} \index{stars!AGB}
long-lived radioactive nuclei discussed in Chapter 6, AGB stars can 
produce some of the radioactive nuclei found to be present in the early 
solar system: $^{26}$Al via hot bottom burning and $^{41}$Ca and $^{60}$Fe 
via neutron captures in the thermal pulse and the 3$^{rd}$ dredge-up. In 
certain mixing conditions the abundances of these nuclei can be produced 
by AGB stars in the same proportions observed in the early solar system 
 \citep[]{wasserburg06,trigo09}. On the other hand, 
$^{36}$Cl cannot be produced in the observed amount. Uncertainties in the 
neutron-capture cross sections of $^{35}$Cl and $^{36}$Cl may play a role 
in this context.

Finally, a characteristic signature of the AGB stars inventory of
long-living radioactive nuclei, is that, unlike supernovae (Chapter 4 and Chapter 5, 
Section 3), 
AGB stars cannot possibly produce $^{55}$Mn, another
nucleus of relevance to early solar system
composition. This is because $^{55}$Mn is a proton-rich nucleus, lying
on the proton-rich side of the valley of $\beta$-stability, and thus it
cannot be made by neutron captures.

\subsection{Long-lived Radioactive Isotopes Heavier than Fe} 
\label{sec:radioneutroncapheavy}
\subsubsection{Predicted Isotopic Abundances} 

The $s$-process in AGB stars produces significant abundances of six
\index{isotopes!81Kr}  \index{isotopes!93Zr}  \index{isotopes!99Tc}  \index{isotopes!107Pd} \index{isotopes!135Cs}  \index{isotopes!205Pb} 
long-lived radioactive nuclei heavier than iron: $^{81}$Kr, $^{93}$Zr,
$^{99}$Tc, $^{107}$Pd, $^{135}$Cs, and $^{205}$Pb. The survival of
$^{135}$Cs and $^{205}$Pb in stellar environments is however very
uncertain and can even be prevented because of the strong and uncertain
temperature and density dependence of their half lives, decreasing by
orders of magnitudes in stellar conditions and determined only
theoretically  \citep[as in the case of $^{41}$Ca. See detailed discussion
by][and also Appendix B]{mowlavi98,wasserburg06}. While $^{93}$Zr,
$^{99}$Tc, $^{107}$Pd, and $^{205}$Pb are on the main $s$-process path
and are produced by neutron captures on the stable isotopes $^{92}$Zr,
$^{98}$Mo\footnote{followed by fast decay of $^{99}$Mo, with a half life
of 66 hours}, $^{106}$Pd, and $^{204}$Pb, respectively, $^{81}$Kr and
$^{135}$Cs are not on the main $s$-process path, but can be reached via
the activation of branching points at $^{79}$Se and $^{80}$Br, and
$^{134}$Cs, respectively (as described in Appendix~B).   

\begin{figure}
  \centering
  \includegraphics[width=0.9\textwidth]{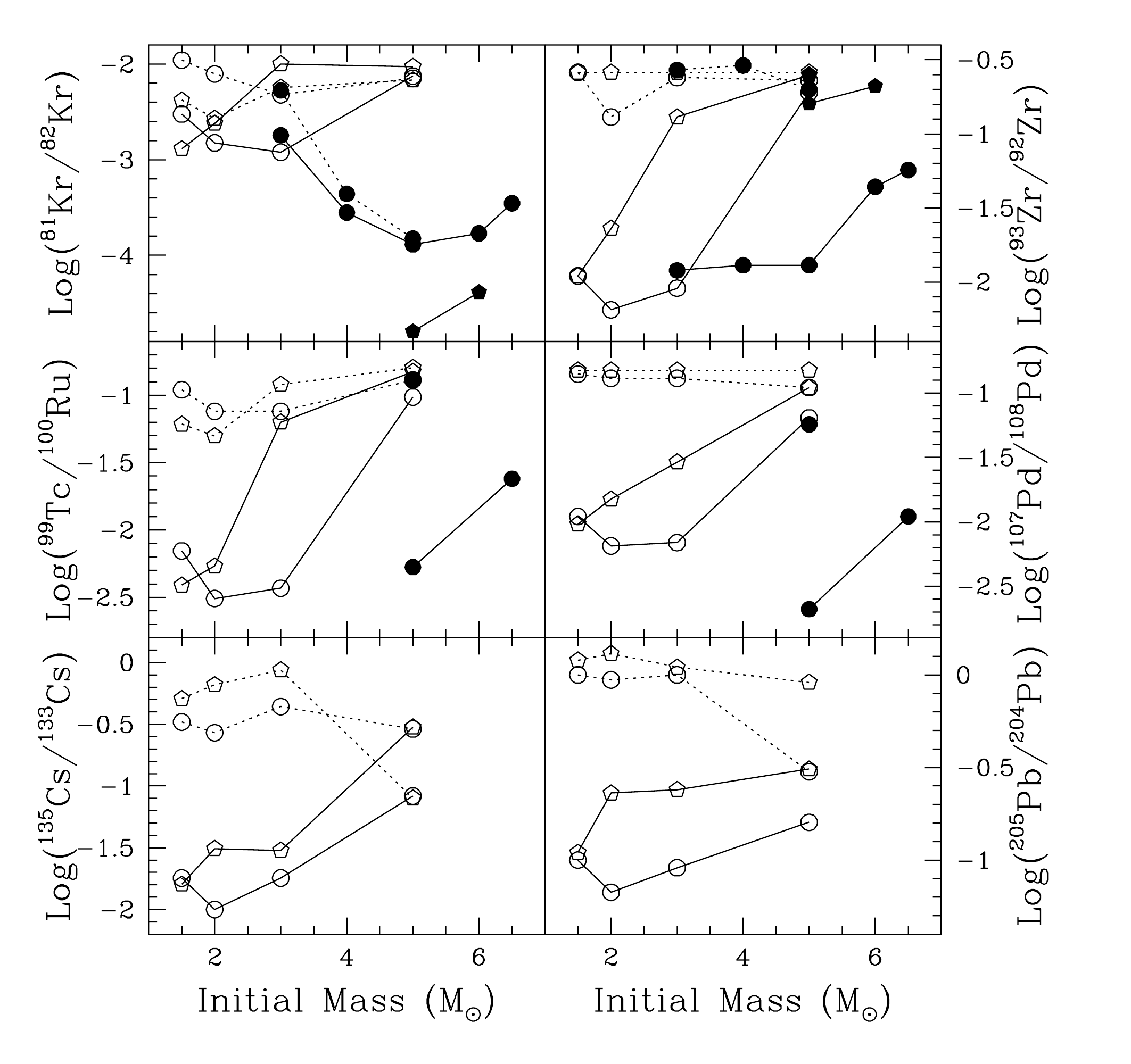}
  \caption{Abundance ratios of long-lived radioactive nuclei heavier than 
iron, with respect to one of their nearest stable isotope, at the end of 
the AGB evolution computed by  \citet{wasserburg06} (open symbols) and 
by van Raai, Lugaro, \& Karakas (unpublished results, full symbols). 
The symbols representing the different metallicities are the 
same as in Fig.~\ref{fig:al26}.
Symbols connected by the solid line 
represent models computed without the inclusion of the $^{13}$C neutron 
source, symbols connected by the dotted lines represent models computed 
with the inclusion of the $^{13}$C neutron source.} 
\label{fig:kr81} 
\end{figure}

Figure~\ref{fig:kr81} presents the abundance ratios of long-living 
radioactive isotopes heavier than iron produced during the $s$-process in 
AGB stars to one of their nearest stable isotopes calculated by 
 \citet{wasserburg06} and by van Raai et al. (unpublished results). For all ratios, except    
$^{81}$Kr/$^{82}$Kr, the inclusion of the $^{13}$C neutron source for 
models of masses lower than $\simeq$ 3 - 4 \Msun, completely changes the results, 
since the $^{22}$Ne source is not significantly activated in these 
low-mass stellar models. It also makes an important difference in the 
absolute abundance of the all isotopes involved, with very low production 
factors with respect to the initial value if the $^{13}$C neutron source 
is not included  \citep[see Table 4 of][]{wasserburg06}. The case of 
$^{81}$Kr/$^{82}$Kr is different in that it does not feel the 
inclusion of the $^{13}$C neutron source as much as the other ratios 
because, even 
for the low-mass stars, the marginal activation of the $^{22}$Ne reaction 
in the latest thermal pulses affects the production of the $s$-process 
elements up to the first $s$-process peak, including Kr, and of $^{81}$Kr in 
particular via the branching point at $^{79}$Se.

For stellar models with initial masses higher than $\simeq$ 3 - 4 \Msun, 
depending on the metallicity, the $^{22}$Ne neutron source is mainly
responsible for the activation of the $s$-process and thus the production 
of the heavy long-lived isotopes. Hence, in these models, the inclusion 
of a $^{13}$C neutron source typically does not make a significant 
difference in the final ratios, except in the case of 
$^{205}$Pb/$^{204}$Pb. This ratio is different in that it always feels the 
effect of the inclusion of the $^{13}$C neutron source because production 
of the element Pb, corresponding to the third and last $s$-process peak, 
is possible only if very large neutron exposures are available ($\sim$ 
mbarn$^{-1}$), which can only be produced by the $^{13}$C neutron source.

It is interesting to discuss in detail the results for the 3 \Msun\ 
stellar model of 1/3 solar metallicity, because this model represents an 
example of the transition between the two regimes of the $s$-process in 
AGB stars: when neutrons are provided by the $^{13}$C or by the $^{22}$Ne 
source. In this model the number of free neutrons produced by the 
$^{22}$Ne source is higher than in the solar metallicity model of the same 
mass partly because the temperature in the thermal pulses is slightly higher, but 
mostly because there is a smaller number of nuclei present to capture 
neutrons. Hence, the neutron flux coming from the $^{22}$Ne neutron source 
affects the production of the long-living isotopes up to $^{99}$Tc, but not 
that of the long-living isotopes of higher masses: for $^{107}$Pd, 
$^{133}$Cs, and $^{205}$Pb, $^{13}$C is still the main neutron source.

In addition to the main effect due to the shift from the $^{13}$C to the 
$^{22}$N regime with changing the initial mass and metallicity of the 
star, smaller variations due to the marginal effect of the $^{22}$Ne 
neutron source in the models of low-mass are always visible in the details 
of the production of the heavy long-living nuclei affected by the 
operation of the branching points activated in thermal pulses: $^{81}$Kr 
and $^{135}$Cs. For example, restricting our view to the solar 
metallicity models of mass lower than 4 \Msun\ and computed with the 
inclusion of the $^{13}$C neutron source, the $^{81}$Kr/$^{82}$Kr ratio 
decreases with the stellar mass as $^{81}$Kr is progressively skipped by 
the branching point at $^{79}$Se at the higher neutron densities 
experienced by the higher mass models. The opposite happens for the 
$^{135}$Cs/$^{133}$Cs ratio, which increases with the stellar mass as the 
branching point at $^{134}$Cs becomes progressively more active.

When considering the effect of branching points on the production of heavy 
\index{branching point} \index{process!s process} \index{stars!AGB}
long-living radioactive nuclei by the $s$-process in AGB stars it is worth 
noting that $^{129}$I and $^{182}$Hf - two long-lived radioactive 
isotopes of special interest for the composition of the early solar system 
- are not significantly produced in AGB stars. Production of $^{129}$I 
is not possible because the half life of $^{128}$I
is only 25 minutes (see Appendix~B), 
while $^{182}$Hf is produced, with
$^{182}$Hf/$^{180}$Hf up to $\simeq$ 0.02 in the AGB envelope, via
activation of the branching point at $^{181}$Hf (Appendix~B). This
results in $^{182}$Hf/$^{180}$Hf down to $\simeq 10^{-6}$ 
\citep[after dilution of AGB
ejecta in the interstellar medium][]{wasserburg94}, which is too low to explain
the early solar system value of $\simeq 10^{-4}$ (see Chapter 6).
\index{isotopes!129I}  \index{isotopes!182Hf} 

The main uncertainties affecting both sets of predictions shown in Fig. 
~\ref{fig:kr81} are the detailed features of the proton diffusion 
leading to the production of the $^{13}$C neutron source
(see end of Section~\ref{subsec:AGB}). Ratios that 
depend on the activation of the $^{22}$Ne neutron source are also 
sensitive to the choice of the mass loss rate and of the 
$^{22}$Ne($\alpha$,n)$^{25}$Mg reaction rate. The treatment of branching 
points is also of importance in the determination of $^{81}$Kr and $^{135}$Cs.
For example, in the case of 
$^{81}$Kr/$^{82}$Kr, the treatment of the temperature dependence of the 
decay rate of the branching point nucleus $^{79}$Se is fundamental to 
the final result, as demonstrated by the fact that including the 
temperature dependence of these decay rate  \citep[as carried out  
by][]{wasserburg06} produce a $^{81}$Kr/$^{82}$Kr ratio two orders of 
magnitude larger than using the terrestrial value as constant (as done 
by van Raai, Lugaro, \& Karakas, unpublished results).

In summary, due to the $s$-process, AGB stars are a rich source of
radioactive elements heavier than Fe. The signature of this production
is confirmed, e.g., by the presence of Tc observed in AGB star and in
meteoritic stardust grains (as discussed further in the next section) and
possibly in primitive meteoritic solar-system materials (as discussed in
detail in Chapter 6).

\subsubsection{Signatures in Stardust}
\label{subsec:heavystardust}

The historical observation of Tc in late type giants 
(Section~\ref{subsec:sprocess}) was confirmed by the presence of $^{99}$Tc in 
single stardust SiC grains at the time of their formation discovered via 
laboratory analysis of the Ru isotopic composition of these grains 
\citep[]{savina:04}.  Since both Tc and Ru are refractory elements, they were 
included in SiC grains as trace elements during grain formation. To match 
the observational stardust data both the contribution of $^{99}$Ru and 
$^{99}$Tc predicted by AGB stellar models to the total nuclear abundance 
at mass 99 must be considered. Radiogenic decay of $^{99}$Tc occurs  
in the intershell in the absence of neutron fluxes, in the stellar 
envelope, and inside the grains.
\index{isotopes!99Tc}  \index{isotopes!135Cs} \index{stardust} 

On the other hand, there is no evidence for a contribution of $^{135}$Cs 
to $^{135}$Ba when comparing AGB model predictions to laboratory data of 
the $^{135}$Ba/$^{136}$Ba ratio in single SiC grains 
\citep[see Fig. 16 of][]{lugaro03b}. 
This is probably because Cs is not as refractory as Ba and 
thus was not included in the grains at the time of their formation.

\bibliographystyle{spbasic}




\backmatter
\appendix
%
%
%
%
%
%

%
\chapauthor{Maria Lugaro\footnote{Monash University, Victoria 3800, Australia} and Alessandro Chieffi\footnote{I.N.A.F., 00133 Roma, Italy}}
\chapter{Radionuclides and their Stellar Origins}
\label{App:Radionuclides} 
\chaptermark{radionuclides and their stellar origins} 
We supplement this book, and in particular 
the discussion of stellar nucleosynthesis presented in Chapter 3, by a list of the unstable isotopes 
at which branching points that become 
relevant in the $s$-process reaction chain in AGB stars are activated. 
For sake of clarity and a better understanding it is advisable to 
go through the list with a chart of the nuclides at hand. For each 
branching point a brief description of its operation and its relevance in 
the study of the $s$-process in AGB stars is presented. All listed 
isotopes suffer $\beta^-$ decay, unless specified otherwise. It should 
also be noted that usually in $s$-process conditions nuclear energy 
metastable levels higher than the ground state are not populated, thus the 
effect of these states does not need to be included in the study of 
branching points, except for the special cases reported in the list 
\citep[see also][]{ward77}.

Branching factors (see Sect.~\ref{subsec:sprocess})
for each branching point can be calculated for 
a given temperature, density, and neutron density conditions
referring to \cite{takahashi87} for the $\beta$ decay rates, and to
\cite{rauscher00} for the neutron capture cross section, unless advised
otherwise in the description below.\footnote{Maria Lugaro thanks Roberto
Gallino and Franz Kaeppeler for communicating the passion for branching
points and for help with this section.}

\begin{description}


\item[{\bf $^{35}$S}]{This branching point may lead to production of the
rare neutron-rich $^{36}$S, whose abundance can be observed in stars via
molecular lines, and may be measured in sulphur-rich meteoritic
materials \cite[for discussion and models see][]{mauersberger04}.}

\item[{\bf $^{36}$Cl and $^{41}$Ca}]{These are both long-living nuclei
produced and destroyed - mostly via ($n,p$) and ($n,\alpha$) channels -
via neutron captures in AGB stars, and discussed in detail in
Section~\ref{sec:cl36ca41} of Chapter 3. While $^{36}$Cl behaves as as stable nucleus
during the $s$-process, the half life of $^{41}$Ca against electron
captures has a strong temperature and density dependence, which could
make it act as a branching point and most importantly prevent its
survival instellar environments as in the case, e.g., of the other
long-living nucleus $^{205}$Pb.}

\item[{\bf $^{45}$Ca}]{This branching point may 
lead to production of the rare neutron-rich 
$^{46}$Ca, which could be measured in Ca-rich meteoritic material.}

\item[{\bf $^{59}$Fe}]{This important branching point leads to the 
production of 
the long-living radioactive nucleus $^{60}$Fe. See Section~\ref{sec:fe60} of Chapter 3 for 
a detailed description and AGB model results.}


\item[{\bf $^{63}$Ni}]{The half life of this nucleus 
decreases from 100 yr to $\simeq$ 12 yr at 
300 MK. The associated branching point affects 
the production of the rare neutron-rich $^{64}$Ni as 
well as the $^{65}$Cu/$^{63}$Cu ratio.}

\item[{\bf $^{64}$Cu}]{The half life of this nucleus 
is short, of the order of a few hours, 
however, this isotope is a branching point on the $s$-process paths as it 
has comparable $\beta^+$ and $\beta^-$ decay rates. The 
branching point may affect the 
production of $^{64}$Ni and $^{65}$Cu.}

\item[{\bf $^{65}$Zn}]{This nucleus suffers $\beta^+$ decay and the 
branching point may affect the production of $^{65}$Cu.}

\item[{\bf $^{71}$Ge}]{This nucleus suffers $\beta^+$ decay and 
the branching point may affect the 
production of $^{71}$Ga.}

\item[{\bf $^{79}$Se}]{This branching point may lead to 
production of the long-living 
radioactive isotope $^{81}$Kr. This production occurs when the temperature 
increases in the AGB thermal pulse, and the half life of $^{79}$Se decreases 
from the terrestrial half life of 65,000 yr to roughly 4 yr at 300 MK due 
to population of the shorter-living isomeric state \citep[][and see 
Section~\ref{sec:radioneutroncapheavy} for model results]{klay88}. Operation 
of this branching point also affects the $^{81}$Br/$^{79}$Br ratio.}

\item[{\bf $^{80}$Br}]{The half life of this nucleus is short, 
of the order of minutes, 
however, it is a branching point on the $s$-process paths as it 
can decay both $\beta^+$ and $\beta^-$, with the $\beta^-$ roughly ten 
times faster than the $\beta^+$ channel. It can affect the production of 
$^{81}$Kr.}

\item[{\bf $^{81}$Kr}]{This nucleus is too long living (T$_{1/2}$ = 0.23
Myr, down to 2300 yr at temperature 300 MK) 
to act as a branching point during the $s$-process and rather
behaves as a stable nucleus. Its production during the $s$-process is
discussed in detail in Section~\ref{sec:radioneutroncapheavy} of Chapter 3. Its
radiogenic decay leads to $^{81}$Br.}

\item[{\bf $^{85}$Kr}]{The relatively long half life of $^{85}$Kr (11 yr) 
allows this branching point to activate already at low neutron densities, 
$> 5 \times 10^{8}$ n/cm$^{3}$. The actual operation of this branching 
point is complicated by the fact that roughly 40\% of 
$^{84}$Kr($n,\gamma$) reactions during the $s$-process result in the 
production of the isomeric state of $^{85}$Kr. Approximately 80\% of these 
nuclei quickly decay into $^{85}$Rb, with a half live of 4.5 hours, while 
the remaining 20\% relax into the ground state. The production of 
$^{87}$Rb, a very long-living isotope with half live of 48 Gyr and a magic 
number of neutrons N=50, has traditionally been attributed to the 
activation of the branching point at $^{85}$Kr \cite[]{lambert95,abia01}. 
\cite{vanraai08a} showed that the activation of the branching 
point at $^{85}$Kr mostly results in the production of $^{86}$Kr, a 
nucleus with a magic number of neutrons N=50 and a very small neutron 
capture cross section of only $\simeq$ 3.4 mbarn. $^{86}$Kr is thus more likely to accumulate than to 
capture the further neutron that would allow the production of $^{87}$Rb. 
The importance of the production of $^{86}$Kr in meteoritic SiC grains and 
the $s$-process is discussed in Section~\ref{subsec:branchSiC} of Chapter 3.}

\item[{\bf $^{86}$Rb}]{The branching point at $^{86}$Rb is activated at 
relatively high neutron densities, above $10^{10}$ n/cm$^{3}$, being the 
half life of this nucleus 18.7 days, and it leads directly to the 
production of $^{87}$Rb. The importance of $^{87}$Rb in $s$-process 
observations and models is discussed in  
Section ~\ref{subsec:branchstars} of Chapter 3.}
 
\item[{\bf $^{89,90}$Sr} and {\bf $^{91}$Y}]{The branching point at 
$^{89}$Sr may produce the unstable $^{90}$Sr, also a branching point 
producing $^{91}$Sr, which quickly decays into unstable $^{91}$Y.  This is 
also a branching point, producing $^{92}$Y, which quickly decays into 
stable $^{92}$Zr. The final result of the operation of this chain of 
branching points is to decrease the production of $^{90}$Zr and 
$^{91}$Zr, with respect to that of $^{92}$Zr. This point is discussed by 
\cite{lugaro03b}, in relevance to the Zr isotopis ratios measured in 
meteoritic silicon carbide (SiC) grains from AGB stars.}

\item[{\bf $^{93}$Zr}]{This nucleus is too long-living (T$_{1/2}$ = 1.5
Myr) to act as a branching point during the $s$-process and rather behaves as a stable nucleus (see
Section~\ref{subsec:sprocess} and Fig.~\ref{fig:sprocesspath} of Chapter 3), with an
experimentally determined neutron-capture cross section \cite[]{macklin85a}.
Its production during the $s$-process is discussed in detail in
Section~\ref{sec:radioneutroncapheavy} of Chapter 3. Its radiogenic decay produces most
of the solar abundance of $^{93}$Nb. (A small fraction of $^{93}$Nb 
is also contributed by the radiogenic decay of $^{93}$Mo, T$_{1/2}$ = 3500 
yr, which is not on the main $s$-process path but can be produced by 
neutron-capture on the relatively abundant $p$-only $^{92}$Mo, 15\% of solar 
Mo).}

\item[{\bf $^{95}$Zr}]{This important branching point can lead to 
production by the $s$-process of $^{96}$Zr 
if N$_n > 5 \times 10^8$ n/cm$^3$. Zr isotopic ratios have been estimated 
in MS and S stars via molecular lines and measured in meteoritic SiC 
grains, providing constraints on the neutron density in the thermal 
pulses. This point is further discussed in Secs.~\ref{subsec:branchstars} 
and \ref{subsec:branchSiC}.}

\item[{\bf $^{94,95}$Nb}]{The half life of $^{94}$Nb decreases from 
terrestrial 20,000 yr to $\simeq$ 0.5 yr at 100 MK and $\simeq$ 9 days 
at 300 MK. This branching point can produce the unstable $^{95}$Nb, which 
is also a branching point producing the unstable $^{96}$Nb, which quickly 
decays into stable $^{96}$Mo. Via the operation of the $^{94}$Nb branching 
point the $^{94}$Mo nucleus is skipped during the $s$-process chain, this 
nucleus is in fact classified as a $p$-only nucleus.}

\item[{\bf $^{99}$Tc}]{The half life of $^{99}$Tc is 0.21 Myr, and
decreases to 0.11 Myr at 100 MK and to 4.5 yr at 300 MK. Thus, the
neutron-capture path of the branching point is mostly open, producing
$^{100}$Tc, which quickly decays into $^{100}$Ru, thus skipping
$^{99}$Ru (Fig.~\ref{fig:sprocesspath}). Then, radiogenic decay of
$^{99}$Tc produces $^{99}$Ru.  The production of $^{99}$Tc is discussed
in detail in Sec~\ref{sec:radioneutroncapheavy}, and mentioned in
Section~\ref{subsec:heavystardust} of Chapter 3 in relation to $^{99}$Ru in meteoritic
SiC grains.}

\item[{\bf $^{107}$Pd}]{This nucleus is too long-living
(T$_{1/2}$ = 6.5 Myr, down to $\simeq$ 700 yr at 300 MK) 
to act as a branching point during the $s$-process
and rather behaves as a stable nucleus, with an
experimentally determined neutron-capture cross section \cite[]{macklin85b}.
Its production during the $s$-process is discussed in detail in
Section~\ref{sec:radioneutroncapheavy}of Chapter 3.  Its radiogenic decay is
responsible for production of $^{107}$Ag.}

\item[{\bf $^{128}$I}]{The decay half life of this nucleus is too short to 
allow for neutron captures, however, there is a marginal branching point 
here due to the fact that $^{128}$I has both $\beta^+$ and $\beta^-$ decay 
channels. The $\beta^+$ channel has significant temperature and density 
dependence and represents only a few percent of the decay rate. Nevertheless, 
this branching point has been investigated in detail because it affects 
the precise determination of relative abundances of the two $s$-only 
isotopes $^{128}$Xe and $^{130}$Xe, and because the timescale for its 
activation of the order of 25 minutes is comparable to that of the 
convective turn-over timescale of the material inside AGB thermal pulses 
of hours \cite[]{reifarth04}.}

\item[{\bf $^{133}$Xe}]{May lead to production of the $^{134}$Xe. Of 
interest in relation to the Xe-S component from SiC grains in primitive 
meteorites, as discussed in Section~\ref{subsec:branchSiC} of Chapter 3.}

\item[{\bf $^{134,135,136,137}$Cs}]{The chain of branching points at the
Cs isotopes is of particular interest because it affects the isotopic
composition of the $s$-process element Ba and in particular the relative
abudances of the two $s$-only nuclei $^{134}$Ba and $^{136}$Ba, as it is
discussed in Section~\ref{subsec:branchSiC} in relation to Ba
data from meteoritic SiC grains. The branching point at $^{134}$Cs
allows production of the long-living isotope $^{135}$Cs (see
Section~\ref{sec:radioneutroncapheavy} of Chapter 3 for model results). The half lifes
of both $^{134}$Cs and $^{135}$Cs have a strong theoretical temperature
dependence, decreasing by orders of magnitude in stellar conditions. 
Specifically for the long-living $^{135}$Cs, T$_{1/2}$ varies from 
terrestrial of 2 Myr down to $\simeq$ 200 yr at 300 MK, while its 
neutron-capture cross section has been
experimentally determined \cite[]{jaag97}. The branching point at
$^{135}$Cs can produce the unstable $^{136}$Cs, which is also a
branching point producing the unstable $^{137}$Cs. With a constant half
life of $\simeq$ 30 yr, this is also a branching point producing the
unstable $^{138}$Cs, which quickly decays into stable $^{138}$Ba.}

\item[{\bf $^{141}$Ce}]{This branching point may lead to production of 
the neutron-rich $^{142}$Ce, thus skipping the $s$-only $^{142}$Nd and 
affecting the Nd isotopic ratios, which are measured in SiC stardust 
grains \cite[]{gallino97}.}
 
\item[{\bf $^{142,143}$Pr}]{The branching point at $^{142}$Pr is affected 
by the temperature dependence of the $\beta^{-}$ 
half life of $^{142}$Pr, which increases to $\simeq$ 4 days at 300 MK from 
the terrestrial 19 hours. The neutron-capture branch may produce the 
unstable $^{143}$Pr, which is also a branching point producing the 
unstable $^{144}$Pr, which quickly decays into $^{144}$Nd. The operation 
of this chain of branching points may affect the isotopic composition of Nd 
because $^{142}$Nd and $^{143}$Nd are skipped by the neutron-capture flux 
and their abundances are decreased.}

\item[{\bf $^{147}$Nd}]{This branching point may lead to the production of 
the neutron-rich ``$r$-only'' $^{148}$Nd, which is of interest in relation 
to stellar SiC grain Nd data \cite[]{gallino97}.}
 
\item[{\bf $^{147,148}$Pm}]{The branching point at $^{147}$Pm is
affected by the strong temperature dependence of the $\beta^{-}$ decay
of this nucleus, where the half life decreases from the terrestrial
value of 2.6 yr down to $\simeq$ 1 yr at 300 MK. The neutron-capture
cross section of this nucleus is experimentally determined
\cite[]{reifarth03}. When the branching is open, it produces the unstable
$^{148}$Pm, a branching point that may lead to production of $^{149}$Pm,
which quickly decays into $^{149}$Sm. The operation of this chain of
branching points affects the isotopic composition of Sm, by skipping
$^{147}$Sm and the $s$-only $^{148}$Sm. This is of interest in relation
to stellar SiC grain Sm data \cite[]{gallino97}.}

\item[{\bf $^{151}$Sm}]{The operation of this branching point is affected 
by the temperature dependence of the $\beta^{-}$ decay rate of $^{151}$Sm, 
where the half life of this nucleus decreases from 93 yr to $\simeq$ 3 yr 
at 300 MK. Its operation changes the $^{153}$Eu/$^{151}$Eu ratio, which 
can be measured in stars (Section~\ref{subsec:branchstars} of Chapter 3) and in SiC 
stardust grains (Section~\ref{subsec:branchSiC} of Chapter 3). Note that 
$^{151}$Sm is one of few radioactive nuclei acting as 
branching points on the $s$-process path for which an experimental 
determination of the neutron capture cross section is available 
\cite[]{abbondanno04,wisshak06}.}

\item[{\bf $^{153}$Sm}]{This branching point can produce the neutron-rich 
$^{154}$Sm and affect the $^{153}$Eu/$^{151}$Eu ratio.}

\item[{\bf $^{152}$Eu}]{This nucleus suffers both $\beta^{-}$ and
$\beta^{+}$ decays, with rates showing a strong temperature dependence
covering several orders of magnitude variation in stellar conditions.
The $\beta^{+}$ decay rate also has a strong dependence on density.  The
operation of this branching point, in combination with that at $^{151}$Sm,
makes possible the 
production of the rare $p$-only isotope $^{152}$Gd by the $s$-process.}

\item[{\bf $^{154,155}$Eu}]{The decay rate of $^{154}$Eu has a strong 
temperature dependence, with its half life decreasing from 8.8 yr down to 
$\simeq$11 days at 300 MK. If activated, it leads to production of the 
unstable $^{155}$Eu, a branching point also with a temperature dependence, and 
an experimentally determined neutron-capture cross section \cite[]{jaag95},
which may produce $^{156}$Eu, which quickly decays into $^{156}$Gd. The 
operation of this chain of branching points affects the isotopic 
composition of Gd, which is a refractory element present in stellar SiC 
grains \cite[]{yin06}.}

\item[{\bf $^{153}$Gd}]{This nucleus suffers $\beta^{+}$ decay with a 
temperature dependence, where the terrestrial half life of 239 days 
increases with increasing the 
temperature by up to an order of magnitude in AGB stars conditions. 
The operation of this branching point may 
affect the $^{153}$Eu/$^{151}$Eu ratio.}

\item[{\bf $^{163}$Dy} and {\bf $^{163,164}$Ho}]{The nucleus $^{163}$Dy is 
stable in terrestrial conditions, but it can become unstable inside stars: 
at 300 MK the half life of this isotope becomes $\simeq$ 18 days. Thus, a 
branching can open on the $s$-process path, leading to the production of 
the unstable $^{163}$Ho via $\beta^{-}$ decay of $^{163}$Dy. In this 
conditions, the $\beta^{+}$ half life of $^{163}$Ho (which also has a 
strong temperature and density dependence) is $\simeq$ 12 yr, so another 
branching can open on the $s$-process neutron capture path. Neutron 
captures on $^{163}$Ho lead to production of the unstable $^{164}$Ho, 
which has fast $\beta^{-}$ and $\beta^{+}$ channels, both temperature 
dependent. The $\beta^{-}$ channel can eventually lead to the production 
of $^{164}$Er, a $p$-only nucleus, which may thus have a $s$-process 
component in its cosmic abundance.}

\item[{\bf $^{169}$Er}]{This branching point may lead to the production of 
the neutron-rich $^{170}$Er.}

\item[{\bf $^{170,171}$Tm}]{The branching point at $^{170}$Tm may produce 
the unstable $^{171}$Tm, which is also a branching point (with a 
temperature dependence) producing the unstable $^{172}$Tm, which quickly 
decays into $^{172}$Yb. By skipping $^{171,172}$Yb during the $s$-process 
flux, these branching points affect the isotopic composition of Yb, which 
is a refractory element present in meteoritic stellar SiC grains 
\cite[]{yin06}.}
 
\item[{\bf $^{176}$Lu}]{A branching point at $^{176}$Lu is activated 
because of the production of the short-living (half life of $\simeq$ 4 
hours) isomeric state of $^{176}$Lu via neutron captures on $^{175}$Lu. 
The situation is further complicated because, at around 300 MK, the 
isomeric and the ground state of $^{176}$Lu are connected via the thermal 
population of nuclear states that can act as mediators between the two. 
Hence, the half life of the $^{176}$Lu system can decrease at such 
temperatures by orders of magnitude. This branching point is of importance 
for the production of the very long-living ground state of $^{176}$Lu 
(half life of 380 Gyr) and of the stable $^{176}$Hf, which are both 
$s$-only isotopes, shielded by $^{176}$Yb against $r$-process production. 
Hence, the relative solar abundances of these two isotopes need to be 
matched by $s$-process in AGB stars. For details and models see 
\cite{heil08,mohr09}.}

\item[{\bf $^{177}$Lu}]{This branching point may lead to production of the 
unstable $^{178}$Lu, which quickly decays into $^{178}$Hf, thus decreasing 
the abundance of $^{177}$Hf.}

\item[{\bf $^{179}$Hf}, {\bf $^{179,180}$Ta}]{A branching point at 
$^{179}$Hf may be activated on the $s$-process path because this stable 
nucleus becomes unstable in stellar conditions (as in the case of 
$^{163}$Dy) with a $\beta^{-}$ half life of $\simeq$ 40 yr at 300 MK. This 
may allow the production of the unstable $^{179}$Ta, which is also a 
branching point with a temperature-dependent $\beta^{+}$ decay rate, which 
may lead to the production of $^{180}$Ta, the least abundant nucleus in 
the solar system \cite[]{kaeppeler04}, as a few percent of neutron captures on 
$^{179}$Ta lead to production of the very long-living isomeric state of 
$^{180}$Ta, instead of the ground state, which suffers fast $\beta^{+}$ 
and $\beta^{-}$ decays. As in the case of $^{176}$Lu, the ground and the 
isomeric states of $^{180}$Ta can be connected via the thermal population 
of nuclear states that act as mediators between the two. It is still 
unclear if the cosmic abundance of $^{180}$Ta is to be ascribed to the $s$-process or to nucleosynthetic 
processes in supernovae connected to 
neutrino fluxes.}

\item[{\bf $^{181}$Hf}]{This branching point may lead to production of
the long-living radioactive nucleus $^{182}$Hf \cite[one of the few
radioactive isotopes with an experimentally determined neutron-capture
cross section available,][]{vockenhuber07} whose decay into $^{182}$W is
of extreme importance for early solar system datation. The half life of
$^{181}$Hf is relatively long in terrestrial conditions (42 days), but
decreases down to 2 days at 300 MK, when the
$^{22}$Ne($\alpha$,n)$^{25}$Mg reaction is activated, so the actual
production of $^{182}$Hf in AGB stars is predicted to be 
relatively marginal (see also
Section~\ref{sec:radioneutroncapheavy} of Chapter 3).}

\item[{\bf $^{182,183}$Ta}]{The branching point at $^{182}$Ta 
is temperature dependent 
and may produce the unstable $^{183}$Ta, also a branching point, producing 
$^{184}$Ta, which quickly decays into the stable $^{184}$W. These 
branching points may affect the isotopic composition of W, which is a 
refractory element that is present in stellar SiC grains \citep{avila08}.}

\item[{\bf $^{185}$W}]{This branching point may produce $^{186}$W, and 
affect the isotopic composition of W as well as the $^{186}$Os/$^{188}$Os 
ratio. Its signature may be seen in data from stellar SiC grains for 
W and Os \cite[]{humayun07}. Note that $^{185}$W is one of few  
radioactive nuclei acting as branching points on the 
$s$-process path for which an experimental determination of the neutron 
capture cross section is available, even thought only via 
indirect ($\gamma, n$) studies, which have rather large
uncertainties of about 30\% \cite[]{sonnabend03,mohr04}.}

\item[{\bf $^{186}$Re}]{This isotope decays in $\simeq$ 89 hours, and
has both $\beta^{-}$ and $\beta^{+}$ decay channel. The $\beta^{-}$ decay
channel is 
faster by one to two orders of magnitude depending on the temperature,
which also affects the $\beta^{+}$ decay rate. This branching point can
affect the production of $^{186}$Os, $^{186}$W, and the very long-living
$^{187}$Re, whose slow decay into $^{187}$Os is used as a cosmological
clock (see discussion in Chapter 2).}

\item[{\bf $^{191}$Os}]{This branching point has a mild temperature dependence 
whereby the half life of $^{191}$Os decreases with the temperature from 
the terrestrial 15 days to $\simeq$8 days at 300 MK. If activated, the 
neutron-capture branch can decrease the $s$-process abundances of 
$^{191}$Ir and $^{192}$Pt and lead to production of $^{192}$Os, thus 
affecting the isotopic composition of Os, which is measured in meteoritic 
materials \cite[]{brandon05}, and $^{193}$Ir.}

\item[{\bf $^{192}$Ir}]{This branching point can produce $^{193}$Ir, and 
affect the $s$-process production of the rare proton-rich $^{192}$Pt. A 
few percent of the decay rate of $^{192}$Ir is made by $\beta^{+}$ 
decays.}

\item[{\bf $^{193}$Pt}]{This isotope decays $\beta^{+}$ with a half life 
of $\simeq$ 50 yr, which may affect the production of $^{193}$Ir.}

\item[{\bf $^{204}$Tl}]{This branching point has a strong temperature 
dependence with its half life decreasing from the terrestrial value of 
$\simeq$ 3.8 yr to $\simeq$ 7 days at 300 MK, leading to production of the 
$s$-only $^{204}$Pb.}

\item[{\bf $^{205}$Pb}]{This nucleus is long-living in terrestrial
conditions (T$_{1/2}$ = 15 Myr), but its half life against electron
captures has a strong temperature and density dependence, which affects
its survival in stellar environments, as in the case of $^{41}$Ca.  Its
production during the $s$-process is discussed in detail in
Section~\ref{sec:radioneutroncapheavy} of Chapter 3. Its radiogenic decay is responsible
for production of $^{105}$Tl.}

\item[{\bf $^{210}$Bi}]{This temperature-dependent branching point may 
lead to production of the unstable $^{211}$Bi, which $\alpha$ decays into 
$^{207}$Tl, which quickly decays $\beta^{+}$ into $^{207}$Pb.}

\item[{\bf $^{210}$Po}]{May produce $^{211}$Po, which quickly $\alpha$ 
decays into $^{207}$Pb. The $\alpha$ decay of $^{210}$Po, and $^{211}$Bi 
above, represent the chain of reactions that terminates the $s$-process 
\cite[]{clayton67,ratzel04}.}

\end{description}

To complete the picture we list nuclei that could be classified as 
potential $s$-process branching points, given that their terrestrial half 
life is greater than a few days, however, they do not open during the $s$-process because their half life 
decreases with the temperature to below a 
few days. These are: $^{103}$Ru, $^{123}$Sn, $^{124}$Sb, $^{156}$Eu, 
$^{160,161}$Tb, $^{175}$Yb, $^{198}$Au, and $^{205}$Hg. Finally, 
we point out the special case of $^{157}$Gd, a stable nucleus which 
becomes unstable at stellar temperatures, but not enough to open a 
branching point on the $s$-process path in AGB stars.

	\bibliographystyle{spbasic}


\end{document}